\documentclass[times,review,preprint,2022]{elsarticle}
\usepackage[margin=2.5cm]{geometry}

\usepackage{framed,multirow}

\usepackage{amssymb}
\usepackage{latexsym}

\usepackage{url}
\usepackage{xcolor}
\definecolor{newcolor}{rgb}{.8,.349,.1}

\usepackage{graphicx,amsmath,amsfonts}
\usepackage{subfigure}
\usepackage{pgfplots}
\usepackage{multicol}
\usepackage{color}
\usepackage{hyperref}
\usepackage{wasysym}
\usepackage{multirow}
\usepackage{hypernat}
\usepackage{hyperref}
\usepackage{setspace}
\hypersetup{
    pdfborder={0 0 0},
    citecolor=black,
    filecolor=black,
    linkcolor=black,
    urlcolor=black
}
\allowdisplaybreaks

\numberwithin{equation}{section}

\newcommand{\indx}[1]{\textrm{#1}}
\newcommand{\pd}[2]{\frac{\partial #1}{\partial #2}}

\newcommand{\abs}[1]{\lvert#1\rvert}

\newcommand{\eq}[1]{Eq.~(\ref{eq:#1})}

\newcommand{\fig}[1]{Fig.~\ref{fig:#1}}
\newcommand{\tab}[1]{Table~\ref{tab:#1}}
\newcommand{\sect}[1]{Section~\ref{sec:#1}}

\newcommand{\overbar}[1]{\mkern 1.5mu\overline{\mkern-1.5mu#1 \vphantom{b} \mkern-1.5mu}\mkern 1.5mu }
\newcommand*{\rom}[1]{\uppercase\expandafter{\romannumeral #1\relax}}
\newcommand{\tabitem}{~~\hspace{0.2in}\llap{-}~~}

\def\fign{2.9in}

\def\fighalf{2in}

\usetikzlibrary{arrows,decorations.markings}
\tikzstyle{solid}=                   [dash pattern=]
\tikzstyle{dotted}=                  [dash pattern=on \pgflinewidth off 2pt]
\tikzstyle{densely dotted}=          [dash pattern=on \pgflinewidth off 1pt]
\tikzstyle{loosely dotted}=          [dash pattern=on \pgflinewidth off 4pt]
\tikzstyle{dashed}=                  [dash pattern=on 3pt off 3pt]
\tikzstyle{densely dashed}=          [dash pattern=on 3pt off 0.6pt]
\tikzstyle{loosely dashed}=          [dash pattern=on 3pt off 6pt]
\tikzstyle{dashdotted}=              [dash pattern=on 3pt off 2pt on \the\pgflinewidth off 2pt]
\tikzstyle{densely dashdotted}=      [dash pattern=on 3pt off 1pt on \the\pgflinewidth off 1pt]
\tikzstyle{loosely dashdotted}=      [dash pattern=on 3pt off 4pt on \the\pgflinewidth off 4pt]

\begin{document}


\begin{frontmatter}

\title{A seven-equation diffused interface method for resolved multiphase flows}

\author[1]{Achyut Panchal}
\author[2]{Spencer H.  Bryngelson}
\author[1]{Suresh Menon\corref{cor1}}
\cortext[cor1]{Corresponding author: suresh.menon@ae.gatech.edu}

\address[1]{School of Aerospace Engineering,\\
Georgia Institute of Technology, \\
Atlanta, GA, 30332.}
\address[2]{School of Computational Science and Engineering,\\
Georgia Institute of Technology, \\
Atlanta, GA, 30332.}

\begin{abstract}
    The seven-equation model is a compressible multiphase formulation that allows for phasic velocity and pressure disequilibrium.
    These equations are solved using a diffused interface method that models resolved multiphase flows.
    Novel extensions are proposed for including the effects of surface tension, viscosity, multi-species, and reactions.
    The allowed non-equilibrium of pressure in the seven-equation model provides numerical stability in strong shocks and allows for arbitrary and independent equations of states.
    A discrete equations method (DEM) models the fluxes.
    We show that even though stiff pressure- and velocity-relaxation solvers have been used, they are not needed for the DEM because the non-conservative fluxes are accurately modeled.
    An interface compression scheme controls the numerical diffusion of the interface, and its effects on the solution are discussed.
    Test cases are used to validate the computational method and demonstrate its applicability.
    They include multiphase shock tubes, shock propagation through a material interface, a surface-tension-driven oscillating droplet, an accelerating droplet in a viscous medium, and shock--detonation interacting with a deforming droplet.
    Simulation results are compared against exact solutions and experiments when possible.
\end{abstract}


\end{frontmatter}


\section{Introduction}

Multiphase flows are characterized by more than one phase, each with its own material properties.
Such flows are encountered in many situations, including liquid fuel-powered devices~\cite{lefebvre_1998, sutton_2006}, solid explosives~\cite{tarver_1996}, pharmaceutical sprays~\cite{vehring_2008}, and more.
The design and engineering of these applications require computational modeling.
From this perspective, there are two sets of methods: resolved approaches where the multiphase entities (MPE) are fully-resolved on the computational grid~\cite{gorokhovski_annrev_2008, saurel_annrev_2018}, and dispersed approaches where the MPEs are unresolved on the grid and instead treated as point particles~\cite{balachandar_annrev_2010, panchal_jcp_2021, bryngelson19_IJMF, bryngelson21}. 

This work focuses on resolved multiphase flow modeling, which can be further segmented into sharp- or diffused-interface methods (SIM or DIM).
The SIM naturally represents materials as infinitely sharp~\cite{sethian_annrev_2003,sussman_jcp_2007}.
Specialized techniques are used to advect the interface, including the level-set method (LS)~\cite{sethian_annrev_2003}, the volume of fluid approach (VOF)~\cite{hirt_jcp_1981}, and the coupled LS--VOF method (CLSVOF)~\cite{sussman_jcp_2000}.
Example methods for handling jumps at the interface are continuum surface forces (CSF)~\cite{brackbill_jcp_1992} and ghost fluid methods (GFM)~\cite{fedkiw_jcp_1999}.
SIM are particularly attractive for their ease in specifying jump conditions and efficient interface tracking.
However, they have historically been limited by low-speed flows, low density--viscosity ratios~\cite{elghobashi_annrev_2019}, and mass-conservation errors~\cite{sethian_annrev_2003}.
We note that some of these drawbacks are changing via bespoke methods~\cite{das_jcp_2020, terashima_jcp_2009, barton_jcp_2011}.

Diffused interface methods (DIM) are hallmarked by the Baer--Nunziato (BN) ``seven-equation'' formulation of a compressible and mixture conservative flow~\cite{baer1986,saurel_annrev_2018}.
The equations follow from an averaging procedure to the phasic governing flow equations.
The derived partial differential equation (PDE) solves for phasic volume fractions and phase-averaged quantities.
This formulation evolves the conserved quantities on the grid, usually via numerical methods that allow for sharp gradients~\cite{liu_jcp_1994, van_jcp_1979}.
As a result, the interface naturally diffuses across multiple grid cells due to numerical diffusion~\cite{saurel_annrev_2018}.
Besides their conservation properties, DIMs also allow for the creation or destruction of interfaces, common during supercavitation and flashing~\cite{petitpas_ijmf_2009, schmidmayer_cpm_2020,saurel_cf_2016, rodio_jcp_2015}.

The first numerical method for a single-component inviscid seven-equation model was developed by~\citet{saurel_jcp_1999}.
They used a two-step approach, where the two-phase equations were first solved using a hyperbolic solver, followed by a relaxation step to handle phase disequilibrium.
After this initial effort, reduced forms of the Baer--Nunziato equations were developed to relieve the otherwise stiff pressure- and velocity-relaxation procedures.
These methods are broadly used today.
For example, \citet{kapila_pof_2001} developed a reduced five-equation model with single velocity and pressure, and its use has been widespread~\cite{allaire_jcp_2002, meng_sw_2015, liu_ast_2019, murrone_jcp_2005, petitpas_jcp_2007}.
This is partly due to the simplicity of including additional physical phenomena like surface tension~\cite{perigaud_jcp_2005}, mass-transfer~\cite{rodio_jcp_2015}, viscous effects~\cite{abgrall_caf_2014}, flashing~\cite{saurel_jfm_2008}, multiple components~\cite{bryngelson_2021}, and cavitation~\cite{schmidmayer_jcp_2020}.

The drawback to the five-equation approach is that a single-pressure requires establishing a mixture EOS with a non-monotonic sound speed~\cite{wood_1930}.
This results in numerical instability near strong shocks, difficulty maintaining volume fraction-positivity, and inaccurate wave transmission through the interface~\cite{saurel_jcp_2009}.
The five-equation model by~\citet{allaire_jcp_2002} is relatively robust, but it cannot model the creation and destruction of the interface due to the absence of a volume fraction source term.
A six-equation model was developed to resolve this, which used a single velocity but phase-independent pressures~\cite{saurel_jcp_2009}.
Extensions of this work include phase-change effects~\cite{zein_jcp_2010, pelanti_ijmf_2019, demou_jcp_2022, rodriguez21}, viscous effects~\cite{dorschner_jfm_2020}, and arbitrary numbers of components~\cite{bryngelson_2021, coralic_jcp_2014}.

Herein, we use the more extensive seven-equation model.
The seven-equation model maintains phase-independent pressures so that it can represent with strong shocks and large density ratios: $\rho^{(2)}/\rho^{(1)}, p^{(2)}/p^{(1)} \gg 1$.
There is no need to formulate a mixture EOS, so one can use phase-independent, arbitrary EOSs.
It also retains phase-independent velocities, which allows representation of dispersed unresolved MPEs~\cite{panchal_jcp_2021}.
Considering this, the various features of SIM and DIM variants are summarized in~\tab{methods}.

After the initial development of the seven-equation model by~\citet{saurel_jcp_1999}, there have been relatively limited extensions and application use-cases~\cite{abgrall_jcp_2003, chang_jcp_2007, tokareva_jcp_2010, nguyen_amc_2015, zein_jcp_2010}.
This appears to be due to the complexity of formulating an efficient numerical method.
In this work, we take a step towards addressing this problem.
We start with the baseline numerical approach of~\citet{abgrall_jcp_2003} and provide novel extensions to develop a robust computational method for the seven-equation model that can handle strong shocks, arbitrary EOS, viscous effects, surface tension, multiple species, and reactions~\cite{panchal2022}.

Novel contributions of this work:
\begin{itemize}
    \item Various numerical methods have been developed for solving the BN PDE~\cite{saurel_jcp_1999, abgrall_jcp_2003, chang_jcp_2007, tokareva_jcp_2010, nguyen_amc_2015}.
    Still, the discrete equations method (DEM)~\cite{abgrall_jcp_2003} accurately models non-conservative terms at the interface.
    For 1D Riemann problems, DEM relieves the need to use a stiff pressure--velocity relaxation solver for pure fluid interfaces~\cite{abgrall_jcp_2003}.
    This effect is further studied in this work for more complex problems.

    \item Surface tension is required for modeling droplet breakup and deformation.
    Past works included it in the five-equation model~\cite{perigaud_jcp_2005} and a path-conservative framework for the seven-equation model~\cite{nguyen_amc_2015} via source terms.
    We include it within the DEM approach via appropriate modifications to the Riemann solver.

    \item Viscous effects are necessary for applications such as drag computation of a deforming droplet or droplets interacting with turbulence.
These have been previously considered in the DEM framework~\cite{abgrall_caf_2014} for the five-equation model.
This work extends it to the seven-equation model.

    \item The developed computational framework can deal with arbitrary and phase-independent EOS due to the pressure non-equilibrium, and this is demonstrated by using calorically perfect gas (CPG), thermally perfect gas (TPG), stiffened gas (SG), and Mie–Grüneisen (MG) EOS.

    \item Being able to handle multi-species mixtures for reactive flows, e.g., droplets in a detonation wave~\cite{dyson_aiaa_2022}.
    Each phase in this model includes an arbitrary number of species and reactions.

    \item The DIM interface can numerically diffuse across multiple cells.
Considering this, an interface compression scheme, initially developed for the five-equation model~\cite{shukla_jcp_2010} is modified to be used with the seven-equation model.
\end{itemize}

\begin{table}
    \caption{Various numerical methods for modeling resolved MPEs and their features. P: Possible; NP: Not possible;  PNA: Possible, not attempted. \textbf{Bold text} refers to cases considered in this work.}
    \begin{center}
        \def\arraystretch{1.2}
        \begin{tabular}{|p{0.89in}|p{0.75in}|p{0.75in}|p{0.75in}|p{0.75in}|}
            \hline
            & SIM & 5-eq DIM & 6-eq DIM & 7-eq DIM \\\hline
            $\rho^{(2)}/\rho^{(1)}$ $\gg 1,\quad$ $p^{(2)}/p^{(1)}$ $\gg 1$ & Limited \hspace{0.3in}\cite{das_jcp_2020, terashima_jcp_2009, barton_jcp_2011, dodd_jfm_2016}  & Limited robustness~\cite{kapila_pof_2001}  & P~\cite{saurel_jcp_2009} & \textbf{P}~\cite{saurel_jcp_1999,  abgrall_jcp_2003} \\\hline
            Arbitrary EOS & P~\cite{fedkiw_jcp_1999} & NP & PNA & \textbf{PNA}\\\hline
            Viscous & P~\cite{dodd_jfm_2016} & P~\cite{abgrall_caf_2014} & P~\cite{bryngelson_2021} & \textbf{PNA} \\\hline
            Surface tension & P~\cite{fechter_jcp_2017} & P~\cite{perigaud_jcp_2005} & P~\cite{schmidmayer_jcp_2017} & \textbf{P}~\cite{nguyen_amc_2015} \\\hline
            Flashing / trans-critical & NP & P~\cite{saurel_jfm_2008, schmidmayer_jcp_2020} & P~\cite{zein_jcp_2010, pelanti_jcp_2014, pelanti_ijmf_2019}  & P~\cite{zein_jcp_2010}  \\\hline
            Evaporation & P~\cite{tanguy_jcp_2007, das_jcp_2020} & P~\cite{rodio_jcp_2015} & P~\cite{zein_jcp_2010, pelanti_jcp_2014, pelanti_ijmf_2019} & P~\cite{zein_jcp_2010} \\\hline
            Dispersed & NP & NP & NP & P~\cite{saurel_jcp_1999} \\\hline
            Multi-species / reaction & P~\cite{das_jcp_2020} & PNA & PNA & \textbf{PNA} \\\hline
            Interface Compression & Not needed & Used~\cite{shukla_jcp_2010, jain_jcp_2020} & Used~\cite{chiapolino_jcp_2017sharpening} & \textbf{PNA} \\\hline
        \end{tabular}
    \end{center}
    \label{tab:methods}
\end{table}

Herein, \sect{eqn}~describes the seven-equation mathematical model, \sect{num}~formulates the numerical method, and \sect{results}~shows verification and validation cases.

\section{Mathematical Formulation}
\label{sec:eqn}

\subsection{Fine-grained description}

We start from a fine-grained description of a multiphase flow before BN averaging.
An indicator function for phase $\kappa=1,2$ is defined as $I^{\kappa} (x_\indx{i},t)$, which is $1$ if the location $x_{\indx{i}}$ is in phase $\kappa$ at time $t$, and $0$ otherwise.
An evolution equation for $I^\kappa$ is
\begin{equation}
     \pd{I^{\kappa}}{t} + u^{I}_{\indx{i}} \pd{I^{\kappa}}{x_\indx{i}} = 0 \quad (\indx{i}=1,2,3),
     \label{eq:indicator}
\end{equation}
where $u^I_{\indx{i}}$ is the interface velocity. 
The single-phase Navier--Stokes equations accompany these for each phase as
\begin{equation}
    \pd{U}{t} + \pd{F_{\indx{i},inv}}{x_{\indx{i}}} +\pd{F_{\indx{i},visc}}{x_{\indx{i}}} = \dot{S}_{\kappa} \quad ({\indx{i}}=1,2,3),
    \label{eq:pdesharp1}
\end{equation}
where, 
\begin{equation}
    U = \begin{pmatrix}1, \\
    \rho,  \\
    \rho u_{\indx{j}},  \\
    \rho E,  \\
    \rho Y_{\indx{m}}
    \end{pmatrix}, \,
    F_{{i},inv}= \begin{pmatrix} 0, \\
    \rho u_{\indx{i}}, \\
    \rho  u_{\indx{i}} u_{\indx{j}} , \\
    \rho E u_{\indx{i}} ,\\
    \rho Y_{\indx{m}} u_{\indx{i}} 
    \end{pmatrix} + 
    \begin{pmatrix} 0, \\
    0,  \\
    p \delta_{\indx{ij}} , \\
    p u_{\indx{i}} , \\
    0
    \end{pmatrix}, \,
    F_{{i},visc} = \begin{pmatrix} 0, \\
    0,  \\
    \tau_{\indx{i}\indx{j}},  \\
    \tau_{\indx{i}\indx{j}} u_{\indx{j}} + q_{\indx{i}},  \\
    \rho Y_{\indx{m}} v_{\indx{im}}
    \end{pmatrix}.
    \label{eq:finegrained_flux}
\end{equation}
where $U$, $F_{\indx{i},inv}$, and $F_{\indx{i},visc}$ are solution variables, inviscid fluxes and viscous fluxes, respectively.
The total flux is $F_{\indx{i}} = F_{\indx{i},inv} + F_{\indx{i},visc}$, and $\rho$, $u_{\indx{j}},\ (\indx{j}=1,2,3)$, $p$, and $Y_{\indx{m}},\ (\indx{m}=1,\dots,N_{s,\kappa})$ represent density, velocity, pressure, and mass-fractions for $\indx{m}^\textrm{th}$ species, respectively.
Note that $U(0) = 1$ is redundant but corresponds to a volume fraction evolution equation after the averaging operation.
 We will show this later.
The number of species that are tracked, $N^{\kappa}_{s}$, can be different for different phases.

The total energy $E$ is related to the internal energy $e = h_{0,mix} + C_{v,mix} T$ as $E = e + 1/2\ u_{\indx{i}} u_{\indx{i}}$, where $C_{v,mix}$ is the mixture constant volume heat capacity and $h_{0,mix}$ is the mixture enthalpy of formation per mass.
Pressure $p$ is computed as $P^\kappa (\rho, e, Y_{\indx{m}})$ via an EOS.
(EOS)  The mixture formation enthalpy per mass is computed as $h_{0,mix} = \sum_{m=1}^{N_s^{\kappa}} h_{0,\indx{m}}^{\kappa} Y_{\indx{m}}$ with $h_{0,\indx{m}}^{\kappa}$ being the formation enthalpy per mass of ${\indx{m}}^{\textrm{th}}$ species.
The mixture heat capacities are computed as 
\begin{gather}
    C_{v,mix} = \sum_{\indx{m}=1}^{N_s^{\kappa}} C_{v,\indx{m}}^{\kappa} Y_{\indx{m}} 
    \quad \text{and} \quad 
    C_{p,mix} = \sum_{\indx{m}=1}^{N_s^{\kappa}} C_{p,\indx{m}}^{\kappa} Y_{\indx{m}}.
\end{gather}
The mixture specific enthalpy is $h_{mix} = h_{0,mix} + C_{p,mix}T$.
The species-specific heat capacities are $C_{p,\indx{m}}^{\kappa}$ and $C_{v,\indx{m}}^{\kappa}$, and they can be curve-fits of the local temperature if needed.
Both phases have independent EOS that determine $P^\kappa$ and thermodynamic properties $h_{0,\indx{m}}^{\kappa}$, $C_{p,\indx{m}}^{\kappa}$, $C_{v,\indx{m}}^{\kappa}$.

The viscous stress term $\tau_{\indx{i}\indx{j}}$ is computed from the shear stress tensor
\begin{gather}
    \mathcal{T}^{\kappa}(u_{\indx{i}})=\mu^{\kappa} \ {\partial u_{\indx{i}}}/{\partial x_{j}} -2/3\mu^{\kappa}\  \delta_{\indx{i}\indx{j}} {\partial u_{\indx{k}}}/{\partial x_{\indx{k}}}.
\end{gather}
The heat flux is modeled as 
\begin{gather}
    q_{\indx{i}} = \lambda^{\kappa} \ {\partial T}/{\partial x_{\indx{i}}} + \sum_{\indx{m}=1}^{N_s^{\kappa}} h_{\indx{m}}^{\kappa} Y_{\indx{m}} v_{\indx{im}}
\end{gather}
where $v_{\indx{im}}$ are species diffusive velocities, and $\mu^{\kappa}$ and $\lambda^{\kappa}$ are viscosity and thermal conductivity.
Similar to the thermodynamic properties, the transport properties $\mathcal{T}^{\kappa}$, $\mu^{\kappa}$, $\lambda^{\kappa}$, and species diffusivities are independent of the phases.
The same form of the shear stress tensor $\mathcal{T}^{\kappa}$ is used for both the phases (corresponding to a Newtonian fluid). 
However, a different form can also be used, for example, to model a solid material.

The source terms $\dot{S}$ are $(0,0,g_{\indx{i}}, g_{\indx{i}} u_{\indx{i}}, \dot{\omega}_{\indx{m}})^{\text{T}}$, where $g_{\indx{i}}$ represents gravity and $\dot{\omega}_{\indx{m}}^{\kappa}$ are the chemical reaction rates.
We do not consider gravity hereon but include it for completeness.
In the presence of chemical reactions, we also have the relations 
\begin{gather}
    \sum_\indx{\indx{m}} \dot{\omega}_{\indx{m}}^{\kappa} = 0 
    \quad \text{and} \quad 
    \sum_\indx{\indx{m}} Y_{\indx{m}}v_{\indx{mi}} = 0.
\end{gather}
The kinetics that models the reaction rates $\dot{\omega}_{\indx{m}}^{\kappa}$ are also different and independent between the phases.

These equations are complemented by the jump conditions across the phases as
\begin{equation}
    \left[ (F_{\indx{i}} - u^{I}_{\indx{i}} U) \right] = \mathcal{J}^{U}_{\indx{i}} \quad (\indx{i}=1,2,3),
    \label{eq:jump}
\end{equation}
where 
\begin{equation}
    \mathcal{J}^{U}_{\indx{i}} = \begin{pmatrix} 0, \\
        0, \\
        \sigma \beta n_{\indx{i}} n_{\indx{j}},  \\
        \sigma \beta u^{I}_{\indx{i}}, \\
        0 
    \end{pmatrix},
    \label{eq:jump2}
\end{equation}
and $[f]$ is the jump across the interface for any arbitrary field variable $f$, e.g., $f^{(2)} - f^{(1)}$.
Here, since surface tension is considered, $n_{i}$ is the interface surface normal pointing from the liquid ($\kappa=2$) to the gas ($\kappa=1$) phase, and $\sigma$ and $\beta = \partial n_{\indx{i}} / \partial x_{\indx{i}}$ are the surface tension and the interface curvature, respectively.
These jump conditions provide boundary conditions at the phase interfaces.

In the absence of any mass transfer across the phase interface and the presence of a thermal equilibrium due to heat transfer, these jump conditions transition into
\begin{equation}
    [u_{\indx{i}}]=0, \quad [p] = \sigma \beta - \left[ \tau_{\indx{ij}} n_{\indx{i}} n_{\indx{j}} \right], \quad [T] = 0.
    \label{eq:jump3}
\end{equation}
One would also maintain a Gibbs free energy equilibrium across the interface if mass transfer across the phase interface had been considered~\cite{zein_jcp_2010, pelanti_jcp_2014, demou_jcp_2022}.
In the absence of mass-transfer  the interface velocity $u^I_{\indx{i}}$ is $u_{\indx{i}}$ because $[u_{\indx{i}}] = 0$.

\subsection{Averaged equations}

To derive the coarse-grained Baer--Nunziato formulation from the fine-grained description, an averaging operator $\langle \cdot \rangle$ is applied.
The coarse-grained quantities are defined as $\overbar{\alpha}^{\kappa} = \langle I^{\kappa} \rangle$ (volume fraction) and $\overbar{f}^{\kappa} \overbar{\alpha}^{\kappa} = \langle I^{\kappa} f \rangle$ (phase-averaged quantities).
Here the operator $\overline{\cdot}$ is spatial filtering as a result of the phase-averaging. However, since both the flow features and the interface are considered resolved in this work, $\overbar{f}^{\kappa}$ is represented as $f^{\kappa}$ from here onward.
The generalized seven-equation formulation is derived from~\eq{indicator} and~\eq{pdesharp1}, by applying $\langle \cdot \rangle$ as
\begin{equation}
    \begin{split}
        \underbrace{\pd{\alpha^{\kappa} {U}^{\kappa} }{t}}_{\textrm{\rom{1}}} + \underbrace{\pd{ {\alpha}^{\kappa}  {F}^{\kappa}_{\indx{i},inv} }{x_\indx{i}}}_{\textrm{\rom{2}}} + \underbrace{\pd{{\alpha}^{\kappa} {F}^{\kappa}_{\indx{i},visc}}{x_\indx{i}}}_{\textrm{\rom{3}}} = &\underbrace{\left\langle \left( F^{\kappa}_{\indx{i},inv} - u_{\indx{i}}^{\kappa} U^{\kappa} \right) \pd{I^{\kappa}}{x_\indx{i}} \right\rangle}_{\textrm{\rom{4}}} \\
        &+ \underbrace{\left\langle F^{\kappa}_{\indx{i},visc} \pd{I^{\kappa}}{x_\indx{i}} \right\rangle}_{\textrm{\rom{5}}} + \underbrace{\left\langle U^{\kappa} (u^{\kappa}_{\indx{i}} - u^{I}_{\indx{i}}) \pd{I^{\kappa}}{x_\indx{i}} \right\rangle}_{\textrm{\rom{6}}} + \underbrace{\alpha^{\kappa} \dot{S}^{\kappa}}_{\textrm{\rom{7}}}.
    \end{split}
    \label{eq:pdediffuse}
\end{equation}
Here, term \rom{1} is the time-derivative of flow-field variables, terms \rom{2} and \rom{3} are conservative flues, terms \rom{4} and \rom{5} are non-conservative terms that are active only at the multiphase interfaces, and they are responsible for interphase exchanges. 
Term \rom{6} is mass transfer across the phase boundary, and they are neglected here but will be a part of future work.
Term \rom{7} are source terms. 

Regardless of whether the size of the MPE is larger or smaller than the filter-size of the averaging operator $\langle \cdot \rangle$ (implicitly same as the finite-volume grid), the same~\eq{pdediffuse} can be used for modeling either dispersed or resolved multiphase flows.
Numerical modeling of terms \rom{1}, \rom{2}, \rom{3}, and \rom{7} is straightforward and they remain the same for either resolved or dispersed flows.
\rom{4} and \rom{5} are unclosed, and their modeling requires a specific focus.
In the absence of mass transfer, these are typically modeled for the seven-equation model as~\cite{saurel_jcp_1999, nguyen_amc_2015, abgrall_caf_2014, zein_jcp_2010}
\begin{equation}
    \begin{gathered}
        \left( \rom{4} + \rom{5}\right) ^{(1)} = I^{(1)} + R^{(1)}_{p} +  R^{(1)}_{u} + R^{(1)}_{T},  \\
        I^{(1)} = \begin{pmatrix}
             -u_{\indx{i}}^{I} \frac{\partial \alpha^{(1)}}{\partial x_{\indx{j}}} , \\ 
             0, \\
             ( (p^{I} - \sigma \beta \delta^{(1)})\delta_{\indx{ij}} + \tau_{\indx{ij}}^{I}) \frac{\partial \alpha^{(1)}}{\partial x_{\indx{j}}}, \\
             \left( (p^{I}-\sigma \beta\delta^{(1)}) u_{\indx{i}}^{I} + \tau_{\indx{ij}}^{I} u^{I}_{\indx{i}} + q_{\indx{i}}^{I} \right) \frac{\partial \alpha^{(1)}}{\partial x_{\indx{i}}}, \\
              0  
        \end{pmatrix},  \\
        R^{(1)}_p = \begin{pmatrix}
             \theta^{p} \Delta p , \\ 
             0 \\
             0, \\
             (p^{I} - \sigma \beta \delta^{(1)}) \theta^{p} \Delta p, \\
             0  
        \end{pmatrix}, 
        R^{(1)}_{u} = \begin{pmatrix}
             0, \\ 
             0, \\
             \theta^{u} \Delta u_i, \\
             u_{\indx{i}}^{I} \theta^{u} \Delta u_{\indx{i}} \\
             0  
        \end{pmatrix}, 
        R^{(1)}_{T} = \begin{pmatrix}
             \frac{1}{\kappa^{I}}\theta^{T} \Delta T , \\ 
             0,\\
             0,\\
             \theta^{T} \Delta T,\\
             0  
        \end{pmatrix}
    \end{gathered}
    \label{eq:noncons_model1}
\end{equation}
where $\Delta p = p^{(1)} - p^{(2)} + \sigma \beta$, $\Delta u_i = u^{(1)}_i - u^{(2)}_i$, and $\Delta T = T^{(1)} - T^{(2)}$ are differences of pressure, velocity,  and temperature between the phases.
Furthermore,  $p^{I}$, $u_{{i}}^{I}$ (modeling discussed in~\cite{saurel_jcp_1999}), $\tau_{{ij}}^{I}$,  $q_{{i}}^{I}$  (modeling discussed in~\cite{abgrall_caf_2014}), and $\kappa^{I}$ (modeling discussed in~\cite{zein_jcp_2010}) are interface-averaged terms within the filter $\langle \cdot \rangle$, and since they are unknown, they typically have to be modeled as listed in the references listed.
For instance, $p^{I}$,$u^{I}$ have been modeled in the past as $p^{(1)}$ and $u^{(2)}$ (with $\kappa=1$ being the gas phase and $\kappa=2$ as the liquid phase) for gas-liquid flows.
In this work, however, the discrete equations method (DEM) is used~\cite{abgrall_jcp_2003}, and its use circumvents the need for this modeling by solving Riemann problems at each discrete interface instead.
A similar form as in~\eq{noncons_model1} is also used for the phase $(2)$, and it is not repeated here for brevity.
As noted before, assuming that the phase $(1)$ is a gas-phase the phase $(2)$ is a liquid-phase,  the surface tension force $\sigma \beta$ acts from the gas-phase towards the liquid-phase, and as a result, in~\eq{noncons_model1}, $\delta^{(1)}=1$ for the gas-phase, whereas $\delta^{(2)}=0$ for the liquid-phase~\cite{nguyen_amc_2015}.

The terms containing $\theta^{p}$, $\theta^{u}$, and $\theta^{T}$ are referred to as the relaxation terms for pressure, velocity,  and temperature, respectively.
The pressure relaxation results in volume exchange between the phases, the velocity relaxation results in interphase momentum exchanges, e.g.,
drag, and the temperature relaxation results in heat exchange between the phases.
For resolved interfaces,  these are typically modeled as $\theta^{p},  \theta^{u}, \theta^{T}, \theta^{f}\rightarrow \infty$, whereas, for dispersed phases, they are modeled using finite-time processes such as drag and heat transfer~\cite{saurel_jcp_1999}.
Enforcing this condition analytically would have resulted in a reduced five-/six-equation model ~\cite{kapila_pof_2001, saurel_jcp_2009}.
Past works~\cite{saurel_jcp_1999, zein_jcp_2010} had to use a stiff relaxation solver with the seven-equation model to enforce $\theta^{p},  \theta^{u} \rightarrow \infty$. However, the use of DEM circumvented this as it was shown for 1D canonical cases in~\cite{abgrall_jcp_2003}.
This is further confirmed for more complex test cases in this work (\sect{results} for more details).

\subsection{Material properties}

The governing equations are accompanied by an equation of state ($P^{\kappa}$) and material properties for each phase.
For demonstration, stiffened gas (SG, includes parameters $p_0^{\kappa}$ and $\gamma^{\kappa}$)~\cite{saurel_jcp_1999}, calorically perfect gas (CPG, has parameter $\gamma^{\kappa}$), thermally perfect gas (TPG, requires property curve-fits for individual species), and Mie–Grüneisen (MG, requires properties for individual species)~\cite{akiki_aiaa_2017} EOS are used in this work.
These and the corresponding parameters are well-described in the available literature, and their forms are not repeated here for brevity.
The corresponding parameters for various materials are described separately for each test as required in \sect{results}.

A Sutherland transport model is used for the gas-phase viscosity ($\mu^{(1)}$).
The liquid- viscosity ($\mu^{(2)}$) and surface tension $\sigma$ are assumed to be constants, although it is possible to use curve-fits for those as well.
The thermal conductivity $\lambda^{\kappa}$ is computed from the viscosity $\mu^{\kappa}$ using a constant Prandtl number ($Pr=0.739$) as $\lambda^{\kappa} = C_{p,mix}^{\kappa} \mu^{\kappa} / Pr$.
The species diffusion is modeled using a unity Lewis number ($Le$) assumption, with $Le^{\kappa}_{\indx{m}}=\lambda^{\kappa} / \left( \rho^{\kappa} D^{\kappa}_{\indx{m}} C_{p,\indx{m}}^{\kappa}\right) = 1$.
For the reactive simulations shown in this work,  following previous work on hydrogen-air detonations~\cite{salvadori_ijhe_2022}, a 7-step, 6-species chemical mechanism~\cite{baurle_jpp_1994} is used to model finite-rate kinetics.

\section{Numerical Method}
\label{sec:num}

A three-dimensional finite-volume computational framework is used to solve~\eq{pdediffuse}.
The equations are implemented in LESLIE, a parallel multi-physics solver, which has been used extensively in the past for multiphase reacting flow simulations~\cite{panchal_jcp_2021, patel_2008,balakrishnan_2010}.
The numerical approach here primarily follows published literature~\cite{abgrall_jcp_2003, saurel_jcp_1999, nguyen_amc_2015, shukla_jcp_2010, abgrall_caf_2014}, and every detail is not repeated for brevity.
However, the following novel contributions are made, and details about them are provided in the following sections.
\begin{itemize}
    \item Implementation of the DEM~\cite{abgrall_jcp_2003} for multi-block structured curvilinear grids.
    \item Modification in HLLC~\cite{toro_2009} Riemann solver within the DEM to include surface tension.
    \item Extension of viscous effects within the DEM~\cite{abgrall_caf_2014} for multiple dimensions.
    \item Improved robustness of the stiff relaxation solver~\cite{saurel_jcp_1999} for dealing with arbitrary EOS.
    \item Application of an interface compression scheme~\cite{shukla_jcp_2010} for the seven-equation model.
\end{itemize}

\subsection{Baseline approach}\label{sec:timeintegration}

A multi-block structured grid is used, wherein $\mathcal{C}_{ijk}$ is a computational cell with $i$, $j$, and $k$ as the indices in the three curvilinear directions.
The corresponding cell volume is denoted as $V_{ijk}$, and $S_{l,ijk}$ is used to indicate the cell faces.
Note that these $i,\ j,\ k$ are different from the i, j, k that we used in \sect{eqn} for the Einstein notation.
Let $(\xi, \eta, \zeta)$ are the computational directions of increasing $i$, $j$ and $k$, and $\partial \xi / \partial x_i$, $\partial \eta / \partial x_i$, $\partial \zeta / \partial x_i$ are the corresponding grid metrics.
Quantities computed at the face-centers are denoted with sub-scripts $i+1/2$, $j+1/2$ and $k+1/2$. For example, $S_{i+\frac{1}{2}jk}$ and ${\mathcal{N}}_{\indx{i},i+\frac{1}{2}jk}$ are the face-area and the face-normal, respectively, between the cells $\mathcal{C}_{i+1jk}$ and $\mathcal{C}_{ijk}$. 

A volume-integrated form of~\eq{pdediffuse} is numerically solved on this grid.
The cell-centered quantity to be updated at each time-step is $\left(\alpha^{\kappa}U^{\kappa}\right)_{ijk}$, and it is also referred to as $Q_{ijk}$. The corresponding time-update is denoted as $d\left(\alpha^{\kappa}U^{\kappa}\right)_{ijk}$ or $dQ_{ijk}$.  The increment $dQ$ (or $d\left(\alpha_{\kappa} U_{\kappa} \right)$) has contributions due to various terms in~\eq{pdediffuse} and they are denoted using subscripts as $dQ_{\textrm{\rom{2}}}, dQ_{\textrm{\rom{3}}}$, etc.

The computation is advanced in time using an explicit two-stage MacCormack approach.
The time-step $\Delta t$ is dynamically computed considering convective, acoustic, and diffusive propagation speeds of both the phases as
\begin{equation}
    \Delta t^{\kappa}_{ijk} = CFL \cdot {V}_{ijk} \bigg/ \left(\frac{1}{6} \sum_{l=1}^{6} {\mid v^{\kappa}_{l,ijk} \mid S_{{l,ijk}} + c^{\kappa} \overline{S}_{ijk} + \frac{2 \gamma^{\kappa} \nu^{\kappa}}{Pr}^{\kappa} \frac{{\overline{S}_{ijk}}^{2}}{V_{ijk}}}\right), \quad \Delta t = \underset{ijk}{\textrm{min}}\ \textrm{min} \left( \Delta t^{(1)}_{ijk}, \Delta t^{(2)}_{ijk} \right)
    \label{eq:dt}
\end{equation}
where, $S_{l,ijk}$ are face area and $v^{\kappa}_{l,ijk} = \sum_{\indx{i}=1}^{3} u^{\kappa}_{\indx{i},ijk}n_{\indx{i},ijk}$ is the normal velocity on the $l^{\textrm{th}}$ face of the cell $\mathcal{C}_{ijk}$.
The averaged face area for cell $\mathcal{C}_{ijk}$ is given by $\overline{S}_{ijk} = \frac{1}{6}\sum_{l=1}^{6} S_{l,ijk}$, and $\gamma^{\kappa}$, $\nu^{\kappa} = \mu^{\kappa}/\rho^{\kappa}$, and $Pr^{\kappa}$ are the heat capacity, the kinematic viscosity and the Prandtl number, respectively.
The $CFL$ is chosen as $0.5$ for stability.

\subsection{Inviscid fluxes}
\label{sec:inviscid}

In the finite-volume scheme, the inviscid increments due to the terms $\rom{2}$ and $\rom{4}$ in~\eq{pdediffuse} are provided as
\begin{equation}
     d \left( \alpha^{\kappa} U^{\kappa} \right)_{ijk,\rom{2}/\rom{4}} = d \left( \alpha^{\kappa} U^{\kappa} \right)_{ijk,\rom{2}/\rom{4}}^{i} + d \left( \alpha^{\kappa} U^{\kappa} \right)_{ijk,\rom{2}/\rom{4}}^{j} + d \left( \alpha^{\kappa} U^{\kappa} \right)_{ijk,\rom{2}/\rom{4}}^{k},
    \label{eq:inv_direction_cons}
\end{equation}
where the superscripts $i,j,k$ denote contributions from each direction. 
The fluxes in each direction are then computed using the DEM. 
Their computation is illustrated here only for the $i$-direction, but the same forms are used in the other directions as well.

\subsubsection{Discrete equations method (DEM)}

\begin{figure}
   \begin{center}
   \subfigure[Riemann sub-problems]{\includegraphics[width=\fign]{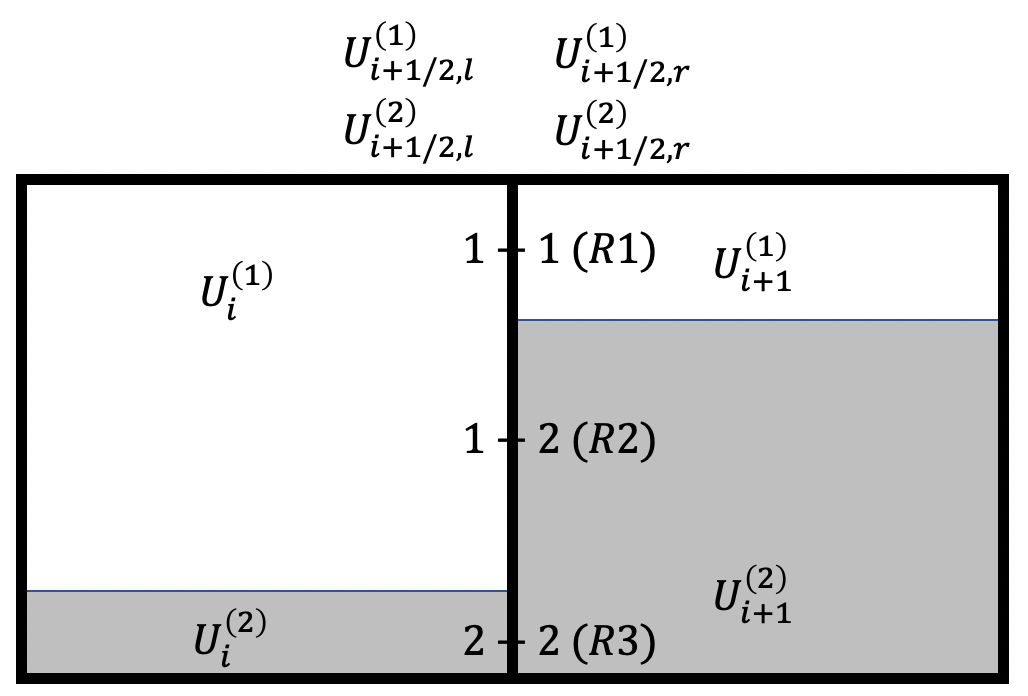}}
   \subfigure[Fluxes via DEM]{\includegraphics[width=\fign]{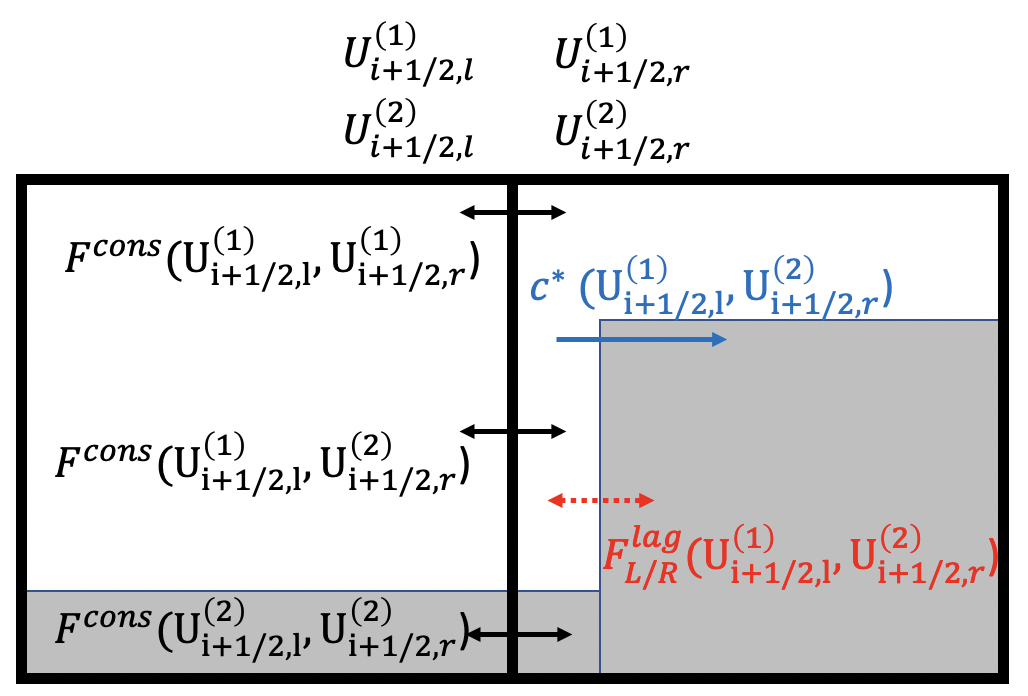}}
   \caption{Schematic showing the three Riemann problems (a) and the flux computation (b) with the DEM. Piecewise constant values are shown within the cell for illustration. However, the numerical scheme allows for second-order accuracy via piecewise linear interpolations. The Gray/white-colored region shows phases~(2)/(1).}
   \label{fig:demschematic}
  \end{center}
\end{figure}

Instead of directly solving for the conservative and the non-conservative terms, DEM divides the problem into three Riemann problems at each cell-face.
A schematic for this is shown in~\fig{demschematic}(a), where the three Riemann problems at each face are shown corresponding to three possibilities of the two-phase boundaries, i.e., (1)--(1) (R1),  (1)--(2) (R2) and (2)--(2) (R3).
Note that piece-wise constant fields are shown in each cell here for illustration. However, we go up to second-order accuracy with piece-wise linear fields for the numerical method implemented.
On each face, left- and right-interpolations are handled using a first/second-order MUSCL approach~\cite{van_jcp_1979}, and the interpolated values are denoted with subscripts $i+1/2,l$, $i+1/2,r$, etc.

\subsubsection{Riemann solution}

\begin{figure}
    \begin{center}
        \begin{tikzpicture}[scale=1]
            \draw [decoration={markings,mark=at position 1 with
                {\arrow[scale=2,>=stealth]{>}}},postaction={decorate}] (0,0) -- (6,0);
            \draw [decoration={markings,mark=at position 1 with
                {\arrow[scale=2,>=stealth]{>}}},postaction={decorate}] (3,0) -- (3,4);
            \draw[] (3,0) -- (1,3);\draw[] (3,0) -- (1,3.1);\draw[] (3,0) -- (1,3.2);\draw[] (3,0) -- (1,3.3);
            \draw[dashed] (3,0) -- (4,3);
            \draw[] (3,0) -- (6,3);\draw[] (3,0) -- (6,3.1);\draw[] (3,0) -- (6,3.2);\draw[] (3,0) -- (6,3.3);
            \draw           (5.8,2) node {$\mathcal{U}^R$};
            \draw           (0.5,2) node {$\mathcal{U}^L$};
            \draw           (4,3) node {$c^{*}(\mathcal{U}^L,\mathcal{U}^R)$};
            \draw           (4.3,2.1) node {$\mathcal{U}^{R*}$};
            \draw           (2.3,2.2) node {$\mathcal{U}^{L*}$};
            \draw           (3.8,3.5) node {Contact};
            \draw           (6.2,1.5) node {Shock/expansion};
            \draw           (0.0,1.5) node {Shock/expansion};
        \end{tikzpicture} 
    \end{center}
    \label{fig:riemann}
    \caption{A  sample Riemann problem is shown here, where $\mathcal{U}^L$ and $\mathcal{U}^R$ are the initial left- and right-states, and $\mathcal{U}^{L*}$, $\mathcal{U}^{R*}$ are the generated intermediate states. The waves on the extreme right/left could be a shock or an expansion wave (solid line), and the wave in the middle is a contact wave (dashed line), whose speed is denoted using $c^{*}$.}
\end{figure}
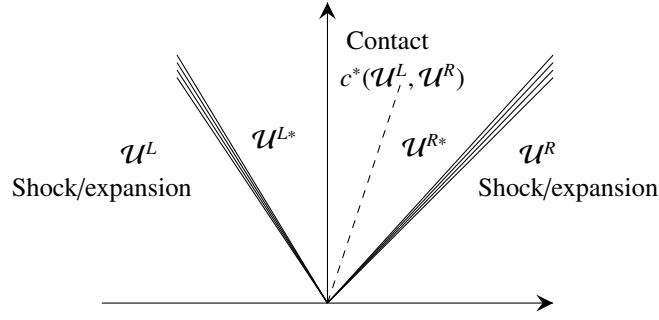

Solutions to the Riemann sub-problems (1)--(1), (1)--(2), and (2)--(2) are now needed.
An HLLC approximate Riemann solver~\cite{toro_2009} has been used traditionally~\cite{saurel_jcp_1999, abgrall_jcp_2003}.
However, modifications to it are required, including the surface tension.
Note that for a (1)--(2) Riemann problem,  a discontinuity in pressure would be maintained across the contact surface such as $p^{(2)} - p^{(1)} = p^{\sigma}$, whereas, the solution should stay unaffected for (1)--(1) or (2)--(2) problems.
A typical 1D Riemann problem is depicted in~\fig{riemann}, which could be either of the three sub-problems R1, R2, or R3.
The state vector $\mathcal{U} = (1,\rho, \rho u_{\indx{i}}, \allowbreak \rho E, \rho Y_{\indx{m}}  )^{T}$ is set equal to either $U^{(1)}$ or $U^{(2)}$ (on both the left and the right side of the face) depending on whether it belong to the phase (1) or~(2).
In the case where the left and the right state vectors $\mathcal{U}^{L}$ or $\mathcal{U}^{R}$ belong to different phases, i.e., (1)--(2) Riemann problem, the contact wave separates the phases as the problem evolves.

A sharp jump is assumed across both the left and the right shock/expansion waves in the HLLC approximate Riemann solver~\cite{toro_2009}, and these waves travel with the speeds $c^L$ and $c^R$, respectively. These are the acoustic propagation speeds corresponding to the state vectors $\mathcal{U}^L$ and $\mathcal{U}^R$, respectively. Considering this, we have the following relations across these jumps.
\begin{equation}
    \mathcal{F}^{K*} - \mathcal{F}^{K} = c^{K} (\mathcal{U}^{K*} - \mathcal{U}^K),\ K\in\{L,R\}, \quad \mathcal{F}^{R*} - \mathcal{F}^{L*} + \mathcal{F}^{\sigma}= c^{*} (\mathcal{U}^{R*} - \mathcal{U}^{L*}).
\end{equation}
where $\mathcal{F} = \left( 0, \rho v_n, \rho u_{\indx{i}} v_n + p n_{\indx{i}},  \rho E v_n + p v_n, \rho Y_{\indx{m}} v_n \right)$ is the flux vector, and that is defined for each state in the Riemann problem.
As noted before, $v_n = \sum_{\indx{i}=1}^{3} u_{\indx{i}} n_{\indx{i}}$ is the  velocity normal to the corresponding cell-face.
The term $\mathcal{F}^{\sigma} = \left(0,0,p^\sigma n_{\indx{i}}, p^\sigma c^{*}, 0 \right)$ is the flux due to the surface tension at the interface boundary (at the contact wave in this case). 

The pressures and the normal velocities across the interface (contact wave) are related as $p^{L*} = p^{R*} + p^{\sigma}, \quad v^{L*}_n = v^{R*}_n$, where, $p^{\sigma} = n^{LR}\sigma \beta $ is the pressure jump across the interface due to the surface tension.
The normal vector $n^{LR}$ is $1$ if the left phase is the liquid phase and the right phase is the gas phase, i.e., (2)--(1), $-1$ if the left phase is the gas phase and the right phase is the liquid-phase, i.e., (1)--(2), and $0$ if the phases on both the side of the interface are the same, i.e., (1)--(1) or (2)--(2).
Combining these equations and following the derivation in~\citet{toro_2009} with these additional modifications due to the surface tension force, the speed of the contact wave and the intermediate states are given as
\begin{equation}
    c^{*}\left( \mathcal{U}^{L}, \mathcal{U}^{R} \right) = \frac{p^{R} - p^{L} + p^{\sigma} + \rho^{L}v^{L}_n(c^{L} - v^{L}_n) - \rho^{R}v^{R}_n (c^{R} - v^{R}_n)}{ \rho^{L}(c^{L} - v^{L}_n) - \rho^{R}(c^{R} - v^{R}_n)}, \quad \mathcal{F}^{*}= \rho^{K} \left( \frac{c^{K} - v^{K}_n}{c^{K} - c^{*}} \right)
\end{equation}
\begin{equation}
    \mathcal{U}^{K*}\left( \mathcal{U}^{L}, \mathcal{U}^{R} \right) = \begin{pmatrix} 
    1,  \\
    \mathcal{F}^{*},  \\
    \mathcal{F}^{*} \left(u_{\indx{i}}^{K} +  (c^{*} - v^{K}_n) n_{\indx{i}}\right) , \\
    \mathcal{F}^{*}E^{K} + \mathcal{F}^{*}\left( c^{*} - v^{K}_n \right) \left( c^{*} + \frac{p^{K}}{\rho^{K} (c^{K} - v^{K}_n)} \right), \\
    \mathcal{F}^{*}Y_{\indx{m}}^{K}
    \end{pmatrix}.
\end{equation}
Once the intermediate states are known, the conservative Riemann flux at the cell face is computed as
\begin{equation}
    \begin{split}
        F^{cons}_{inv} \left( \mathcal{U}^{L}, \mathcal{U}^{R} \right) = 
        \begin{cases}
        \mathcal{F}(\mathcal{U}^{L}) \quad \textrm{if } 0 \le c^{L} \\
        \mathcal{F}(\mathcal{U}^{L*}) \quad \textrm{if }c^{L} \le 0 \le c^{*} \\
        \mathcal{F}(\mathcal{U}^{R*}) \quad \textrm{if }c^{*} \le 0 \le c^{R} \\ 
        \mathcal{F}(\mathcal{U}^{R}) \quad \textrm{if }c^{R} \le 0\\
        \end{cases}
    \end{split}
    \label{eq:consflux}
\end{equation}
Note that the conventional single-phase Riemann solution is recovered by letting $p^{\sigma} = 0$.

\subsubsection{Inviscid conservative flux (Terms $\rom{2}$)}

Based on the Riemann solutions that are constructed for the sub-problems (1)--(1), (1)--(2), and (2)--(2), we can now compute the inviscid conservative fluxes.
To illustrate this process with DEM, a schematic is shown in~\fig{demschematic}~(b).
For this example, during the Riemann problem evolution, the interface that was initially at $i+1/2$ has now moved within the cell on the right based on the speed of the contact wave $c^{*}$.
$F^{cons}_{inv}$ are the conservative fluxes for the Riemann problems~(1)--(1), (1)--(2), and (2)--(2) that act at the $i+1/2$ cell-face.

Considering the solutions from the three Riemann problems, the averaged i-directional conservative fluxes (term \rom{2} in~\eq{pdediffuse}) at face $i+1/2$ for phase~(1) are given using the DEM as 
\begin{equation}
    \begin{split}
    \left\langle I^{(1)} \left( F^{(1)}_{inv} \right) \right\rangle_{i+\frac{1}{2}} =& \quad \textrm{min} \left( \alpha^{(1)}_{i+\frac{1}{2},l},\alpha^{(1)}_{i+\frac{1}{2},r} \right) F^{cons}_{inv} \left( U_{i+\frac{1}{2},l}^{(1)}, U_{i+\frac{1}{2},r}^{(1)} \right) \\
    &+\textrm{max} \left( \alpha^{(1)}_{i+\frac{1}{2},l}-\alpha^{(1)}_{i+\frac{1}{2},r},0 \right) \left(\beta_{i+\frac{1}{2}}^{(1,2)} \right)^{+} F^{cons}_{inv} \left( U_{i+\frac{1}{2},l}^{(1)}, U_{i+\frac{1}{2},r}^{(2)} \right) \\
    &+ \textrm{max} \left( \alpha^{(2)}_{i+\frac{1}{2},l} - \alpha^{(2)}_{i+\frac{1}{2},r},0 \right) \left( -\beta_{i+\frac{1}{2}}^{(2,1)} \right)^{+} F^{cons}_{inv} \left( U_{i+\frac{1}{2},l}^{(2)}, U_{i+\frac{1}{2},r}^{(1)} \right),
    \end{split}
    \label{eq:demcons1}
\end{equation}
where $F^{cons}_{inv}$ are the fluxes based on the Riemann solutions as discussed earlier, and their coefficients are the probabilities of encountering a gas-gas,  a liquid-liquid,  a gas-liquid, or a liquid-gas interface on a fine-grained discrete level.
Further details and justifications are provided in the original DEM work~\cite{abgrall_jcp_2003}, and they are not repeated here for brevity.
Coefficients $\beta_{i+{1}/{2}}^{(p,q)}$ are computed as $\textrm{sign} \left( c^{*} \left( U_{i+1/2,l}^{p}, U_{i+1/2,r}^{q} \right) \right)$, and $\left( \beta \right)^{+} = \mathrm{max} \left( \beta, 0 \right)$.

After computing the fluxes, the increment $d \left( \alpha^{\kappa} U^{\kappa} \right)$ due to the $i-$directional conservative inviscid flux (term \rom{2} in~\eq{pdediffuse}) is computed as
\begin{equation}
    \begin{split}
         V d \left( \alpha^{\kappa} U^{\kappa} \right)_{ijk,\rom{2}}^{i}
         =& S_{i+\frac{1}{2}jk} \left\langle I^{\kappa} \left( F^{\kappa}_{inv}  \right) \right\rangle_{i+\frac{1}{2}jk} -  S_{i+\frac{1}{2}jk} \left\langle I^{\kappa} \left( F^{\kappa}_{inv} \right) \right\rangle_{i-\frac{1}{2}jk}.
    \end{split}
    \label{eq:demcons2}
\end{equation}

\subsubsection{Inviscid non-conservative flux (Terms $\rom{4}$)}

The non-conservative fluxes are referred to as the ones that act at a~(1)-(2) interface (contact) but within a finite-volume cell once the multiphase interface has moved into it, e.g., the flux $F^{lag}_{L/R}$ acting on the interface that has moved into the cell $i+1$ in~\fig{demschematic}~(b).
The subscripts $L/R$ correspond to whether the flux is acting on the material on the left or right side of the interface.
Following~\cite{abgrall_jcp_2003}, second-order $i-$directional update for phase $(1)$ due to the inviscid non-conservative flux (term \rom{4} in~\eq{pdediffuse}) is given as
\begin{equation}
    \begin{split}
        V d \left( \alpha^{(1)} U^{(1)} \right)_{ijk,\rom{4}}^{i} &= 
            \textrm{max} \left( \alpha^{(1)}_{i+\frac{1}{2},l}-\alpha^{(1)}_{i+\frac{1}{2},r},0 \right) \left( -\beta_{i+\frac{1}{2}}^{(2,1)} \right)^{+} F^{lag}_{L, inv} \left( U_{i+\frac{1}{2},l}^{(1)}, U_{i+\frac{1}{2},r}^{(2)} \right) S_{i+\frac{1}{2}}\\
            &- \textrm{max} \left( \alpha^{(2)}_{i+\frac{1}{2},l} - \alpha^{(2)}_{i+\frac{1}{2},r},0 \right) \left( -\beta_{i+\frac{1}{2}}^{(2,1)} \right)^{+} F^{lag}_{R, inv} \left( U_{i+\frac{1}{2},l}^{(2)}, U_{i+\frac{1}{2},r}^{(1)} \right) S_{i+\frac{1}{2}} \\ 
            &+ \textrm{max} \left( \alpha^{(1)}_{i-\frac{1}{2},l}-\alpha^{(1)}_{i-\frac{1}{2},r},0 \right) \left( \beta_{i+\frac{1}{2}}^{(1,2)} \right)^{+} F^{lag}_{L,, inv} \left( U_{i+\frac{1}{2},l}^{(1)}, U_{i+\frac{1}{2},r}^{(2)} \right) S_{i-\frac{1}{2}} \\ 
            &- \textrm{max} \left( \alpha^{(2)}_{i-\frac{1}{2},l} - \alpha^{(2)}_{i-\frac{1}{2},r},0 \right) \left( \beta_{i-\frac{1}{2}}^{(2,1)} \right)^{+} F^{lag}_{R, inv} \left( U_{i-\frac{1}{2},l}^{(2)}, U_{i-\frac{1}{2},r}^{(1)} \right) S_{i-\frac{1}{2}}\\
            &+\textrm{max} \left( \alpha^{(1)}_{i-\frac{1}{2},r}-\alpha^{(1)}_{i+\frac{1}{2},l},0 \right) F^{lag}_L \left( U_{i}^{(1)}, U_{i}^{(2)} \right) \frac{1}{2}\left( S_{i+\frac{1}{2}} + S_{i-\frac{1}{2}}\right) \\ 
            &- \textrm{max} \left( \alpha^{(2)}_{i-\frac{1}{2},r} - \alpha^{(2)}_{i+\frac{1}{2},l},0 \right) F^{lag}_{R} \left( U_{i}^{(2)}, U_{i}^{(1)} \right) \frac{1}{2}\left( S_{i+\frac{1}{2}} + S_{i-\frac{1}{2}}\right) \\
            &+  \langle N_{int} \rangle \left( F_{R}^{lag} \left( U_{i}^{(2)}, U_{i}^{(1)} \right) - F_{L}^{lag} \left( U_{i}^{(1)}, U_{i}^{(2)} \right) \right) \frac{1}{2}\left( S_{i+\frac{1}{2}} + S_{i-\frac{1}{2}}\right).
    \end{split}
    \label{eq:dem_noncons}
\end{equation}
where $F^{lag}_{L/R, inv}(U^{L}, U^{R})$ are fluxes between the phases at the interface, and they are computed as part of the previously defined~(1)-(2) Riemann solution as 
\begin{equation}
    \begin{split}
        F^{lag}_{L,inv}(U^{L}, U^{R}) = \mathcal{F}(\mathcal{U}^{L*}) - c^{*}\left( \mathcal{U}^{L}, \mathcal{U}^{L} \right) \mathcal{U}^{L*}, \\
        F^{lag}_{R,inv}(U^{L}, U^{R}) = \mathcal{F}(\mathcal{U}^{R*}) - c^{*}\left( \mathcal{U}^{L}, \mathcal{U}^{R} \right) \mathcal{U}^{R*}
    \end{split}
\end{equation}
Note that based on the approximate Riemann solution derived earlier, we have $F^{lag}_L - F^{lag}_R = F^{\sigma}$ for flows with surface tension.  

The first two terms in~\eq{dem_noncons} correspond to the non-conservative fluxes that act within the cell $i$ as a result of a multiphase interface that would have entered through the cell-face $i+1/2$.
The following two terms correspond to an interface that would have entered through the cell-face $i-1/2$.
The subsequent two terms are for second-order accuracy when piecewise linear interpolations are used within a cell.
In the last term, $\langle N_{int} \rangle$ is the average number of internal interfaces within the cell, and this term was related to the relaxation processes within a cell, which would lead to local equilibrium.
For resolved interfaces, this could be modeled via the stiff relaxation solvers as described later in \sect{relax}.
The reader can find more details about the DEM elsewhere~\cite{abgrall_jcp_2003}, and they are not repeated here for brevity.

\subsection{Viscous fluxes}
\label{sec:viscous}

Terms \rom{3} and \rom{5} in~\eq{pdediffuse} are the viscous conservative and non-conservative updates. 
\citet{abgrall_caf_2014} extended the original DEM framework~\cite{abgrall_jcp_2003} for viscous fluxes.
However, they only did this for 1D flows, and here it is also used for 2D/3D.
The viscous updates $d \left( \alpha^{\kappa} U^{\kappa} \right)_{ijk,\rom{3}}$ and $d \left( \alpha_{\kappa} U_{\kappa} \right)_{ijk,\rom{5}}$ follow the same formula as~\eq{inv_direction_cons} for their directional contributions from $i/j/k$.
The flux computations are described here only for the $i$-direction, but the other directions use a similar form.

\subsubsection{Single-phase viscous fluxes}

A Riemann problem is inherently defined for 1D inviscid flows, and therefore, it is not possible to completely replicate the inviscid computations shown earlier for the viscous terms.
The computation of single-phase viscous fluxes $F^{\kappa}_{i,visc}$ is discussed first in this section, and their compilation into the two-phase conservative and non-conservative terms is described in the following sections.

Centered differencing and centered averaging are used for computing the terms $\partial_{x_{\indx{i}}} f \, \vline_{\ i+1/2jk}$, $f_{i+1/2jk}$, etc.
The physical derivatives $\partial_{x_{\indx{i}}} f$, e.g., $\partial_{x_{\indx{j}}} u_{\indx{i}}$, $\partial_{x_{\indx{i}}} T$, etc., are obtained from the computational derivatives as
\begin{equation}
    \frac{\partial f}{\partial x_{\indx{i}}} = \frac{\partial f}{\partial \zeta} \frac{\partial \zeta}{\partial x_{\indx{i}}} + \frac{\partial f}{\partial \eta} \frac{\partial \eta}{\partial x_{\indx{i}}} + \frac{\partial f}{\partial \xi} \frac{\partial \xi}{\partial x_{\indx{i}}},
    \label{eq:deriv1}
\end{equation}
where $\partial_\zeta f$, $\partial_\eta f$, and $\partial_\xi f$ are computed from the flow-variables as
\begin{equation}
    \begin{split}
        &\quad\quad \frac{\partial f}{\partial \zeta} \vline_{\ i+\frac{1}{2}jk} = f_{i+1jk} - f_{ijk},\\ 
        &\frac{\partial f}{\partial \eta} \vline_{\ i+\frac{1}{2}jk} = \frac{1}{2} \left( f_{i+\frac{1}{2}j+1k} - f_{i+\frac{1}{2}j-1k} \right),\\
        &\frac{\partial f}{\partial \xi} \vline_{\ i+\frac{1}{2}jk} = \frac{1}{2} \left( f_{i+\frac{1}{2}jk+1} - f_{i+\frac{1}{2}jk-1} \right).
    \end{split}
\label{eq:deriv2}
\end{equation}
In addition, any face-averaged quantities, e.g., $\alpha^{\kappa}$, $\mu^{\kappa}$, etc., are computed as
\begin{equation}
f_{i+\frac{1}{2}jk} = \frac{1}{2} (f_{i+1jk}+ f_{ijk}).
\label{eq:center1}
\end{equation}
With these, the $i$-directional single-phase viscous fluxes are computed as
\begin{equation}
    \begin{split}
        F^{\kappa}_{visc, i+\frac{1}{2}jk} &=F^{\kappa}_{\indx{i},visc} \left(\mu^{\kappa}_{i+\frac{1}{2}jk},\dots,\frac{\partial u^{\kappa}_{\indx{i}}}{\partial x_i} \vline\ _{i+\frac{1}{2}jk}, \dots \right) \cdot n_{\indx{i},i+\frac{1}{2}jk}.
    \end{split}
    \label{eq:1phasevisc}
\end{equation}

\subsubsection{Viscous conservative flux (Terms $\rom{3}$)}

For the DEM computation of the viscous conservative fluxes, similar forms as~\eq{demcons1} and~\eq{demcons2} are used.
However, instead of using the $F^{cons}_{inv}$ from the Riemann solutions as before, the terms $F^{\kappa}_{visc}$ at the cell-faces (e.g., $i+1/2,j,k$) are now computed using~\eq{1phasevisc}.
The DEM viscous fluxes are then given as
\begin{equation}
    \begin{split}
         \left\langle I^{(1)} \left( F^{(1)}_{visc} \right) \right\rangle_{i+\frac{1}{2}} =& \textrm{min} \left( \alpha^{(1)}_{i+\frac{1}{2},l},\alpha^{(1)}_{i+\frac{1}{2},r} \right) F_{visc}^{(1)} \\
         &+\textrm{max} \left( \alpha^{(1)}_{i+\frac{1}{2},l}-\alpha^{(1)}_{i+\frac{1}{2},r},0 \right) \left(\beta_{i+\frac{1}{2}}^{(1,2)} \right)^{+} F_{visc}^{(1)} \\
         &+ \textrm{max} \left( \alpha^{(2)}_{i+\frac{1}{2},l} - \alpha^{(2)}_{i+\frac{1}{2},r},0 \right) \left( -\beta_{i+\frac{1}{2}}^{(2,1)} \right)^{+} F_{visc}^{(1)},
     \end{split}
    \label{eq:demcons1_visc}
\end{equation}
and a similar form is used for $\kappa=2$ as well. 

\subsubsection{Viscous non-conservative (Terms $\rom{5}$)}

The non-conservative viscous terms follow the same form as~\eq{dem_noncons}, however, instead of $F^{lag}_{L/R, inv}$ coming from the 1D inviscid Riemann problems, $F^{lag}_{L/R, visc}$ are modeled as a function of the single-phase viscous terms $F^{\kappa}_{visc}$ computed at cell-faces (e.g., $i+1/2,j,k$) as
\begin{equation}
    \begin{gathered}
        F^{lag}_{L, visc}(U^{(1)}_{i+\frac{1}{2},l}, U^{(2)}_{i+\frac{1}{2},l}) = F^{lag}_{R, visc}(U^{(2)}_{i+\frac{1}{2},l}, U^{(1)}_{i+\frac{1}{2},l}) = F^{(1)}_{visc,i+\frac{1}{2}}, \\
        F^{lag}_{L, visc}(U^{(1)}_{i}, U^{(2)}_{i}) = F^{lag}_{R, visc}(U^{(2)}_{i}, U^{(1)}_{i}) = \frac{ S_{i+\frac{1}{2}}F^{(1)}_{visc,i+\frac{1}{2}} + S_{i-\frac{1}{2}}F^{(1)}_{visc,i-\frac{1}{2}}}{S_{i+\frac{1}{2}} + S_{i-\frac{1}{2}}}.\\
    \end{gathered}
\label{eq:dem_noncons_visc}
\end{equation}
where $\kappa=1$ is assumed as the gas-phase, and therefore, it is assumed that $F^{I}_{visc} = F^{(1)}_{visc}$.
This is similar to the previous viscous DEM approach of~\citet{abgrall_caf_2014}.

\subsection{Curvature computation}
\label{sec:curvature}

The interface curvature $\beta$ is required to compute $p^{\sigma} = \sigma \beta$ at the cell-centers $(i,j,k)$.
Once the curvature is computed at the cell centers, it is then interpolated at the cell faces using the MUSCL approach in the same way as the other field quantities for flux computation.
The curvature for a field $f$ is defined as
\begin{equation}
    \beta = \frac{\partial}{\partial x_i} \left( \frac{\partial f}{\partial x_i} \bigg / \bigg | \frac{\partial f}{\partial x_i} \bigg | \right).
    \label{eq:curv1}
\end{equation}
A previous approach for curvature computation is followed that used a vertex-based staggering~\cite{nguyen_amc_2015, perigaud_jcp_2005}, which is illustrated in 2D.
The gradient of $f$ at a grid vertex is first computed as
\begin{equation}
    \begin{split}
        \frac{\partial f}{\partial x_{1}}_{i+\frac{1}{2},j+\frac{1}{2}} &= \frac{1}{2} \left(  \frac{f_{i+1,j+1} - f_{i,j+1} }{\Delta x} +  \frac{f_{i+1,j} -f_{i,j} }{\Delta x} \right), \\
        \frac{\partial f}{\partial x_{2}}_{i+\frac{1}{2},j+\frac{1}{2}} &= \frac{1}{2} \left(  \frac{f_{i+1,j+1} - f_{i+1,j} }{\Delta x} +  \frac{f_{i,j+1} - f_{i,j} }{\Delta x} \right).
    \end{split}
    \label{eq:curv1a}
\end{equation}
where $\Delta x$ and $\Delta y$ are the grid sizes in the $i$ and the $j$ directions.
Based on this gradient, a unit vector normal to the interface is computed as
\begin{equation}
    n_{\indx{i}, i+\frac{1}{2}, j+\frac{1}{2}}  = \begin{cases} 
    \frac{\frac{\partial f}{\partial x_{\indx{i}}}_{i+\frac{1}{2},j+\frac{1}{2}}} { \bigg| \bigg| \frac{\partial f}{\partial x_{\indx{i}}}_{i+\frac{1}{2},j+\frac{1}{2}} \bigg| \bigg|}, \ \ \ \ \ \text{if}\ \ \ \ \ \bigg|\bigg| \frac{\partial f}{\partial x_{\indx{i}}}_{i+\frac{1}{2},j+\frac{1}{2}} \bigg|\bigg| > 0,  \\
    0, \ \ \ \ \ \ \text{otherwise},
    \end{cases}
    \label{eq:curv2}
\end{equation}
where, $||\cdot||$ is the magnitude of the vector.  The normal vector is then interpolated at the cell-faces as 
\begin{equation}
    n_{\indx{i}, i+\frac{1}{2}, j}  = \frac{1}{2} \left( n_{\indx{i},  i+\frac{1}{2}, j+\frac{1}{2},} + n_{\indx{i}, i+\frac{1}{2}, j-\frac{1}{2}} \right),
    \label{eq:curv3}
\end{equation}
and finally, the curvature is computed at the cell-centers as
\begin{equation}
    \kappa_{i,j} = \frac{n_{1, i+\frac{1}{2}, j} - n_{1, i-\frac{1}{2}, j,}} {\Delta x} + \frac{ n_{2, i, j+\frac{1}{2},} - n_{i, j-\frac{1}{2}}} {\Delta y}.
    \label{eq:curv4}
\end{equation}

The previous approach~\cite{nguyen_amc_2015} used $f=\alpha^{(1)}$ in~\eq{curv1} as that would result in the interface curvature at the droplet surface.
However, $\alpha^{(1)}$ varies sharply from 0 to 1 across an interface, and this can result in numerical errors.
To address this, in this work,  $f=\phi$ is used, which is defined as
\begin{equation}
    \phi = \frac{ \left( \alpha^{(2)} \right) ^{m}}{ \left(\alpha^{(2)} \right)^{m} + \left( 1-\alpha^{(2)} \right)^{m}} \quad \textrm{for}\ m < 1.
    \label{eq:phicompute}
\end{equation}
\citet{shukla_jcp_2010} used this form of $\phi$  for their interface compression scheme, and it is used for the curvature computation in this work.
The field $\phi$ is smoother as compared to $\alpha^{(2)}$, and $m$ is a parameter that decides its smoothness.
Note that $m = 1$ would result in $\phi = \alpha^{(2)}$.
Finally, one would compute the curvature everywhere in the flow-field using~\eq{curv4}. However, to reduce any numerical noise, it is only used in the regions where $0.99 > \alpha^{(1)} > 0.01$, and it is set to zero otherwise.

\begin{figure}
    \begin{center}
        \begin{subfigure}\centering{\includegraphics[width=0.7\textwidth]{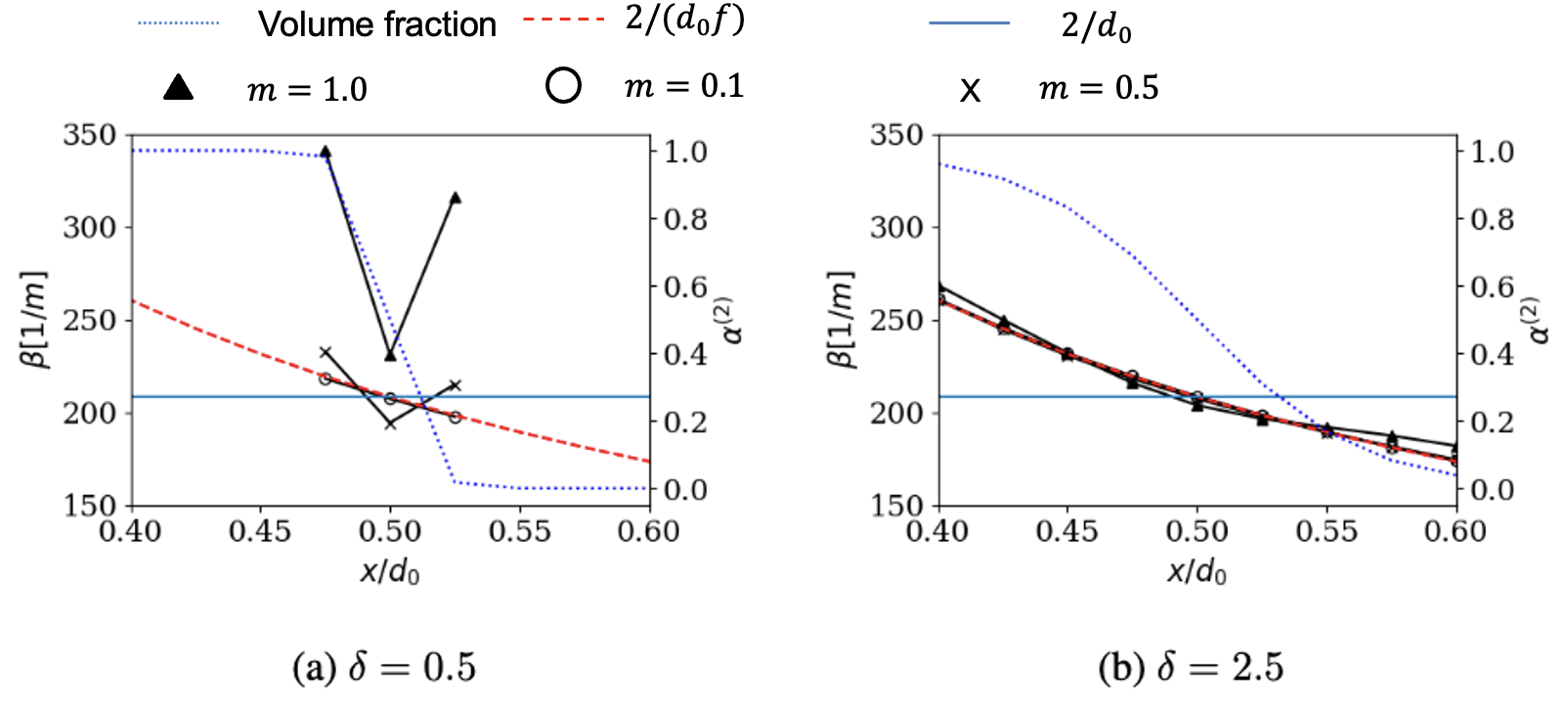}}\end{subfigure}
        \caption{
            {The curvature and the volume fraction are shown for a circular drop along the centerline initialized as~\eq{circle_init}.
        The computed curvature using~\eq{curv4} and~\eq{phicompute} is shown for $m=0.1$, $m=0.5$ and $m=1.0$.
        The exact curvature along the droplet radius is shown as $2/d_0 f$, and the exact droplet curvature at the interface is shown as $2/d_0$.}
        }
    \label{fig:standalone_curvature}
    \end{center}
\end{figure}

To verify the curvature computation, a circular drop with diameter $d_0=9.6$ mm is placed at the center ($x_0 = 0, y_0 = 0$) of a square domain of size $[6 d_0, 6 d_0]$.
The computational domain is uniformly divided into $240\times240$ cells.
The volume fraction for the droplet is initialized via
\begin{equation}
    f = \sqrt{ \left(\frac{x-x_0}{d_0/2} \right)^2 + \left(\frac{y-y_0}{d_0/2} \right)^2 }, 
    \quad \text{and} \quad 
    \alpha^{(2)} = \frac{1}{2} \tanh \left( \frac{(f-1) d_0} {\delta 2 \Delta x } \right) + \frac{1}{2}.
    \label{eq:circle_init}
\end{equation}
where  $\delta$ is the thickness over which the numerical interface is diffused initially.
The corresponding volume fraction fields for $\delta=0.5$ and $\delta=2.5$, as well as the numerically computed curvature are shown in~\fig{standalone_curvature} along the center-line.
The exact curvature for a droplet with diameter $d_0$ would be $2/d_0$ (shown by the solid horizontal line in~\fig{standalone_curvature}).
However, since the interface is diffused, if we were to compute the curvature using~\eq{curv1} and~\eq{circle_init}, then it would be $2/(d_0 f)$ along the radius of the droplet (shown via dashed line in~\fig{standalone_curvature}).

A comparison of the curvature computed using~\eq{curv4} and~\eq{phicompute} at $m$ in~\fig{standalone_curvature} shows that for $\delta = 2.5$, when the interface is diffused across 8-10 cells, $m=0.1, 0.5, 1$, provide a good match against $2/d_0 f$.
The surface tension only acts at the sharp interface where the curvature is $2/d_0$ (horizontal line).
Indeed, a numerically diffused volume fraction could result in further errors.
Therefore, the curvature computation must work for sharp interfaces and should compute the curvature as $2/d_0$.
For $\delta = 0.5$, the interface is diffused over 2-3 cells.
For this case, choosing $m=1$ (i.e., $f=\alpha^{(2)}$, which was used in earlier works~\cite{nguyen_amc_2015}) results in about $50$\% errors, but with $m = 0.5$, the errors are $<5$\%.

Choosing $m = 0.1$ leads to slightly lower errors as compared to $m = 0.5$.
However, the test considered here is an ideal scenario where the volume fraction is defined using~\eq{circle_init}, and as a result, $f$ is monotonic.
In reality, there could be numerical oscillations in the $\alpha^{(2)}$ field, and those, when coupled with the high gradients of $\partial f / \partial \alpha$ near $\alpha=0,1$, could result in a non-monotonic behavior of $f$ for a lower $m$ such as 0.1.
This could lead to further errors in the computation, and considering this, $m = 0.5$ is chosen as a balance between the two effects in this work.

\subsection{Relaxation process}
\label{sec:relax}

The quantities $R_u^{\kappa}$, $R_p^{\kappa}$ and $R_T^{\kappa}$ in~\eq{noncons_model1} are relaxation terms.
Physically, they represent the interface jump conditions of~\eq{jump2} and~\eq{jump3} for a resolved interface.
Since $\theta^{p}, \theta^{u}, \theta^{T} \rightarrow \infty$ for the seven-equation model, the relaxation processes are stiff, and they are applied separately from the inviscid and the viscous updates through a Strang splitting~\cite{saurel_jcp_1999}.
Correspondingly, the flow solution is updated as $Q^{n+1} = \mathcal{R}\mathcal{L}^{c}\mathcal{R}\mathcal{L}^{p} (Q^{n})$, where $\mathcal{R}$ represents the stiff relaxation operator,  $\mathcal{L}^{p}$ and $\mathcal{L}^{c}$ correspond to the predictor and the corrector steps of the time-update, and the superscript $n$ is used to represent the temporal iteration number.
The pressure, velocity and the temperature relaxation operators are represented as $\mathcal{R}_p$, $\mathcal{R}_u$,  $\mathcal{R}_T$, respectively, and $\mathcal{R} = \mathcal{R}_T \mathcal{R}_u \mathcal{R}_p$.

In terms of the numerical modeling of the interphase exchanges through the DEM, the terms with $\langle N_{int} \rangle $ in~\eq{dem_noncons} correspond to the relaxation terms as shown by~\citet{abgrall_jcp_2003}.
Here, $\langle N_{int} \rangle $ represents the number of unresolved interfaces within a finite-volume cell, and this would typically be finite for dispersed flows where the droplets are unresolved.
Still, it is supposed to be negligible for the resolved interfaces such as those considered here.
As a result of this, while using the DEM, there should be no need to use a separate stiff relaxation solver for resolved interfaces, and this is further confirmed in this work (see~\sect{results}).

Interestingly, a recent work~\cite{schmidmayer_2021} showed that it is not necessary to model the non-conservative terms $I^{\kappa}$ in~\eq{noncons_model1} if the relaxation terms $R^{\kappa}$ are modeled consistently.
This suggests that both the non-conservative terms $I^{\kappa}$ and the relaxation terms $R^{\kappa}$ model the same interphase exchange effects but differently. 
While~\citet{schmidmayer_2021} were able to model the interphase exchanges only via $R^{\kappa}$, our results here show that it is possible to model them only via the non-conservative terms $I^{\kappa}$ with the use of the DEM.

However, to provide a consistent comparison against the previous works that did use a relaxation solver~\cite{saurel_jcp_1999}, and because the seven-equation model would asymptote to the widely used five-equation model in the stiff pressure and velocity relaxation limit, the operators $\mathcal{R}_p$ and $\mathcal{R}_u$ are implemented in the current computational framework.
The stiff temperature relaxation operator $\mathcal{R}_T$ is not included in this work, so the thermal exchanges through the interface are only via the resolved interphase exchange terms $I^{\kappa}$ in~\eq{dem_noncons} via the DEM.

The stiff velocity relaxation operator $\mathcal{R}_u$ modifies the phase-averaged momentum and energy ($\rho^{\kappa} \alpha^{\kappa} u_i^{\kappa} $, $\rho^{\kappa} \alpha^{\kappa} E^{\kappa} $) while keeping the mixture momentum and energy conserved.
The algorithm for this is described elsewhere~\cite{saurel_jcp_1999} and it is not repeated here.
The stiff pressure relaxation $\mathcal{R}_p$ results in local expansion/compression of the phases via a change in the volume fraction $\alpha^{\kappa}$, and corresponding pressure work terms are included in the energy $\rho^{\kappa} \alpha^{\kappa} E^{\kappa} $ (see~\eq{dem_noncons}).
The algorithm matches that of~\cite{saurel_jcp_1999}, but with the following improvements:
\begin{itemize}
    \item The viscous and the surface tension forces are included in the pressure jump condition as shown in~\eq{jump3}.
    As identified earlier~\cite{abgrall_caf_2014}, $\tau_{\indx{i}\indx{j}}^{(1)}n_{\indx{i}} n_{\indx{j}} = \tau_{\indx{i}\indx{j}}^{(2)}n_{\indx{i}} n_{\indx{j}}$ is maintained by the viscous DEM, and therefore, the pressure jump that would have to be applied at the interface via the relaxation process is $p^{(2)} - p^{(1)} = \sigma \beta$.
    We can combine the volume fraction and the energy equations, resulting in
    \begin{equation}
        \frac{\partial \alpha^{(1)} \rho^{(1)}E^{(1)}}{\partial t} = -\left(p^{(1)} - \sigma\beta \right) \frac{\partial \alpha^{(1)}}{\partial t} 
        \quad \text{and} \quad 
        \frac{\partial \alpha^{(2)} \rho^{(2)}E^{(2)}}{\partial t} = p^{(1)} \frac{\partial \alpha^{(1)}}{\partial t}.
        \label{eq:relax1}
    \end{equation}
    An iterative method similar to~\cite{saurel_jcp_1999} is used to solve these equations and compute an updated volume fraction $\alpha^{(1)}$ and energies $E^{\kappa}$ to obtain $p^{(2)} - p^{(1)} = \sigma \beta $.
    Unlike previous works without the surface tension, the energy exchange terms do not sum to zero, and the difference between the two, $\sigma \beta \theta^{p} \Delta p$, is a source/sink due to the interfacial surface tension energy.

    \item Gradient descent is used as an iterative scheme.
    However, an adaptive stepping makes the algorithm robust for the various EOSs considered in this work and for a wide range of pressures.
    Specifically, it is used for the gradient descent that would reduce the step size if either the pressures computed via the EOS (either $\kappa = 1,\, 2$) become negative during the iterations or the updated volume fraction goes beyond $[\epsilon, 1-\epsilon]$ with $\epsilon=10^{-10}$.
\end{itemize}

\subsection{Interface compression}
\label{sec:compress}

The interface numerically diffuses across multiple cells as the simulation evolves.
Several techniques have been proposed to alleviate this.
Modifications to the MUSCL face-reconstruction~\cite{shyue_jcp_2014} or the limiter~\cite{chiapolino_jcp_2017sharpening} and applications of higher-order schemes~\cite{gryngarten_2013, coralic_jcp_2014} have been useful for DIM.
However, the DEM used in this work is a specialized technique, and we cannot directly apply it.
Considering this, an interface compression scheme is borrowed here~\cite{shukla_jcp_2010} which works independently of the other parts of the solver.
~\citet{jain_arxiv_2021} recently conducted a one-to-one comparison of various interface compression techniques and concluded that the choice would often depend on the modeling problem of interest.
Their coupling with the seven-equation model can be considered in the future, but it is not addressed here.
The employed iterative technique for interface compression requires solving the following PDE in a pseudo-time $\tau$:
\begin{equation}
    \frac{\partial \alpha^{(1)}}{\partial \tau} =  n_{\indx{i}} \frac{\partial}{\partial x_{\indx{i}}} \left( \epsilon_h\  \vline \frac{\partial \alpha^{(1)}}{\partial x_{\indx{i}}} \vline - \alpha^{(1)} (1-\alpha^{(1)}) \right) = n_{\indx{i}} \frac{\partial f_{comp}}{\partial x_{\indx{i}}}.
    \label{eq:compressionpde}
\end{equation}
The right-hand side of this equation contains compression and diffusion terms and the interface normal, which balance each other when a steady-state is reached in the pseudo-time. 
This would result in a consistent thickness across the diffused interface, which would be a function of $\epsilon_h$. 
\citet{shukla_jcp_2010} used this approach for a five-equation DIM, where they had to use an additional correction for the mixture density.
This approach is not required for the seven-equation model as the densities for the phases are computed separately.

In curvilinear coordinates~\eq{compressionpde} is written as
\begin{equation}
    \frac{\partial \alpha^{(1)}}{\partial \tau} = \left( n_{\indx{i}} \pd{\xi}{x_{\indx{i}}} \pd{f_{comp}}{\xi} + n_{\indx{i}} \pd{\eta}{x_{\indx{i}}}\pd{f_{comp}}{\eta} + n_{\indx{i}} \pd{\zeta}{x_{\indx{i}}}\pd{f_{comp}}{\zeta} \right),
    \label{eq:compressioneta}
\end{equation}
where, $\alpha^{(1)}$ is defined at the cell-centers and updates to it in the pseudo-time are denoted as $d\alpha_{ijk}^{\tau}$.
First, $f_{comp}$ in~\eq{compressionpde} is computed at the cell-faces.
To compute $f_{comp,i+\frac{1}{2}jk}$ at the cell-faces, a centered averaging is used for $\alpha^{(1)}$ following~\eq{center1}, and the derivatives $\partial \alpha^{(1)} / \partial x_{\indx{i}}$ are computed using~\eq{deriv1} and~\eq{deriv2}.
Instead of computing the derivatives $\partial \alpha^{(1)} / \partial x_{\indx{i}}$ directly with $\alpha^{(1)}$, for numerical stability, they are computed as $\partial \phi / \partial x_{\indx{i}}$  with the smooth field $\phi$ defined as previously in~\eq{phicompute}~\cite{shukla_jcp_2010}.

As the next step, $\partial {f_{comp}} / \partial \xi$, etc., are obtained at the cell-centers via a centered differencing such as 
\begin{gather}
    \frac{\partial f_{comp}}{\partial \xi}\ \vline_{\ ijk} = f_{comp,i+\frac{1}{2}jk} - f_{comp,i-\frac{1}{2}jk}.
\end{gather} 
Similar forms are used for the $j$ and the $k$-directions. 
Last, for the updates $d\alpha_{ijk}^{\tau}$ in~\eq{compressioneta}, the interface normal 
\begin{gather}
    n_{\indx{i}} = \frac{\partial \phi / \partial x_{\indx{i}}}{\abs{\partial \phi / \partial x_{\indx{i}}}}
\end{gather} 
is computed at the cell-centers via~\eq{deriv1}, and the corresponding cell-centered derivatives are computed as ${\partial \phi}/{\partial \zeta}\  \vline_{\ i+\frac{1}{2}jk} = {1}/{2} \left( \phi_{i+1jk} - \phi_{i-1jk} \right)$ (shown here representatively in the $i$-direction).

\begin{figure}
   \begin{center}
       \subfigure[Initial $\delta=0.2$]{\includegraphics[width=\fign]{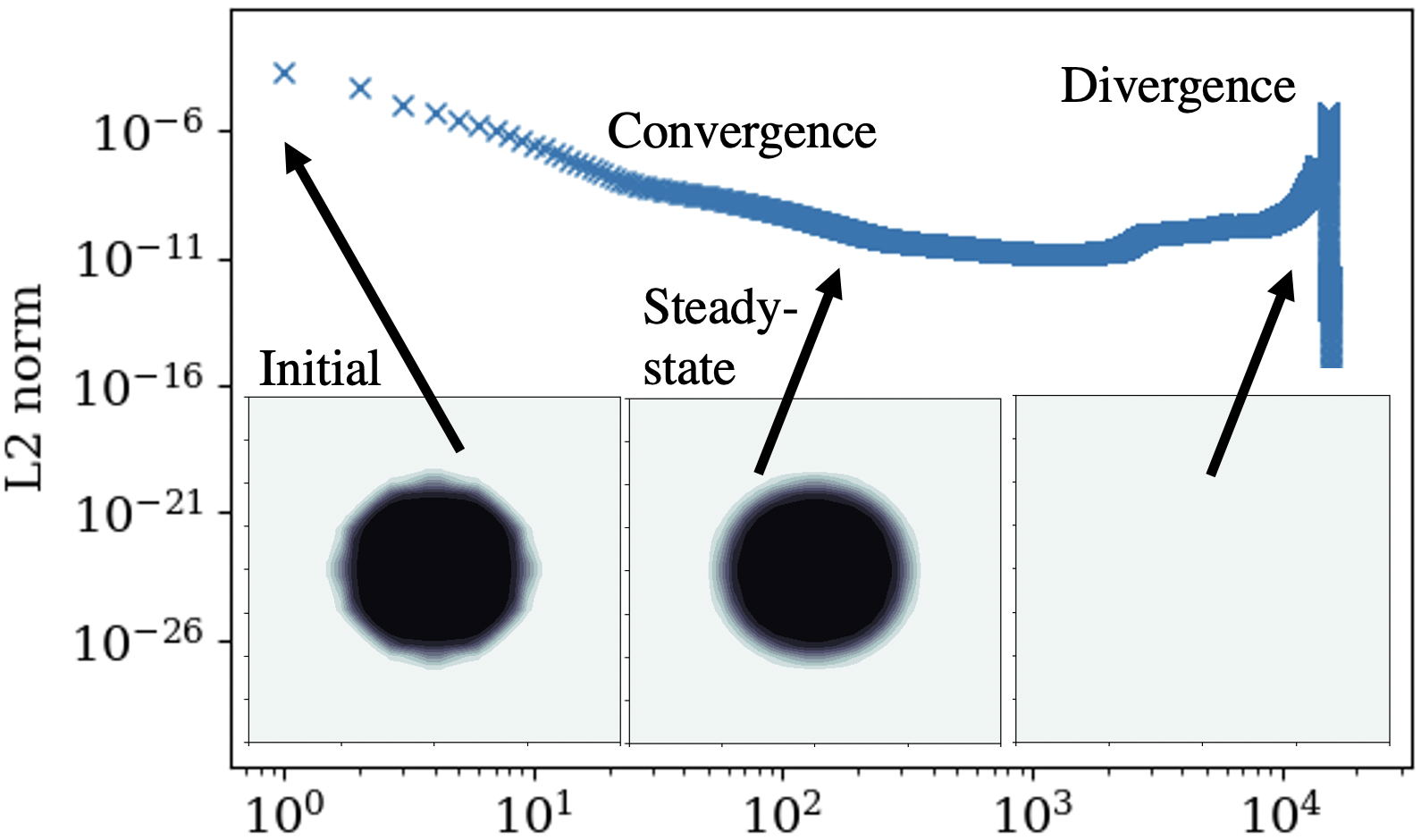}}
       \subfigure[Initial $\delta=2.5$]{\includegraphics[width=\fign]{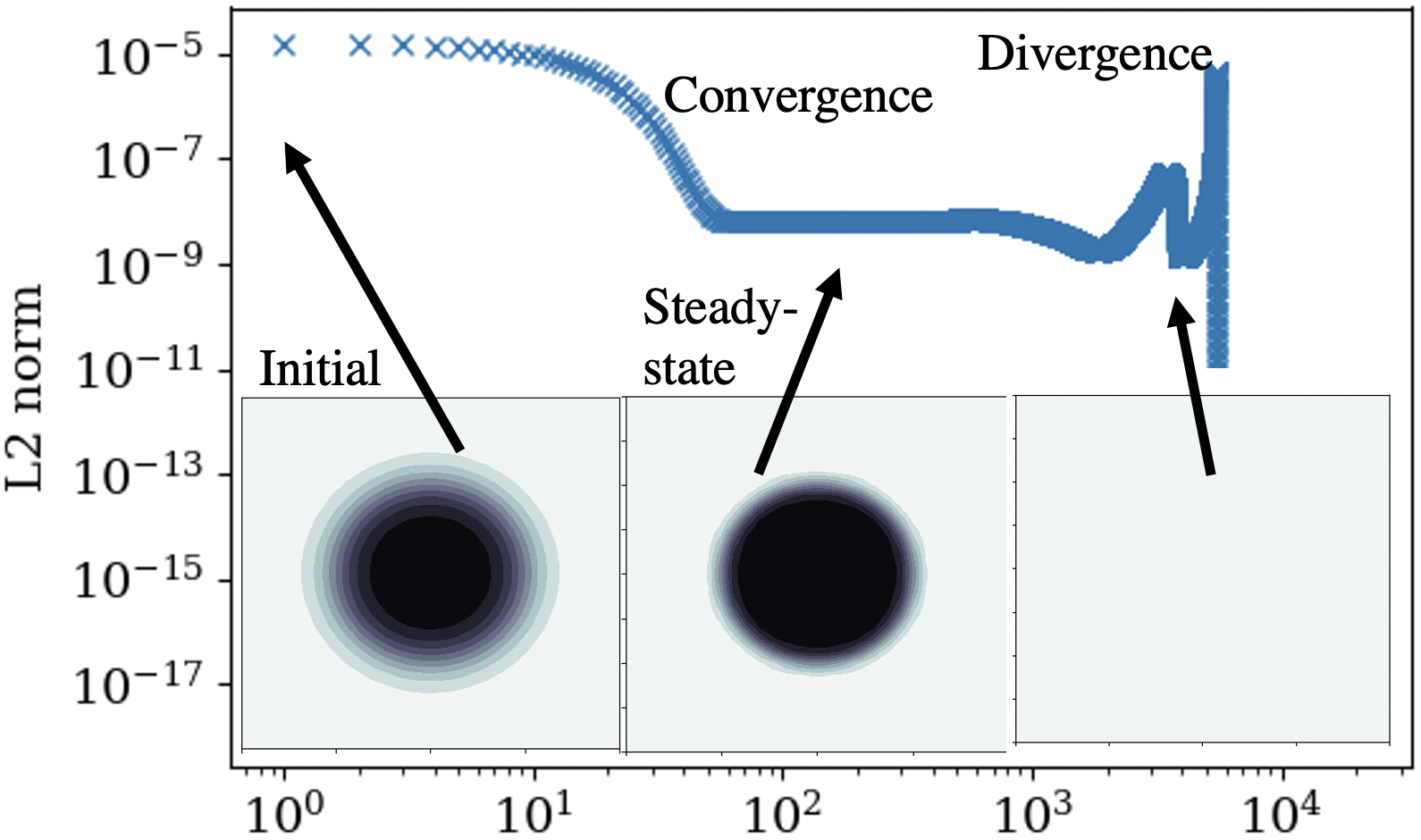}}
       \caption{L2 norms and volume fraction fields are shown with the interface compression algorithm iterations. }
       \label{fig:standalone_compression}
  \end{center}
\end{figure}

While applying the interface compression,  the solution is evolved in the pseudo-time using explicit Euler until a steady-state solution is reached.
The pseudo-time-step $d\tau$ is taken to be 0.2 for numerical stability~\cite{shukla_jcp_2010}.
The compression parameter $\epsilon_h$ is taken as $V_{ijk}^{1/3}/2$, which would result in the interface diffusing across 2-3 cells.

\eq{compressionpde} is not in a strictly conservative form.
As a result, it was observed that when these pseudo-iterations are conducted for a very long time, the solution initially converges to a steady-state, but later it diverges.
This is shown in~\fig{standalone_compression} for example.
For this test,  a circular drop with diameter $d_0=9.6$ mm is placed at the center ($x_0 = 0, y_0 = 0$) of a square domain of size $[2 d_0, 2 d_0]$.
The computational domain is uniformly divided into $30\times30$ cells.
The volume fraction for the droplet is initialized using~\eq{circle_init} with $\delta=0.2$ and $\delta=2.5$.
The convergence is measured using an $L_2$ norm of $\alpha^{(1)}$ between the current and the previous iteration.
For both the cases, initially, the iterations converge (in $\approx$10-30 steps), and a steady-state solution is reached.
However, after running this for a longer time, i.e.
$>10^{3}$ iterations, the solution diverges.
The intermediate steady-state solution shows the interface diffused across 2--3 cells.

The solution to this problem in a standalone computation of the interface compression technique is straightforward.
We could use an absolute or a relative tolerance limit to stop the iterations.
However, while the compression scheme is being used as part of the DIM simulations,  this is not possible.
In addition to the interface compression, the flow fields are also being updated every time step, and even if the interface compression update was to be applied only once after every certain number (called $n^{comp}$) of simulation time-steps, the multiphase simulations can go on for an arbitrarily long number of iterations that could eventually cause this divergence.
As a solution to this, after every $n^{comp}$ simulation time-steps, the interface compression scheme is iterated for at least 3 pseudo-time-steps, and the volume fraction field is updated only if the $L_2$ norms during this monotonically decrease.
If not, the solution is reverted to its original state without any interface compression updates.
The numerical results show that the interface compression did not cause any divergence by using this approach even when we ran the simulations for an arbitrarily long time (up to $10^7$ flow-field iterations).
The interface compression was still active but acted only when necessary, and it maintained the interface thickness diffused only across 2-3 cells.
$n^{comp}=100$ is used in this work.
The results for only a circular drop are shown here, but the interface compression scheme has also been validated for more complex shapes, e.g., ellipse, two lobes, etc.
Those results are not shown here for brevity.

\subsection{Boundary Conditions}
\label{sec:bcs}

Seven different treatments are used in this work at the computational domain boundaries: slip walls, no-slip walls, subsonic inflow, subsonic outflow, supersonic inflow, supersonic outflow, periodic.
For the no-slip walls, the wall-normal and the tangential velocities of both the phases are set to zero at the boundary face.
Only the wall-normal component is set to zero for the slip walls, whereas a zero gradient Neumann condition is applied for the tangential component.
All the other quantities, i.e., the temperature, the species mass-fractions,  and the volume fraction, are set using a zero gradient Neumann condition at the walls.

The eigenvalues of the seven-equation system being solved here are $\{u^{(1)}-c^{(1)}, u^{(1)}, u^{(1)}+c^{(1)}, u^{(2)}-c^{(2)}, u^{(2)}, u^{(2)}+c^{(2)}, u^{I}\}$~\cite{saurel_jcp_1999}.
Correspondingly, seven waves are propagating in this hyperbolic system of equations.
The inflow/outflow treatment is dependent on the wave-propagation directions at the boundaries.
For supersonic flows, i.e., $u^{(1)}, u^{(2)} > c^{(1)}$ and $u^{(1)}, u^{(2)} > c^{(2)}$, all the waves propagate in the same direction.
Therefore, a supersonic inflow, where all the waves enter the domain, is set using a Dirichlet condition. A supersonic outflow, where all the waves leave the domain, is set using a zero-gradient Neumann condition.

For subsonic inflow/outflows, where certain waves are entering the domain and certain that are leaving the domain, characteristic-based boundary conditions~\cite{poinsotlele_1992} are used.
These were initially developed for multi-species single-phase flows, but the same form is used here for the two-phase flows.
Since the wave-propagation speeds considered within this single-phase framework only correspond to the gas-phase wave speeds $\{ u^{(1)}-c^{(1)}, u^{(1)}, u^{(1)}+c^{(1)}\}$,  it is implicitly assumed that the other phase properties, as well as the volume fraction, are propagating at speed $u^{(1)}$ as if they are passive scalars.
This is not a valid approach in general.
Still, it is retained here, considering that only a single phase (typically the gas) is present at the inflow/outflow boundaries for the configurations studied.
The development of a characteristic-based boundary condition for fully compressible two-phase flows is the subject of future work.

\section{Results}
\label{sec:results}

Various features of the seven-equation DIM framework are evaluated in this section.
\tab{tests} lists the tests that are considered in this work and the features that they validate.
For the seven-equation model, the phase-averaged flow fields for both the phases (i.e., $\rho^{\kappa}$, $p^{\kappa}$, etc.) are defined everywhere in the computational domain.
For instance, even in a region where only the phase~(1) is  present,  $\rho^{(2)}$, $p^{(2)}$ etc.
are still mathematically defined.
The corresponding volume fraction of the phase~(2) would be very small, i.e., $\alpha^{(2)} = \epsilon$, and this would result $f^{mix} = \sum_{\kappa=1}^{2} \alpha^\kappa f^\kappa \approx f^{(1)}$, so the phase~(2) values, i.e., $\rho^{(2)}$, $p^{(2)}$, etc., aren't expected to affect the mixture flow-fields with $\epsilon \rightarrow 0$, but they still have to be defined and initialized.
To avoid a division by zero and to make sure that the numerical solution does not get corrupted,  the corresponding $\epsilon$ has to be larger than the machine precision (double precision used here, $10^{-15}$), and in this work it is chosen as $10^{-6}$ unless explicitly mentioned otherwise.

\begin{table}
\caption{List of verification and validation test cases for the seven-equation DIM. }
\begin{center}
\begin{tabular}{p{35mm}|p{65mm}}
\hline
    Feature & Test-case and approach \\\hline \vspace{0.2in}
    Baseline solver & Periodic pulse convection \newline~\tabitem Order of scheme \newline~\tabitem Non-disturbing condition~\cite{saurel_jcp_1999, abgrall_jcp_2003}  \\\hline
    $p^{(2)}/p^{(1)}, \rho^{(2)}/\rho^{(1)} >> 1$ & Multiphase shock tube \newline~\tabitem Comparison against exact solution~\cite{abgrall_jcp_2003} \\\hline
    Arbitrary EOS \& shocks & Shock transmission through \newline~\tabitem Air-Aluminum (CPG/SG)~\cite{sridharan_jap_2015} \newline~\tabitem Air-HMX (CPG/MIEG) interface \newline~\tabitem Comparison against an exact solution \\\hline
    \multirow{2}{*}{Surface tension} & $\Delta p$ in a circular drop~\cite{nguyen_amc_2015, perigaud_jcp_2005} \newline~\tabitem Comparison against exact jump \newline~\tabitem Analysis of parasitic currents \\\cline{2-2}
    & Oscillating elliptical droplet \newline~\tabitem Compared against exact frequency~\cite{nguyen_amc_2015, perigaud_jcp_2005} \\\hline
    \multirow{2}{*}{Viscous effects} & Two-phase lid-driven cavity  \newline~\tabitem Comparison against exact solution~\cite{ghia_jcp_1982} \\\cline{2-2}
    & Cylindrical and deforming droplet drag \newline~\tabitem Comparison against $C_d$ correlations~\cite{white_2006} \\\hline
    Droplet deformation & Shock-droplet interaction \newline~\tabitem Comparison against experiments~\cite{igra_as_2002, chen_aiaa_2008, meng_sw_2015} \\\hline
    Reactive flows & Detonation-droplet interaction demonstration \\\hline
\end{tabular}
\end{center}
\label{tab:tests}
\end{table}

\subsection{Convection}
\label{sec:convection}

A 1D periodic computational domain of length $L = 1$ m is used for this test.
The volume fraction $\alpha^{(1)}$ is initialized (at $t = t_0$) as a sine function varying between 0 to 1 as $\alpha^{(1)}(x)  = (1/4) \sin\left(x/(2 \pi L) \right)+1/2$.
It is capped between $\epsilon$ and $1-\epsilon$ to prevent any division by zero in the code.
Volume fraction for the other phase is set as $\alpha^{(2)} = 1 - \alpha^{(1)}$.
Uniform pressure and velocity are specified as $p^{(1)} = p^{(2)} = 10^5$ Pa and $u^{(1)} = u^{(2)}=100$ m/s.
The densities are set as $\rho^{(1)} = 1.0$ kg/m$^3$ and $\rho^{(2)} =  10^3$ kg/m$^3$.
The EOS for phase~(1) is specified as CPG, with $\gamma^{(1)}=1.4$, and for phase~(2) it is specified as SG with $\gamma^{(2)}=4.4$ and $p_0^{(2)}=6 \times 10^8$.
The surface tension, the viscous terms, and the interface compression are absent for this test, and a single specie is used for both the phases.
Four different grids are tested with the number of cells varying as $N_x=20,\, 50,\, 100,\, 200$ as well as the first- and the second-order accurate schemes.
Simulations are conducted both with and without stiff relaxation.

\begin{figure}
   \begin{center}
   \subfigure{\includegraphics[width=2.5in]{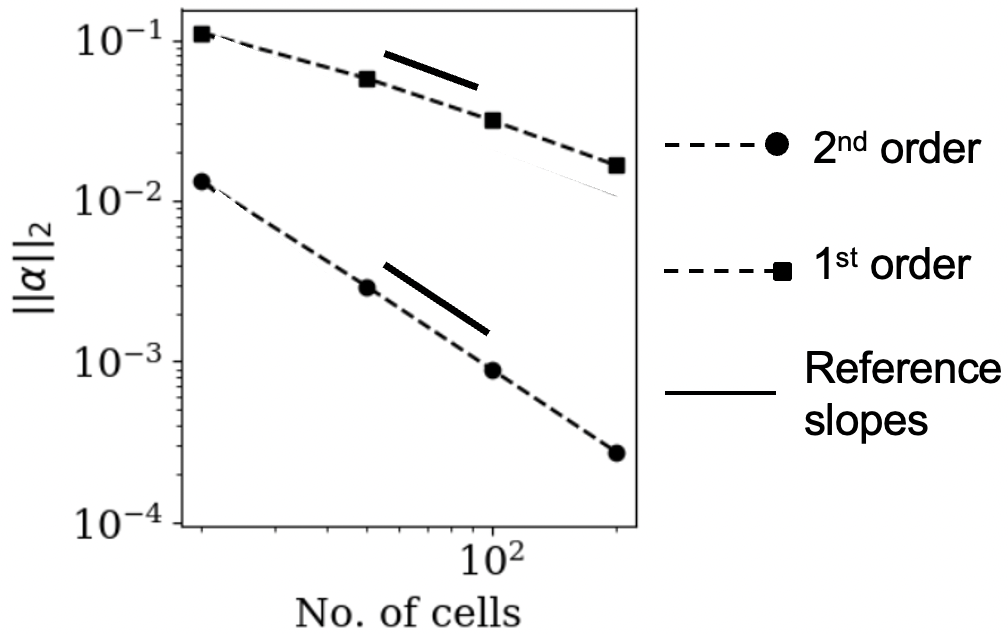}}
   \caption{Normalized $L_2$ error norms for the 1D pulse convection test are shown with different grid sizes and numerical scheme orders.  The solid lines have $1/1$ and $1/2$ slopes for reference. }
   \label{fig:1d-pulse}
  \end{center}
\end{figure}

The volume fraction pulse is allowed to convect in the periodic domain for one flow-through time ($t=t_f$). The solution at the end is compared against the initial solution ($t=t_0$), which serves as a `truth' solution for this case.
The `truth' solutions are denoted with a superscript `$truth$'.
The $L_2$  error norms for any field-variable $f$ are computed as
\begin{equation}
    ||f||_2 =  \sqrt{\frac{1}{N_x} \sum_{i_x=1}^{N_x} \left({f(i_x,  t=t_f) - f^{truth}(i_x)} \right) }.
    \label{eq:l2norm}
\end{equation}
As noted in earlier studies~\cite{abgrall_jcp_2003},  a ``non-disturbing'' condition for this case means that regardless of the volume fraction in the domain, since the simulation starts with a uniform pressure/velocity, one should maintain a uniform pressure/velocity at all times with machine accuracy.
Both the pressure and the velocity maintain uniform values with $||p^{\kappa}||_2/p^{(1)} < 10^{-15}$ and $||u^{\kappa}||_2/u^{(1)} < 10^{-15}$.
The volume fraction $L_2$ error norms are plotted in~\fig{1d-pulse}.
The volume fraction suffers numerical diffusion. However, the scheme order of accuracy is maintained as expected.
The differences in the reference and the numerical order of accuracy slopes are less than 5\% suggesting a reasonable match.
The results shown here are without using the stiff relaxation solver. Since the uniform pressure/velocity condition is maintained between both the phases anyway, using the stiff relaxation solver did not make any difference. Those results are not shown here for brevity.

\subsection{Multiphase shock tube}
\label{sec:srl_abg_t2}

The objective of this test is to evaluate the method's ability to deal with sharp jumps across a material interface~\cite{abgrall_jcp_2003}.  A 1D computational domain of length $L=1$ m is used.
A phase interface is initialized at $x/L=0.8$. The flow variables at the initial time ($t_0$) are set as $\alpha^{(1)} = \epsilon$, $p^{(2)} = p^{(1)} = 2 \times 10^{8}$ for $x/L<0.8$,  and $\alpha^{(2)} = \epsilon$, $p^{(2)} = p^{(1)} = 10^{5}$ for $x/L \ge 0.8$. The initial velocities are $u^{(1)} = u^{(2)} = 0$ everywhere.
The initial densities are set as $\rho^{(1)} = 50$ and $\rho^{(2)} = 1000$ everywhere.
Based on this, the pressure and the velocity equilibrium are already maintained at $t_0$. The phase~(1) is modeled as CPG with $\gamma^{(1)} = 1.4$ to mimic compressed air, whereas the phase~(2) is modeled as SG with $\gamma^{(2)}=4.4$, $p_0^{(2)}=6 \times 10^8$ to mimic liquid water.
Supersonic outflows are used at both ends of the tube to let the waves pass without reflection.
Surface tension, viscous terms, and interface compression are absent for this test.
Four different grids with the number of cells $N_x = 200,\, 500,\, 1000,\, 2000$ are tested with both the first- and the second-order-accurate schemes.  

\begin{figure}
   \begin{center}
   \subfigure{\includegraphics[width=4in]{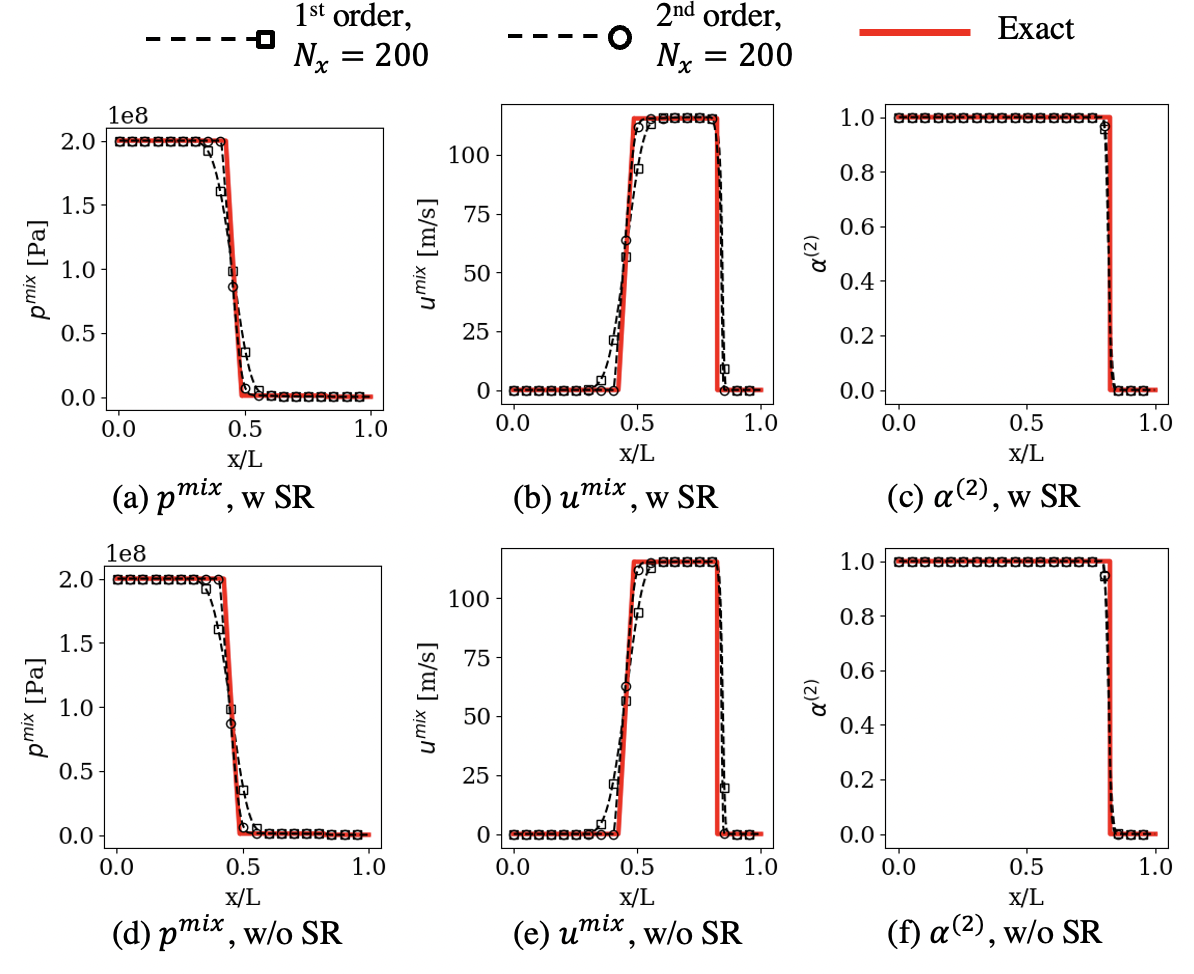}}
   \caption{Results for a multiphase shock tube using the seven-equation DIM are compared against an exact solution.}
   \label{fig:srl_abg_t2}
  \end{center}
\end{figure}

The simulation is conducted for 0.2 ms, and the final solution is compared against an analytical `truth' solution computed via an exact Riemann solver as shown in~\fig{srl_abg_t2}. 
A shock propagates on the right side of the interface (in air), and a rarefaction travels towards the left (in water).
The contact wave, and therefore, the interface, also moves slightly towards the right.
Note that the flow variables for both phases ($\kappa = 1,\, 2$) are evolved for the seven-equation model. However, for the resolved interface simulations that are considered here, only the mixture properties computed as $f^{mix}$ are of relevance, and only those are compared.
The simulation results match the exact solution with increasing accuracy with an increase in the grid resolution and the order of the scheme (see~\fig{srl_abg_t2}).
For instance, the normalized $L_2$ error norm of the pressure ($||p^{mix}||_2/p^{(2)}_{x/L<0.8}$) reduces from $0.07$ to $0.03$ going from $N_x = 200$ to $N_x = 1000$ for the first order scheme, and almost a similar improvement in the accuracy is obtained for $N_x = 200$ by going from the first to the second order scheme. 

Results with and without the stiff relaxation (SR) solver match each other closely at all grid resolutions and scheme orders.
For example, the normalized $L_2$ norm of the difference in $p^{mix}$ with and without the SR for $N_x=200$ and the second-order-accurate scheme is 0.002.
This shows that the use of the DEM relieves the need to use a separate stiff relaxation solver.
The DEM accurately models the interphase exchanges. The pressure and the velocity equality between the phases are obtained within the numerically diffused interface (\cite{abgrall_jcp_2003} for further details) even without enforcing it.
When a stiff pressure relaxation is applied the pressure and the velocity equality are enforced everywhere, and therefore, $p^{mix}=p^{1}=p^{2}$, $u^{mix} = u^{1} = u^{2}$. However, without it, $p^1$, $p^2$ and $u^1$, $u^2$ can be different away from the numerically diffused interface, but $f^{mix} \approx f^{\kappa}$ is still be maintained in the pure fluid regions as $\alpha^{\kappa} \approx 1$.

\subsection{Shock interaction with material interface}
\label{sec:1dshock}

The passage of a shock through a material interface is simulated in this test.
A 1D computational domain of length $L=1$ m is used.
The region $x/L < 0.5$ is filled with air ($\alpha^{(2)}=\epsilon$, CPG, $\gamma^{(1)}=1.4$), whereas, $x/L > 0.5$ is Aluminum ($\alpha^{(1)}=\epsilon$, SG, $p_0^{(2)} = 21.13 \times 10^{9}$, $\gamma^{(2)}=3.8$).
A shock of strength Mach 2 is initialized in the air at $x/L=0.3$ to propagate towards the material interface.
The pre-shock conditions are set to $p^{(1)} = p^{(2)} = 10^{5}$ Pa, $u^{(1)} = u^{(2)} = 0$ m/s,  and $\rho^{(1)} = 1.2$ kg/m$^3$,  $\rho^{(2)} = 2784$ kg/m$^3$.
The post-shock conditions for $x/L<0.3$ are computed using the Rankine-Hugoniot relations for air as $p^{(1)} = 4.56 \times 10^{5}$ Pa, $u^{(1)} = 429.0$ m/s, and $\rho^{(1)} = 3.211$ kg/m$^{3}$.
  
Only pure air is present for $x/L<0.3$ in reality. However, we also need to specify the flow fields for phase~(2) for the seven-equation numerical modeling.
The different initial conditions (IC) attempt to understand their effects on the final solution.

\begin{itemize}
    \item[IC-1]: $\epsilon=10^{-3}$, $p^{(2)} = p^{(1)}$ and $u^{(2)} = u^{(1)}$ for $x/L<0.3$
    \item[IC-2]: $\epsilon=10^{-6}$, $p^{(2)} = p^{(1)}$ and $u^{(2)} = u^{(1)}$ for $x/L<0.3$
    \item[IC-3]: $\epsilon=10^{-9}$, $p^{(2)} = p^{(1)}$ and $u^{(2)} = u^{(1)}$ for $x/L<0.3$
    \item[IC-4]: $\epsilon=10^{-3}$, $p^{(2)} = 10^5$Pa and $u^{(2)} = 0$ for $x/L<0.3$.
\end{itemize}

For the first three IC, the pressure and the velocity equilibria are already attained between the phases at $t_0$.
However, this can be a potential problem since the pre/post-shock conditions that are specified correspond to a Mach 2 shock in the air, and they are not for the liquid, which still has a volume fraction of $\epsilon$ in this region.
This would create a Riemann problem for the liquid resulting in artificial waves even before the shock hits the material interface.
IC-4 aims at solving this problem by initializing phase~(2) with uniform pressure/velocity even though phase~(1) is initialized as a shock.
This means that the SR cannot be enforced for this, but there would not be any artificial waves at the start.

\begin{figure}
   \begin{center}
   \subfigure{\includegraphics[width=4.2in]{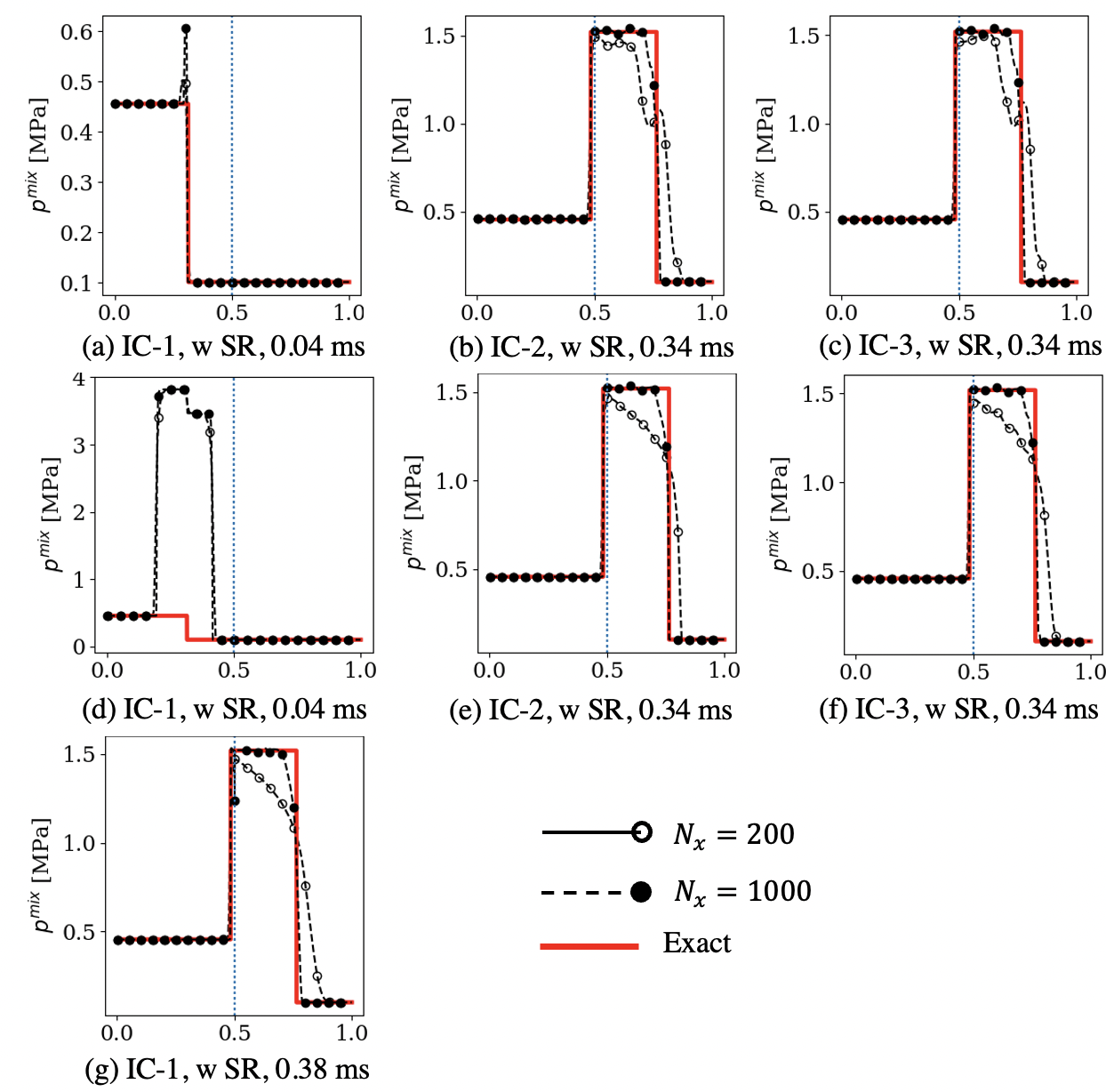}}
   \caption{The $p^{mix}$ is shown for the 1D shock interacting with Air/Al material interface. The vertical dashed line shows the location of the material interface. The results shown here are with the second-order scheme.}
   \label{fig:1d-shock}
  \end{center}
\end{figure}

The results at $t=0.3$ ms, after the shock has imparted the material interface, are analyzed and compared against an analytical solution~\cite{sridharan_jap_2015} in~\fig{1d-shock}.
Depending on the impendence of the materials and the shock strength,  the imparted shock would reflect/transmit as either a shock or a rarefaction.
In this case, both the reflected and the transmitted waves are shocks.
Due to the shock impact, the material interface also moves, although for this case, its velocity is only 0.095 m/s, and it is not noticeable from these plots.

In terms of the effect of the IC, due to the inconsistent pre/post-shock conditions as discussed before and a higher $\epsilon$, as expected, IC-1 shows significant errors resulting from the artificial waves created in the phase~(2) near the initial shock location ($x/L=0.3$).
The errors are already present at t=0.04 ms, well before the shock hits the interface, and therefore, the results at later times are also corrupted (not shown here).
Both IC-2 and IC-3 show a significant improvement over IC-1.
Since $\epsilon$ is smaller for them, even though the artificial waves are created in phase~(2) initially, their effects on the mixture quantities, i.e., $p^{mix}$, $u^{mix}$ seem to be minimal.
IC-4 also seems to be a valid initialization approach for the seven-equation model and captures the shock transmission/reflection through the material interface, even with $\epsilon=10^{-3}$ as there are no artificial waves due to the initialization anymore.
However, as noted before, this IC cannot be used with an SR solver since it is not enforced at the $t_0$.

The results shown here are only for the second-order scheme. They improve accuracy with grid refinement while comparing against the reference exact solution (except for IC-1).
Like the multiphase shock tube test earlier,  the results with and without a stiff relaxation (SR) show a similar trend and reasonably match.
Another test, where a shock initiated in the Aluminum enters the air, was also simulated. For this, the transmitted wave is a shock, whereas the reflected wave is a rarefaction due to the lower impendence of the air.
These results are not shown here for brevity.

\begin{figure}
   \begin{center}
   \subfigure{\includegraphics[width=2.8in]{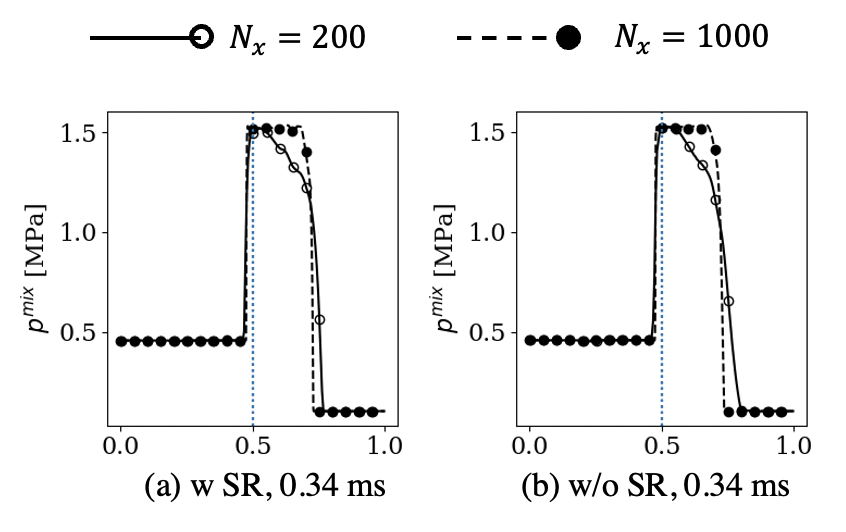}}
   \caption{The $p^{mix}$ is shown for the 1D shock interacting with Air/HMX material interface.  The results are at $t=0.38$ ms.  The vertical dashed line shows the location of the material interface. The results shown here are with the second-order scheme.}
   \label{fig:1d-shock-HMX}
  \end{center}
\end{figure}

Next, to demonstrate the ability of this method to handle arbitrary EOS, the Al that was previously modeled using SG is now replaced with HMX, which is modeled using an MG EOS with the same parameters as used in an earlier work~\cite{akiki_aiaa_2017}.
Unfortunately, an analytical solution to compare is not available here, but the obtained results are shown in~\fig{1d-shock-HMX}.
The results look qualitatively similar to the Al/air test, where the imparted shock resulted in a transmitted and a reflected shock at the interface.
These results are obtained with the IC-2 approach, and they prove the robustness of this method to handle arbitrary EOS and large pressure and density ratios.

\subsection{Surface tension effects}
\label{sec:surften}

Several tests are conducted for validating the surface tension modeling following~\citet{nguyen_amc_2015}.
A 2D square computational domain of width $L = 1$ m is filled with air for the first test.
A circular liquid droplet of radius $r_0 = 0.1549$ m is placed at its center.
Slip walls are used as boundary conditions.
As a result of the droplet curvature, higher pressure should be maintained within the droplet.
The objective of the first test is to evaluate the ability of the DEM surface tension modifications to handle and maintain this pressure jump.
Instead of computing the curvature from the volume fraction field, it is set to $\beta=1/r_0$ everywhere.
The volume fractions inside and outside the droplet are initialized as $\alpha^{(1)}=\epsilon$ and $\alpha^{(2)}=\epsilon$.
The velocities are initialized as zero everywhere.
The pressures are set as $p^{(1)}=10^{3}$ Pa and  $p^{(2)}=p^{(1)}+\sigma \beta$ everywhere, and this would result in a jump in $p^{mix}$ at the droplet interface.
In terms of the DEM flux computation, this is appropriate since this means that the pressure jump is only present at a phase~(1)--(2) interface, where we included modifications for the surface tension.
Whereas the pressure is uniform across the~(1)--(1) and the (2)--(2) interfaces.
The gas and the liquid phases are modeled using CPG and SG, respectively, with $\gamma^{(1)}=1.4$, $\gamma^{(2)}=4.4$, $p_0^{(2)}=6\times 10^{8}$.
The surface tension is set to $\sigma=342$ N/m.
The gas-phase and the liquid-phase densities are $\rho^{(1)}=1.0$ kg/s and $\rho^{(2)}=100$ kg/s, respectively.
This test is conducted both with and without the stiff relaxation solver.
However, the pressure difference $p^{(2)}-p^{(1)}$ is already maintained to $p_{\sigma}= \sigma \beta$ from the start, which would mean that the relaxation solver does not have to act.

Three different grids, $200 \times 200$, $400 \times 400$, and $800 \times 800$ and both first and the second-order schemes are tested.
The simulations evolve until a physical time of 0.085 s, and at the end of it, the $L_2$ error norms for $p^{mix}$, $u^{mix}$ and $\alpha^{(2)}$ are computed considering the initial conditions as the truth.
For all the grids, for all the orders of the scheme, with or without the stiff relaxation solver, and with or without the interface compression, the normalized errors are maintained to machine accuracy ($<10^{-15}$), suggesting that the DEM modifications for the surface tension can handle the surface jump condition exactly.

\begin{figure}
   \begin{center}
       \subfigure{\includegraphics[trim={0.1in 0.1in 0.1in 0.1in}, clip, width=\fighalf]{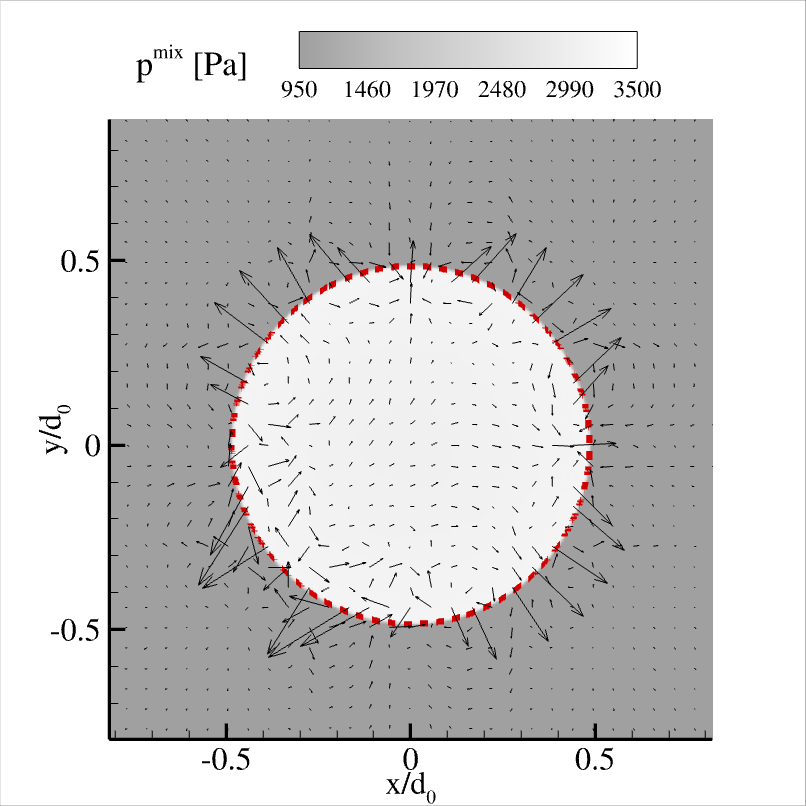}}
       \caption{Velocity vectors and $p^{mix}$ contours are shown for the 800$\times$800 cylindrical droplet case without the stiff relaxation and interface compression at t=0.085 s. The droplet interface $\alpha^{(2)}=0.5$ is shown via a dashed line.}
       \label{fig:surftens_1}
  \end{center}
\end{figure}

For the second test, the same setup is used.
However,  the curvature is computed from the volume fraction field as described in \sect{curvature} instead of setting it as a constant.
The pressure within the domain is initially set  as $p^{(1)} = p^{(2)} = 10^{3}$ Pa, and as a result of the surface tension force, it is expected to rise within the droplet (ideally to $\sigma \beta = 2207$ Pa).
The simulations are run for 0.034 s ($\sim$40,000 steps for the 800$\times$800 grid), and the pressure jump within the droplet at the end of it is measured by integrating over the entire computational domain $\Omega$ as 
\begin{equation}
    \Delta p = \int_{\Omega} p^{(2)} \alpha^{(2)} \bigg/ \int_{\Omega} \alpha^{(2)} - \int_{\Omega} p^{(1)} \alpha^{(1)} \bigg/ \int_{\Omega} \alpha^{(1)}.
\end{equation}
This is compared against the exact pressure difference based on Young-Laplace relation as $\Delta p^{err} = |\Delta p - \sigma \beta|$.
Even though the background flow should have been at rest, parasitic currents are observed due to the numerical errors in the curvature computation.
They are consistent with the previous works~\cite{nguyen_amc_2015}.
These are quantified using the kinetic energy as 
\begin{equation}
    KE^{mix} = \int_{\Omega} \left( \frac{1}{2} \alpha^{(1)} \rho^{(1)} {u^{(1)}}^{2} +  \frac{1}{2} \alpha^{(2)} \rho^{(2)} {u^{(2)}}^{2}   \right) \bigg/ \int_{\Omega} \left( \alpha^{(1)} + \alpha^{(2)} \right) .
\end{equation} 
Values of both the $\Delta p^{err}$ and the $KE^{mix}$ are reported in~\tab{surftens_1} for various grids and with/without the stiff relaxation (SR) and with/without the interface compression (comp.).
Only the second-order scheme results are shown here.
However, the results with the first-order scheme are also consistent with the conclusions here.
The lowest errors are observed for the case without the SR but with the interface compression.
The errors in the computed pressure jump are less than 2\% for all the grids for this case, and the parasitic velocities are $<0.2$ m/s.
The parasitic currents and the $p^{mix}$ within the droplet for this case are shown in~\fig{surftens_1} for reference.

Regardless of the interface compression, the errors with the SR are at least an order of magnitude higher than those without the SR.
Further insights into this are provided below.
The errors do not necessarily reduce with the grid, and this could be due to the contrasting effects, as we would compute the curvature more accurately for the finer grids.
However, the coarser grids would help the parasitic currents to diffuse numerically.
The errors with the interface compression are also significantly lower than those without it since the droplet interface becomes irregular due to the parasitic currents.
Still, the interface compression helps keep it regular, resulting in a more accurate curvature computation.

\begin{table}
\def\arraystretch{1.2}
\caption{Errors $\Delta p^{err}$ [Pa] and $KE^{mix}$ [kg m$^2$/s$^2$] for the circular droplet surface tension test are shown for various grids and various simulation options.}
\begin{center}
\begin{tabular}{c|c|c|c|c|c|c}
    \hline
    Case & \multicolumn{2}{c}{200$\times$200} & \multicolumn{2}{c}{400$\times$400} & \multicolumn{2}{c}{800$\times$800}\\
     &  $\Delta p^{err}$ & $KE^{mix}$ & $\Delta p^{err}$ & $KE^{mix}$ & $\Delta p^{err}$ & $KE^{mix}$ \\\hline
    w/o SR, w/o comp. &  $1.06 \times 10^{2}$ & $7.99 \times 10^{-6}$ & $2.64 \times 10^{2}$ & $5.92 \times 10^{-5}$ & $2.09 \times 10^{2}$ & $4.35 \times 10^{-5}$ \\\hline
    w SR, w/o comp. & $8.84 \times 10^{2}$ & $1.58 \times 10^{-4}$ & $1.18 \times 10^{3}$ & $1.56 \times 10^{-3}$ & $1.25 \times 10^{3}$ & $2.18 \times 10^{-3}$ \\\hline
    w/o SR, w comp. & $3.52 \times 10^{1}$ & $2.26 \times 10^{-6}$ & $1.49 \times 10^{1}$ & $2.06 \times 10^{-6}$ & $5.86 \times 10^{1}$ & $2.10 \times 10^{-6}$ \\\hline
    w SR, w comp. & $1.44 \times 10^{2}$ & $1.68 \times 10^{-4}$ & $1.01 \times 10^{3}$ & $4.59 \times 10^{-4}$ & $4.66 \times 10^{2}$ & $1.49 \times 10^{-3}$
     \\\hline
\end{tabular}
\end{center}
\label{tab:surftens_1}
\end{table}

\begin{figure}
    \begin{center}
        \subfigure{\includegraphics[width=4.2in]{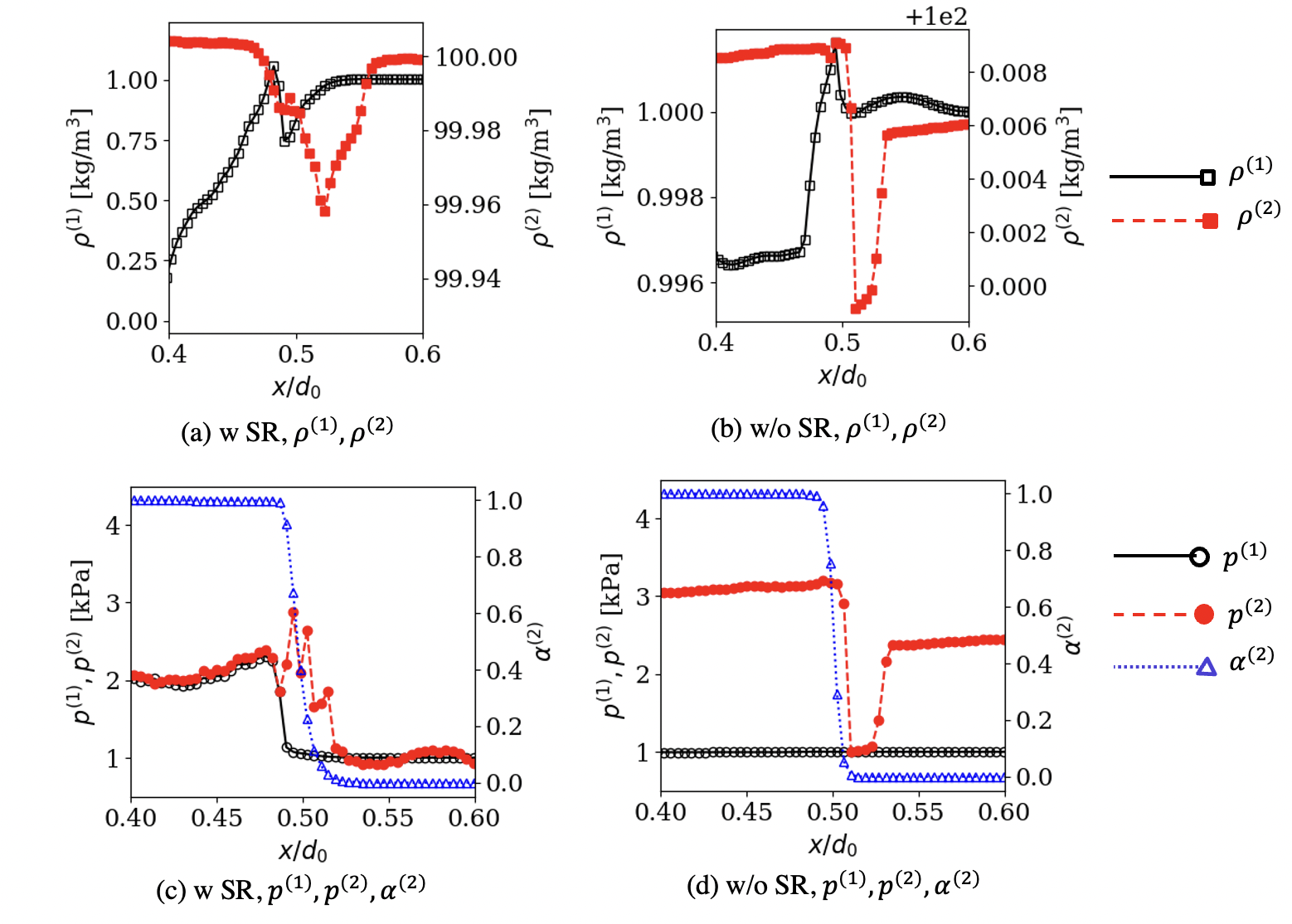}}
        \caption{$p^{(1)}$, $p^{(2)}$, $\rho^{(1)}$, $\rho^{(1)}$, and $\alpha^{(2)}$ are plotted along the centerline for the 800$\times$800 test with and without the SR.}
        \label{fig:surftens_2}
    \end{center}
\end{figure}

To understand the reasons for the large errors with the SR, $p^{\kappa}$, $\rho^{\kappa}$, and $\alpha^{(2)}$ are plotted along the centerline for the 800$\times$800 test with and without the SR in~\fig{surftens_2}.
As noted before, the DEM modifications in the surface tension are designed so that a pressure jump is handled across the~(1)-(2) interface, whereas the~(1)--(1) and the (2)--(2) interfaces should behave as before.
As expected, the case with the SR maintains pressure equality away from the interface and enforces a pressure jump between $p^{(1)}$ and $p^{(2)}$ within the numerically diffused interface.
However, because of this,  pressure jumps are present in both $p^{(1)}$ and $p^{(2)}$ near the boundaries of the diffused interface.
They result in pressure jumps at the~(1)--(1) and the (2)--(2) interfaces within the DEM, possibly causing these larger errors.

On the other hand, for the case without the SR, $p^{(1)}$ stays constant, and only the $p^{(2)}$ changes within the numerically diffused interface.
The liquid pressure $p^{(2)}$ reduces almost to the same value as $p^{(1)}$ right before the interface ($x/d_0 > 0.5$) and then increases back due to the surface tension terms at the interface.
The initial reduction in it is still in the gas phase (i.e., $\alpha^{(2)}$ is small), and therefore, it does not adversely affect the overall mixture ($mix$) flow fields.
The surface tension is accurately captured without the SR within the numerically diffused interface due to the DEM modifications in the~(1)--(2) Riemann problem.
The errors mentioned above due to the SR are also apparent in the density fields, as $\rho^{(1)}$ in the presence of the SR increases by almost $100$\% within the droplet.
In contrast, the oscillations in both $\rho^{(1)}$ and $\rho^{(2)}$ are $<0.5$\% for the case without the SR.

For the third test, instead of starting from a circular droplet, an elliptical drop is used with the following shape~\cite{perigaud_jcp_2005, nguyen_amc_2015}
\begin{equation}
    \frac{(x-x_0)^2}{0.2^2} + \frac{(y-y_0)^2}{0.12^2} = 1.
\end{equation}

The ellipsoidal droplet starts converting into a circular shape because of the initial higher/lower curvature on the horizontal/vertical ends of the droplet.
The potential energy stored in the droplet's surface tension converts into kinetic energy. The droplet oscillates from an initial ellipse into an intermediate circular shape and back to an ellipse.
The kinetic energy oscillates during this process, and we can compute its frequency against an analytical value based on the Rayleigh formula~\cite{rayleigh_1879}
\begin{equation}
    \omega = (o^3 - o) \frac{\sigma}{(\rho^{(1)} + \rho^{(2)})r_0^{3}}.
\end{equation}
where $o$ is the oscillation mode.
Corresponding to $o=2$ and $r_0 = 0.1549$ m, an oscillation frequency is determined as $t_{osc} = 2 \pi / \omega = 0.085$ s.

\begin{figure}
  \begin{center}
   \subfigure{\includegraphics[width=5.5in]{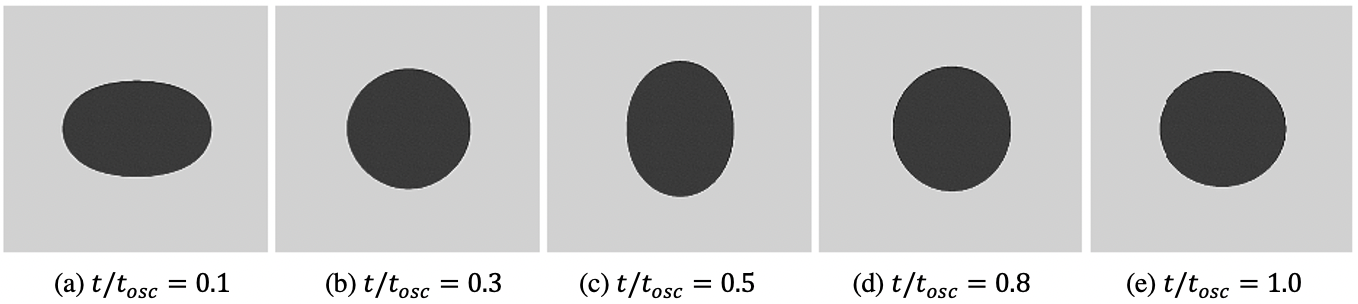}}
     \caption{Oscillating ellipsoidal droplet at various times for the 800$\times$800 grid without the interface compression.}
   \label{fig:surftens_4}
  \end{center}
\end{figure}

\begin{figure}
   \begin{center}
   \subfigure{\includegraphics[width=6.0in]{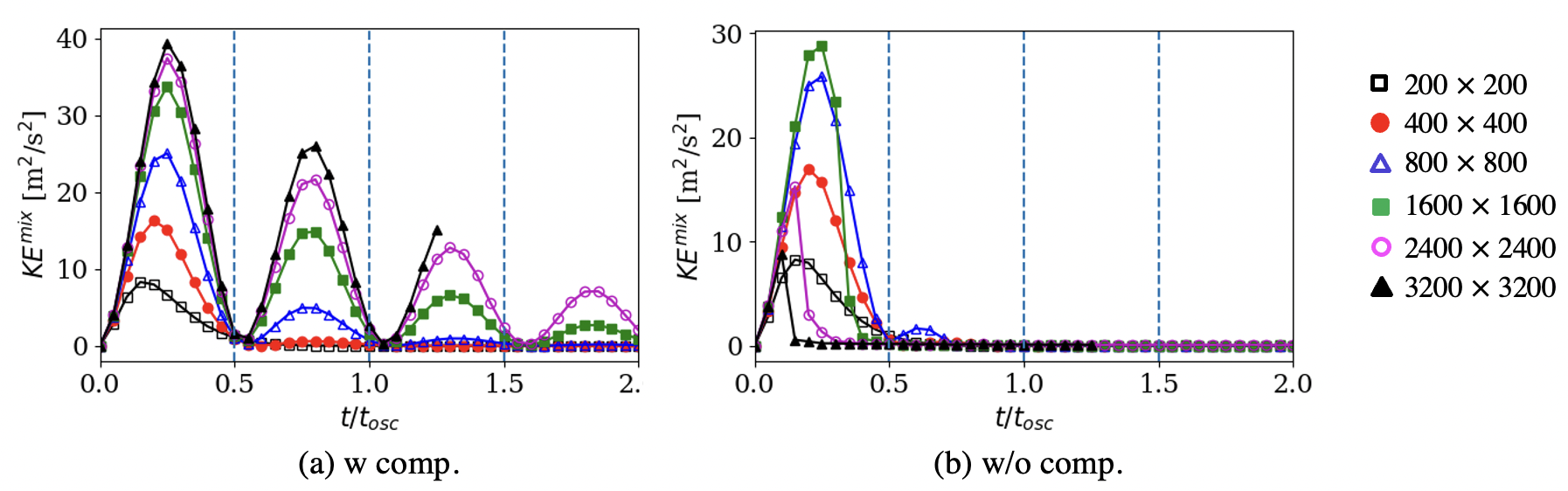}}
   \caption{$KE^{mix}$ [kg m$^2$/s$^2$] for an oscillating ellipsoidal droplet. The dashed vertical lines correspond to the exact oscillation period.}
   \label{fig:surftens_3}
  \end{center}
\end{figure}

Based on the conclusions from the previous test, this test is simulated without the SR.
The evolution of the droplet shape at various times is shown~\fig{surftens_4}, and corresponding $KE^{mix}$ is shown in~\fig{surftens_3}.
Similar to the previous test, interface compression is necessary for accurate curvature and surface tension computation.
As expected, the initial elliptical droplet converts to a circular droplet by $t/t_{osc}=0.3$.
This is accompanied by an increase in the $KE^{mix}$.
The droplet regains an elliptical shape by $t/t_{osc}=0.5$ and $KE^{mix}$ shows reduces to zero.
Ideally, this process should continue in the absence of any viscous effects.
However, because of the numerical diffusion here, the peaks of $KE^{mix}$ reduce with time.
The 400$\times$400 grid only captures two peaks of $KE^{mix}$, whereas the finer grids can capture more.
The computed oscillation frequency matches the exact value with $<$15\% accuracy.

\subsection{Viscous effects}
\label{sec:visc}

\begin{figure}
   \begin{center}
   \subfigure{\includegraphics[width=4.5in]{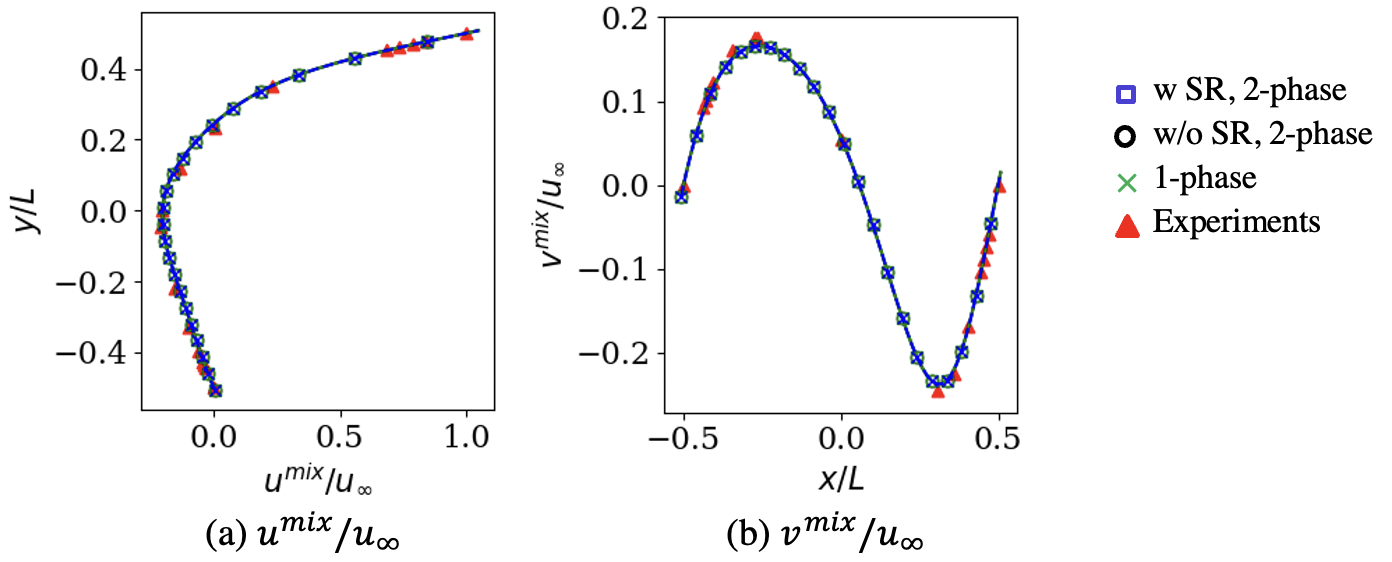}}
   \caption{Steady-state velocities are compared against an exact solution for the lid-driven cavity test.}
   \label{fig:liddriven}
  \end{center}
\end{figure}

To verify the implementation of the viscous terms,  a lid-driven cavity with a Reynolds number (Re) of 100 is simulated first.
Initially, the fluid is at rest in a square domain of size $L$.
The top wall is accelerated to a velocity $u^{\infty}$ at $t=0$, and all the other walls stay at rest.
The flow within the domain starts moving and rotating due to the no-slip walls and the viscous effects.
The simulations are run until a steady-state solution is achieved.
Single-phase simulations~\cite{gallagher_2017} for this case have been established before, and the results match with an exact solution~\cite{ghia_jcp_1982}.
Two-phase results are obtained here with the DIM, where both the phases are set as CPG and with the same properties.
The volume fraction $\alpha^{(2)}$ is initialized corresponding to a circular droplet within the domain using~\eq{circle_init}.
Both the single-phase and the two-phase simulation results with and without the SR match the exact solution as shown in~\fig{liddriven}, verifying the implementation of the viscous terms.
These results are with the second-order scheme.

Next, to validate the viscous effects in two-phase flows, the drag of a droplet in a viscous medium is computed.
A 2D domain of length $L_x=15 r_0$ and width $L_y=10 r_0$ is used, where $r_0=0.5$ mm is the radius of an initially circular droplet initially placed at $(x,y)=(4 r_0, 5 r_0)$.
The gas and the liquid phases are modeled using CPG and SG, respectively, with $\gamma^{(1)}=1.4$, $\gamma^{(2)}=4.4$, $p_0^{(2)}=6\times 10^{8}$.
The velocity field surrounding the droplet is initiated using a potential flow solution, using a stream function 
\begin{equation}
    \psi = U_0 r \rm{sin}(\theta) [1-R_0^{2}/r^2], \quad u^{(1)}_1 = u^{(2)}_1 = - \partial \psi / \partial y, \quad u^{(1)}_2 = u^{(2)}_2 = \partial \psi / \partial x.
\end{equation} 
where $U_0$ is the free-stream velocity, and $\theta=1^{0}$ is used to trigger vortex shedding.
The velocities inside the droplet are initialized as zero.
The dynamic viscosity for the gas phase is set as $\mu^{(1)} = 1.8 \times 10^{-5}$ kg/(m-s), and for the liquid it is $\mu^{(1)} = 1.14 \times 10^{-3}$ kg/(m-s).
The surface tension is specified as $\sigma=0.0728$ N/m.
The Reynolds number ($Re$) is defined as $Re =2 \rho^{(1)} r_0 U_0/\mu_g$, and the Weber number ($We$) is defined as $We=2 \rho^{(1)} r_0 U_0^{2}/\sigma$.

Two conditions are simulated following a previous work using the five-equation model~\cite{patel_thesis_2007}.
One with $U_0=5$ m/s, that corresponds to $Re=81.3$ and $We=0.1$, and another  with $U_0=50$ m/s, that corresponds to $Re=813$ and $We=10$.
A time-scale for normalization is defined as $t_0 = r_0/U_0$.
Three grids are used with the number of cells across the droplet diameter $N_x=20,50,100$.
The simulations are conducted with and without the SR and with and without the interface compression.
The interface diffuses irregularly without interface compression, similar to the previous surface tension test.
The boundary and shear layers that are supposed to grow surrounding the sharp gas-liquid interface are wrongly predicted.
Therefore, only the results with the interface compression are discussed further in this section.

\begin{figure}
   \begin{center}
       \subfigure[$Re=81.3$, $We=0.1$, $t/t_0=15$]{\includegraphics[trim={0.1in 0.1in 0.1in 0.1in}, clip, width=\fign]{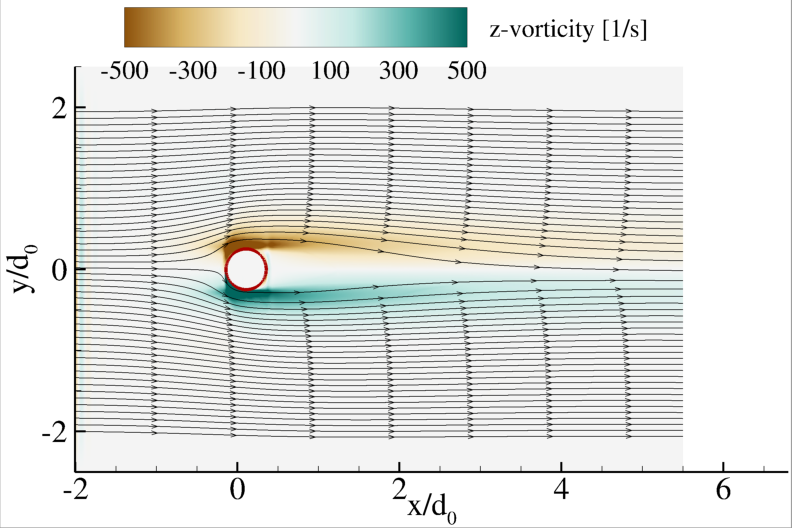}}
       \subfigure[$Re=813$, $We=10$, $t/t_0=39$]{\includegraphics[trim={0.1in 0.1in 0.1in 0.1in}, clip, width=\fign]{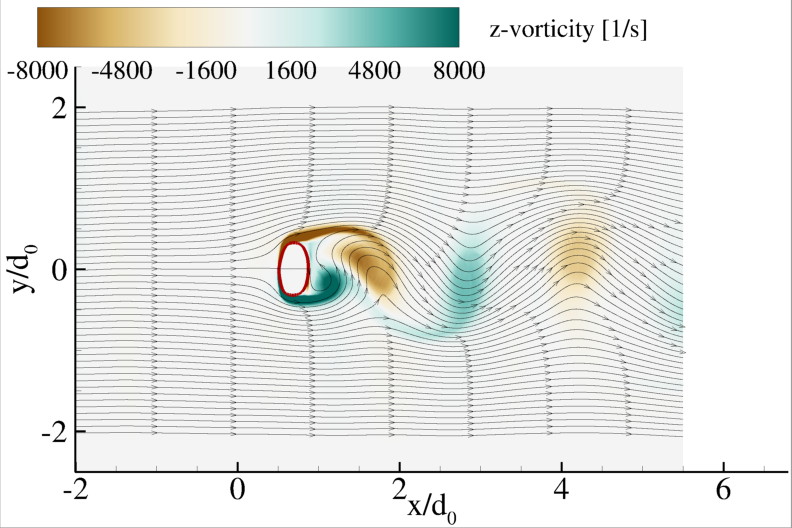}}
       \caption{Streamlines and vorticity contours are shown surrounding a droplet for the viscous computations. These results are without the SR with the interface compression and for the grid with $N_x = 100$}
   \label{fig:drag_2}
  \end{center}
\end{figure}

\begin{figure}
   \begin{center}
       \subfigure{\includegraphics[width=4.0in]{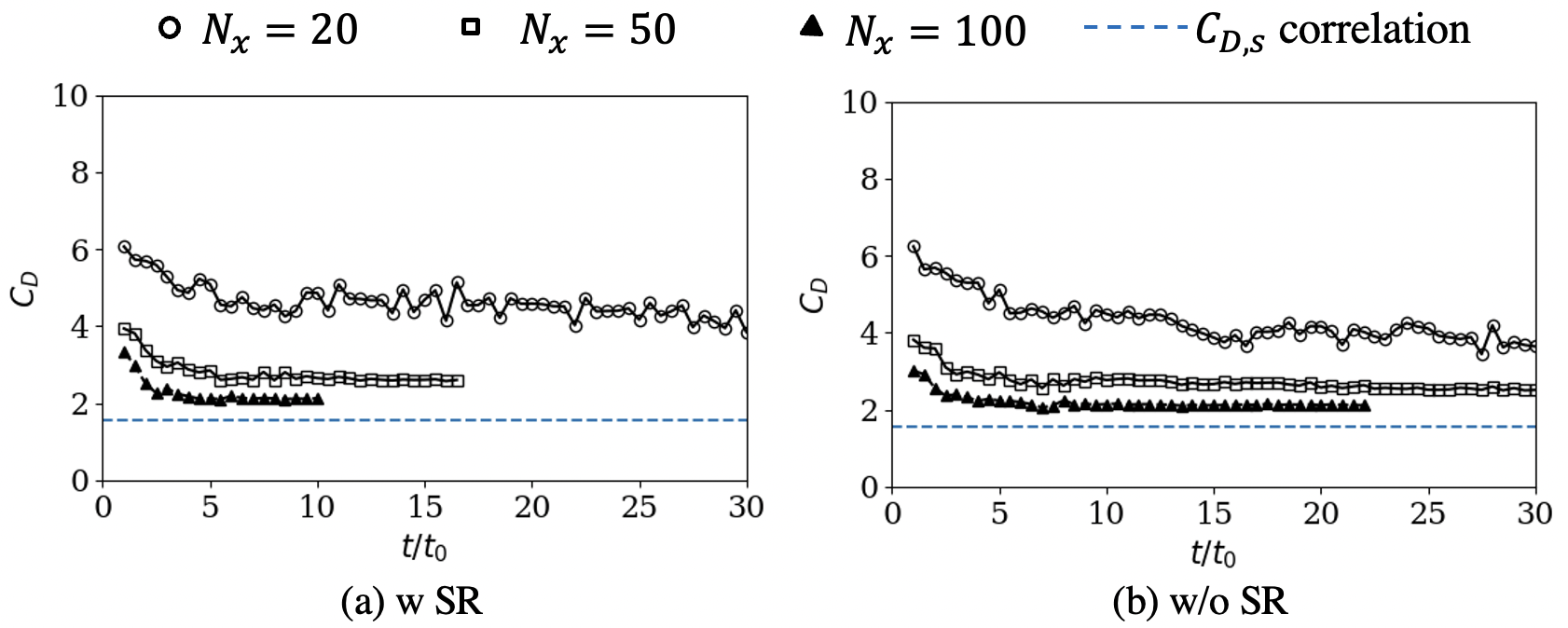}}
       \caption{Computed drag coefficient $C_D$ and its evolution with time is shown for $Re=81.3$, $We=0.1$.  }
   \label{fig:drag_1}
  \end{center}
\end{figure}

The flow fields for the two simulated conditions are shown in~\fig{drag_2}. 
The droplet at $We = 0.1$ maintains a circular shape, but it deforms for $We = 10$. 
The low $Re$ case develops a separation bubble behind the droplet, and the flow field remains relatively steady. 
On the other hand, the high $Re$ case demonstrates a vortex shedding process. 
The positive and negative velocities surrounding the droplet are due to the viscous effects. For further processing, the droplet velocity and a drag coefficient $C_D$ are computed as 
\begin{equation}
    U_c = {\int_{\Omega} \alpha^{(2)} u^{(2)}_1} \bigg/ {\int_{\Omega} \alpha^{(2)}}, \quad C_D = \frac{\rho^{(2)}}{\rho^{(1)}} \frac{dU_c}{dt} \frac{\pi r_0}{(U_0 -U_c)^2}.
\end{equation}
The computed $C_D$ is compared against a steady-state drag correlation for a rigid cylinder $C_{D,S}(Re)$~\cite{white_2006} for $We=0.1$ in~\fig{drag_1}.
The droplet with $We=10$ starts deforming, and therefore, a comparison against the correlation is not possible for that.
These results show that establishing the viscous flow surrounding the droplet takes approximately $5t_0$, and $C_D$ reaches a steady value after that.
The $C_D$ is overpredicted for the coarser grids. 
However, a better match against the correlation $C_{D,S}(Re)$ is achieved with grid-refinement.
The $C_D$ for the finest grid is also $\approx$10\% higher than the correlation $C_{D,S}(Re)$, but this could be attributed to the relative acceleration effect~\cite{temkin_jfm_1980} that is not included in the correlation $C_{D,S}(Re)$, but are present in the current simulations since the relative velocity $U_r = U_c - U_0$ decreases with time.
These results also show that even though the $C_D$ with and without the SR do not match exactly, they are very similar, and this further indicates the efficacy of the viscous DEM in accurately modeling the non-conservative interface terms such that there is no need to use the SR with the seven-equation model.

\subsection{Shock-droplet interaction}
\label{sec:2dshock}

This test simulates the early-stage deformation of a cylindrical water droplet when hit by a shock wave.
The setup follows the experiments by~\citet{igra_as_2002} and other numerical works that used a five-equation DIM~\cite{chen_aiaa_2008, meng_sw_2015}.
A circular droplet of diameter ($d_0$) 4.8 mm is placed in a 2D computational domain of size 23$d_0$ $\times$ 12$d_0$.
The background gas-phase is initialized with a shock of strength $M=1.47$ that is located at a distance $d_0$ ahead of the droplet center  The pre-shock (background) pressure and velocity are $p^{(1)}= p^{(2)}=101325$ Pa and $u^{(1)}=u^{(2)}=0$ m/s, respectively.
The corresponding post-shock conditions are subsonic, and they are set as $p^{(1)}=p^{(2)}=238558$ Pa and $u^{(1)}=u^{(2)}=u_0=226$ m/s.
The water density is set as $\rho^{(2)}=1000$ m/s.
The pre- and the post-shock gas-phase densities are $\rho^{(1)}=1.204$ kg/m$^3$ and $\rho^{(1)}=2.18$  kg/m$^3$, respectively.
Note that these initial conditions are equivalent to the IC-2 as described earlier in \sect{1dshock}.
The liquid water is modeled with a SG EOS with $\gamma^{(2)}=6.12$ and $p_0^{(2)}=343.44\times10^{6}$ Pa.
The gas phase is modeled with a CPG EOS with $\gamma^{(1)}=1.4$.
The volume fraction is initialized using~\eq{circle_init} with $\delta=0.5$.

The computational domain is discretized using a uniform grid $N_x = 17,\, 42,\, 83,\, 170,\, 255,\, 340$ cells across the droplet diameter.
The top and the bottom boundaries are set as slip walls.
The left boundary behind the shock is set as a non-reflective subsonic inflow at the post-shock velocity $u_0$.
The right boundary, ahead of the shock, is set as a supersonic outflow to let the incoming shock wave pass through without any reflections.
The Reynolds number ($Re$) and Weber number ($We$) corresponding to this condition are $Re = 4860$ and $We = 72320$, respectively.
Under these conditions, the droplet deformation is primarily driven by inertia during the early stages. Therefore, the surface tension and viscous effects can be neglected, as confirmed in the earlier numerical studies~\cite{chen_aiaa_2008, meng_sw_2015}.
Note that these effects would still be relevant near the final stages of the breakup and at much smaller length scales that are not resolved here.
The simulations are conducted and compared with and without the surface tension and with and without the interface compression.

\begin{figure}
   \begin{center}
   \subfigure[$\tau=0.03$, w SR]{\includegraphics[trim={0.1in 0.1in 1.4in 2.0in}, clip,width=\fign]{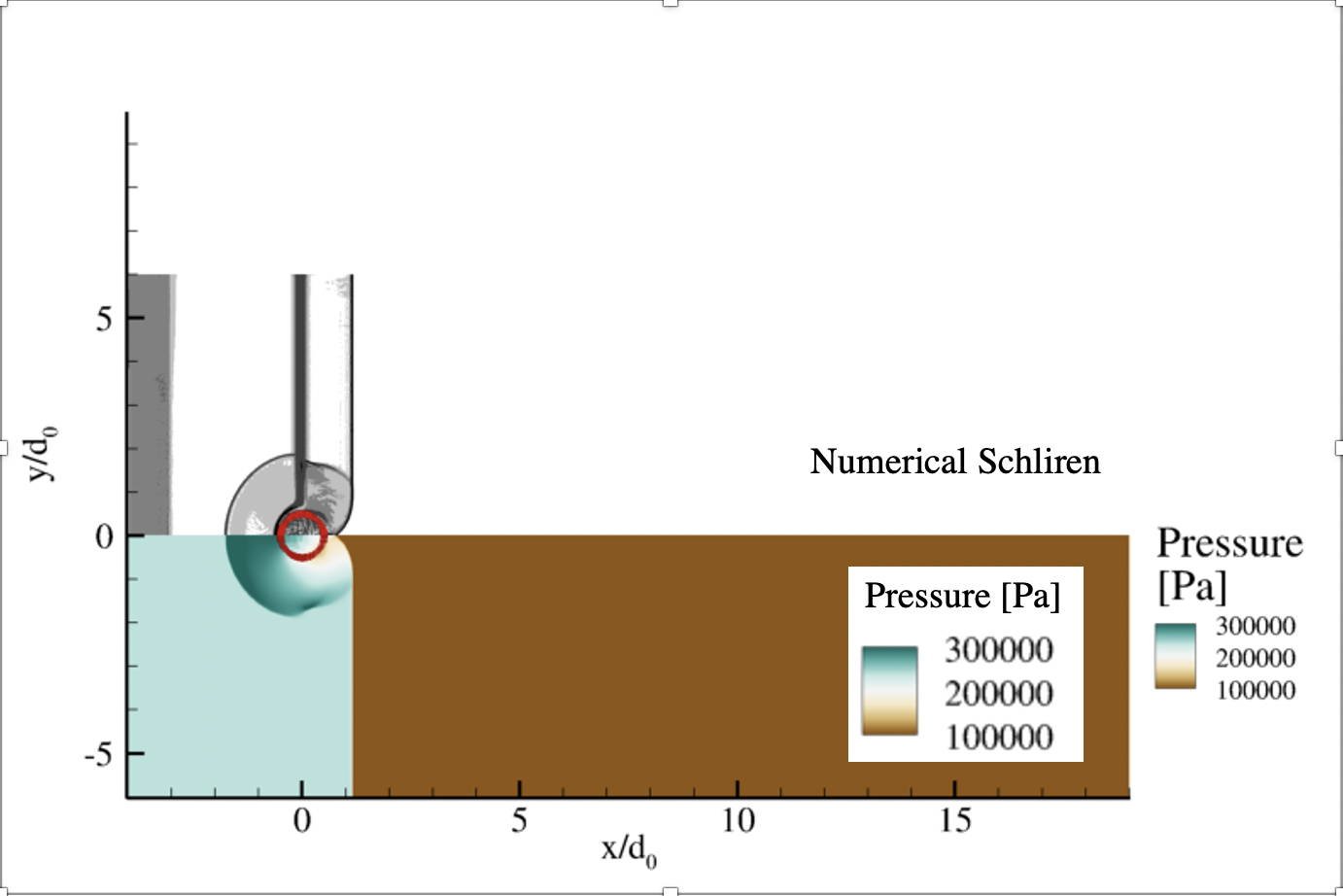}}
    \subfigure[$\tau=0.03$, w/o SR]{\includegraphics[trim={0.1in 0.1in 1.55in 2.0in}, clip,width=\fign]{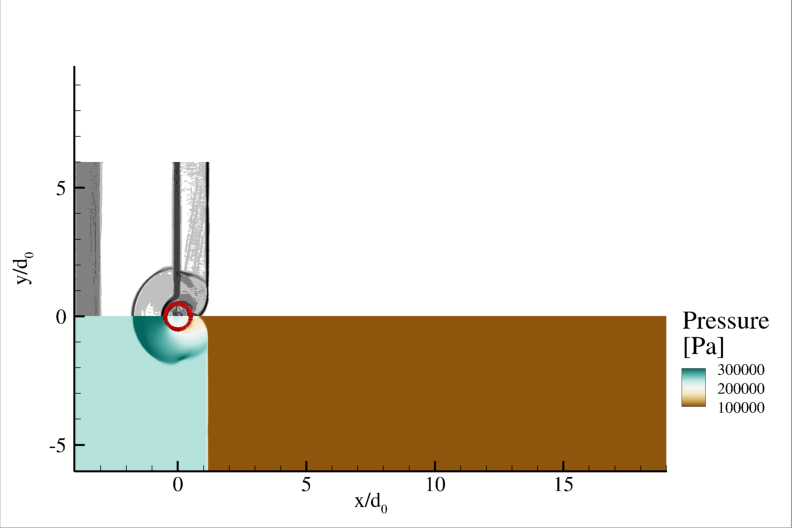}}\\
   \subfigure[$\tau=0.17$, w SR]{\includegraphics[trim={0.1in 0.1in 1.55in 2.0in}, clip,width=\fign]{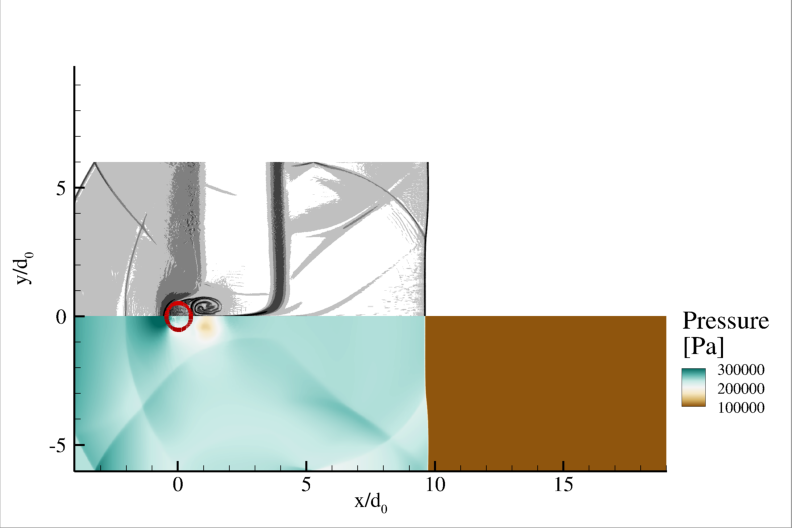}}
    \subfigure[$\tau=0.17$, w/o SR]{\includegraphics[trim={0.1in 0.1in 1.55in 2.0in}, clip,width=\fign]{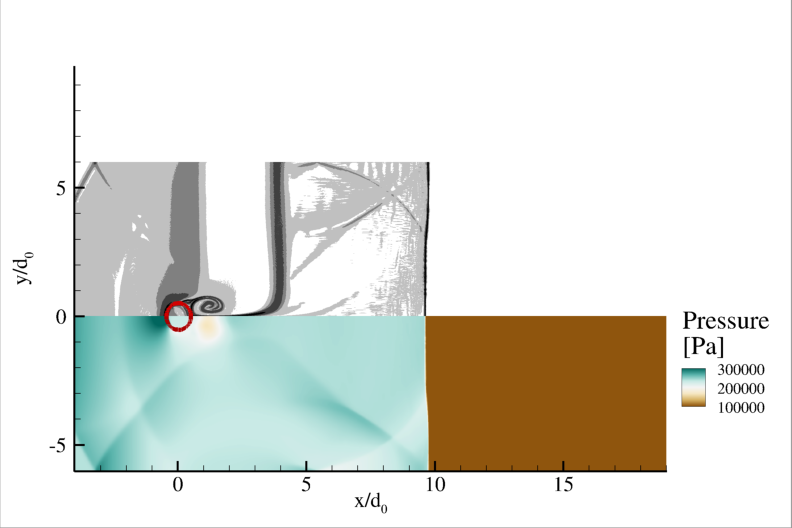}}\\
   \subfigure[$\tau=0.50$, w SR]{\includegraphics[trim={0.1in 0.1in 1.55in 2.0in}, clip,width=\fign]{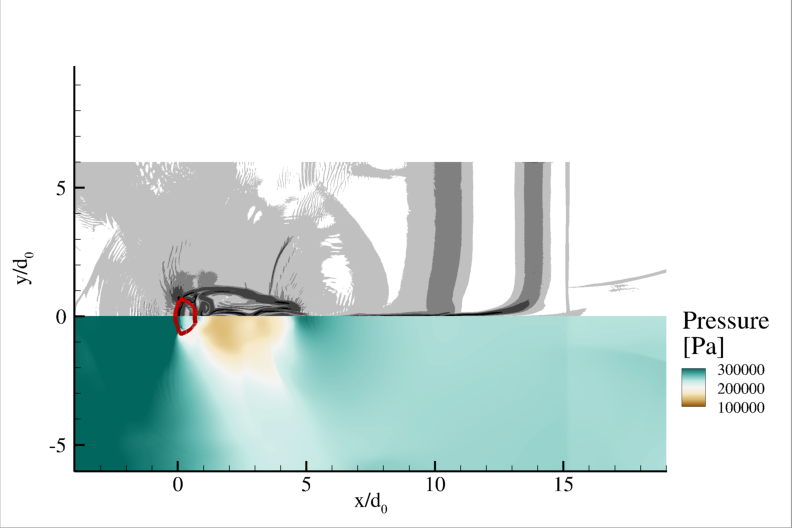}}
    \subfigure[$\tau=0.50$, w/o SR]{\includegraphics[trim={0.1in 0.1in 1.55in 2.0in}, clip,width=\fign]{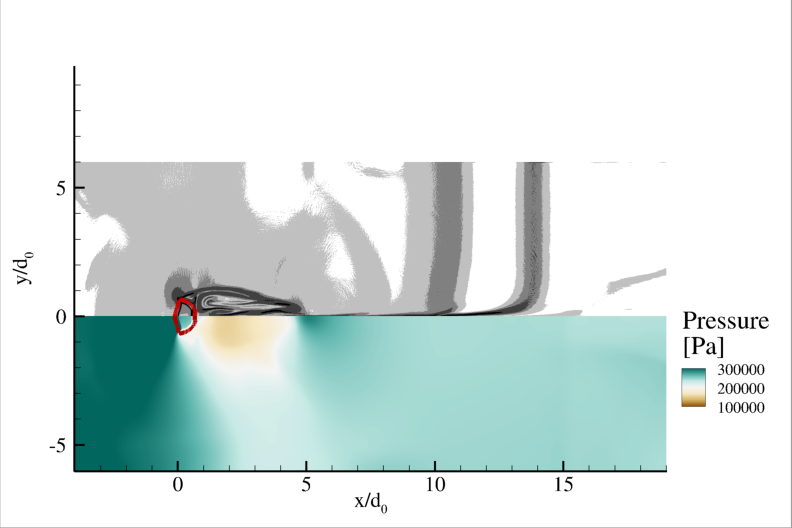}}  \\  
   \subfigure[$\tau=0.83$, w SR]{\includegraphics[trim={0.1in 0.1in 1.55in 2.0in}, clip,width=\fign]{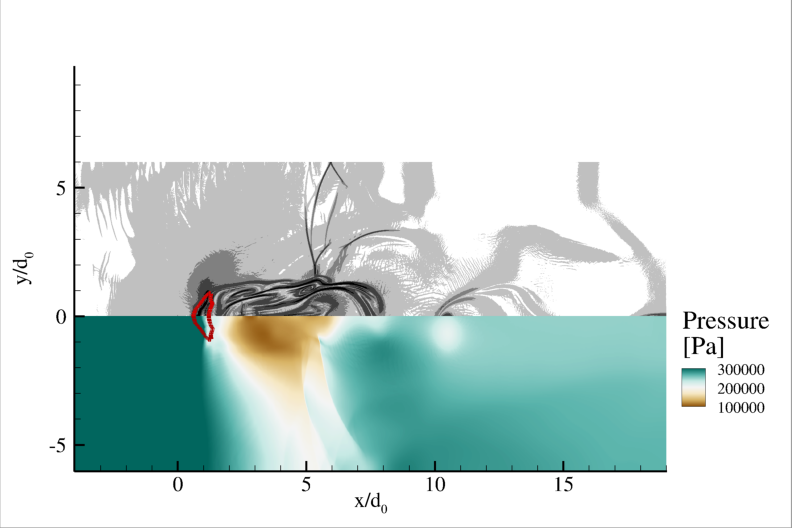}}
    \subfigure[$\tau=0.83$, w/o SR]{\includegraphics[trim={0.1in 0.1in 1.55in 2.0in}, clip,width=\fign]{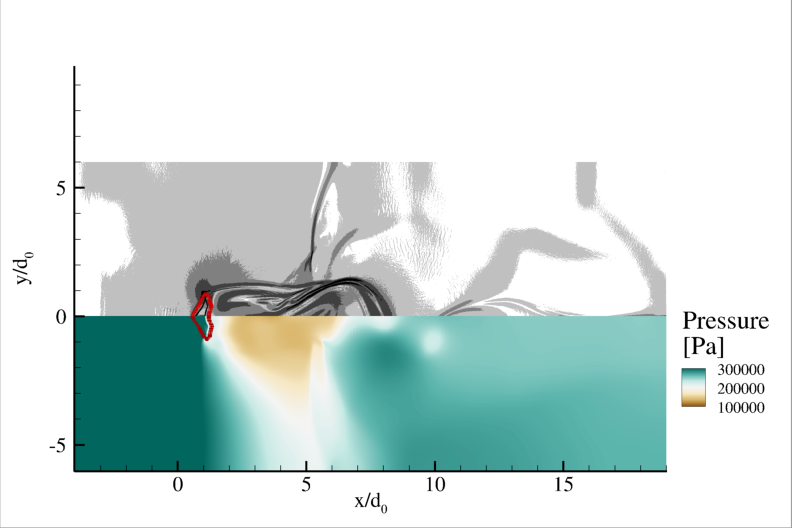}}
    \caption{Pressure contours (bottom) and numerical Schlieren (top) image are shown for the 2D shock-droplet interaction with $N_x=83$, with interface compression, and with/without the stiff relaxations.  The droplet is identified using a $\alpha^{(2)}=0.5$ contour (thick red line).}
   \label{fig:2d-shock1}
  \end{center}
\end{figure}

A non-dimensional time is defined as $\tau=t/ \left(u_0/d_0 \sqrt{\rho^{(1)} / \rho^{(2)}} \right)$, where the denominator characterizes time for breakup via Rayleigh–Taylor or Kelvin–Helmholtz instabilities~\cite{pilch_ijmf_1987}.
The simulation results at various times are shown in~\fig{2d-shock1} for the case with $N_x=83$.
These are with the interface compression and with/without the stiff relaxation.
Pressure contours and Numerical Schliren ($|\partial \rho^{(1)}/\partial x_{\indx{i}}|$) are used to identify the shock structure.
The droplet is using a $\alpha^{(2)}=0.5$ contour.
As the initial shock hits the droplet, there are transmitted and reflected waves.
Unlike the previous 1D case, the shock structures are no longer planar because of the circular droplet.
The waves that propagate inside the cylinder transmit and reflect through the material interface.
The droplet remains circular during the earlier stages (up to $\tau\approx0.17$). However,  it starts to deform due to the vortical structures created around it later.
These features are qualitatively similar to the experimental observations~\cite{igra_as_2002} and previous numerical results~\cite{chen_aiaa_2008, meng_sw_2015}.
The results with and without the SR match closely, reconfirming the ability of the DEM to alleviate the need to use an SR.

The experiments measured the droplet motion and the droplet deformation in four geometrical parameters.
1) Leading edge drift ($x_L$), that is, the x-location of the droplet leading edge measured with respect to its initial location, 2) droplet centerline width ($w$), which measures the width of the cylinder along the $x$-axis, 3) droplet spanwise diameter ($d$), that is the width of the droplet in the $y$-direction, and 4) droplet area ($A$).
All these are computed from the simulation flow-fields using a threshold volume fraction $\alpha^{(2)}_T$, considering that $\alpha^{(2)} > \alpha^{(2)}_T$ is the droplet shape whose geometrical parameters have to be determined.

Because of the uncertainty in the experimental measurements, it is unclear
what value $\alpha^{(2)}_T$ should take to best match the data.
Even though the gas-liquid interface is supposed to remain sharp in reality, this region may appear diffused depending on the measurement technique because of the small droplets strip surrounding the parent droplet surface.
In addition to this,  the interface also numerically diffuses for the simulations.
The results are analyzed for $\alpha^{(2)}_T = 0.1,\, 0.5,\, 0.9$, but only those with $\alpha^{(2)}_T = 0.9$ are shown here.
Sensitivity of five-equation DIM results in the choice of $\alpha^{(2)}_T$ can be found elsewhere~\cite{meng_sw_2015} and our observations are similar for the cases without the interface compression.
The simulation results with the interface compression do not show a sensitivity to $\alpha^{(2)}_T$ since a regular interface thickness is maintained. However, those results are considered erroneous for this test, and further details are discussed later.

The simulations with the SR and the interface compression take longer to run, e.g., the simulations are 1.6 $\times$ slower with the interface compression and 5.6-times slower with the SR.
Because of the cost of these simulations, the finer grids are not simulated until completion with the SR and the interface compression.
As discussed next, the results are still considered sufficient to draw valuable conclusions.

\begin{figure}
   \begin{center}
     \subfigure{\includegraphics[width=4in]{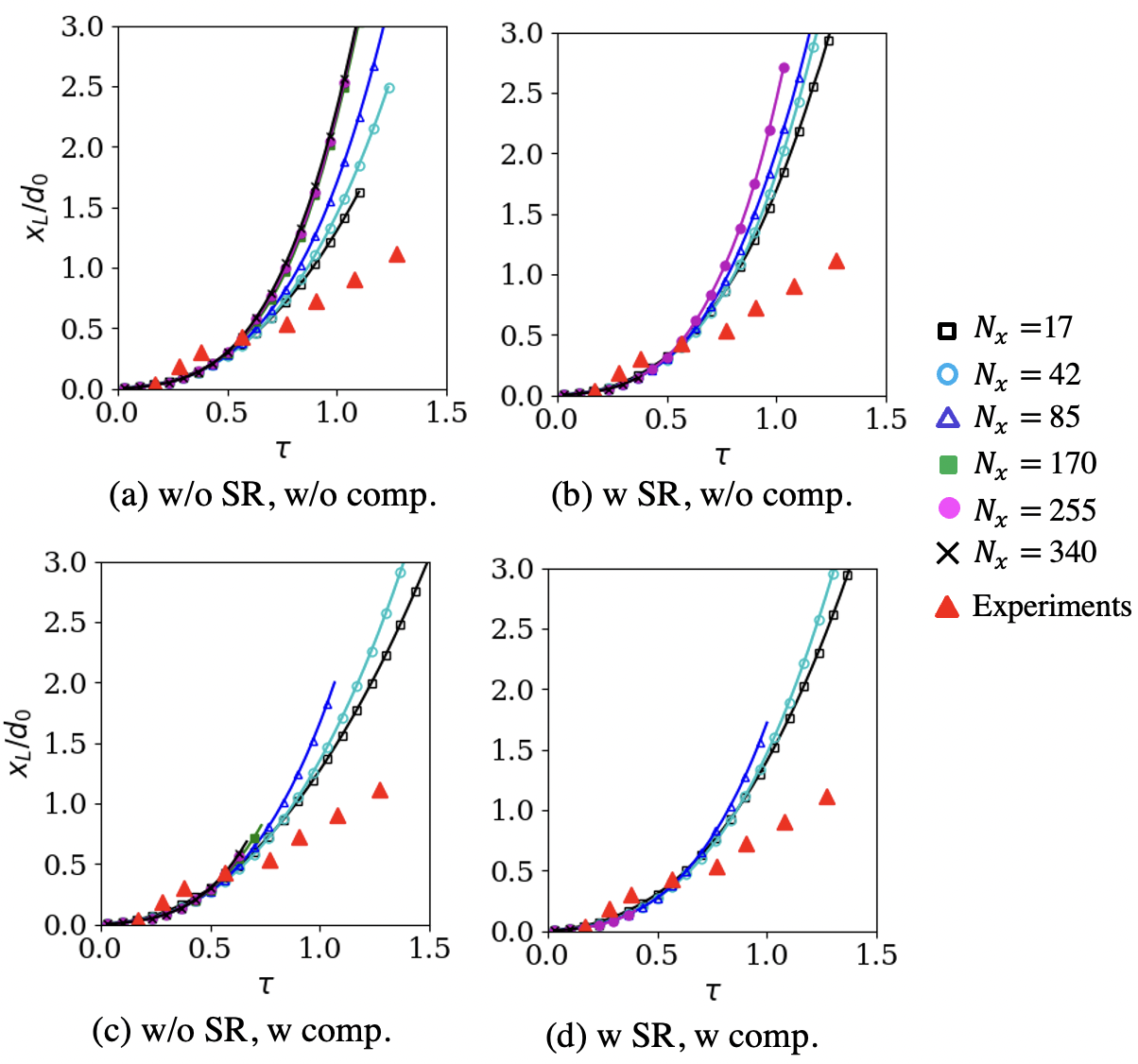}}
       \caption{Time-evolution of the droplet leading edge ($x_L$) compared against the experimental measurements~\cite{igra_as_2002}. }
   \label{fig:2d-shockx}
  \end{center}
\end{figure}

The droplet leading edge ($x_L/d_0$) evolution is shown in~\fig{2d-shockx}.
All grid resolutions show a similar trend, but a grid convergence is achieved for $N_x \ge 170$.
Stiff relaxation or interface compression does not result in a noticeable difference in leading-edge predictions.
The results match the experimental data for $\tau<0.7$, however,  the difference eventually grows to almost 100\% by $\tau=1.3$.
This difference could be due to the subsonic inflow boundary conditions used on the left boundary for the simulations to maintain the post-shock velocity. The experiments were conducted within a longer shock tube where We did not need this.

\begin{figure}
   \begin{center}
     \subfigure{\includegraphics[width=4in]{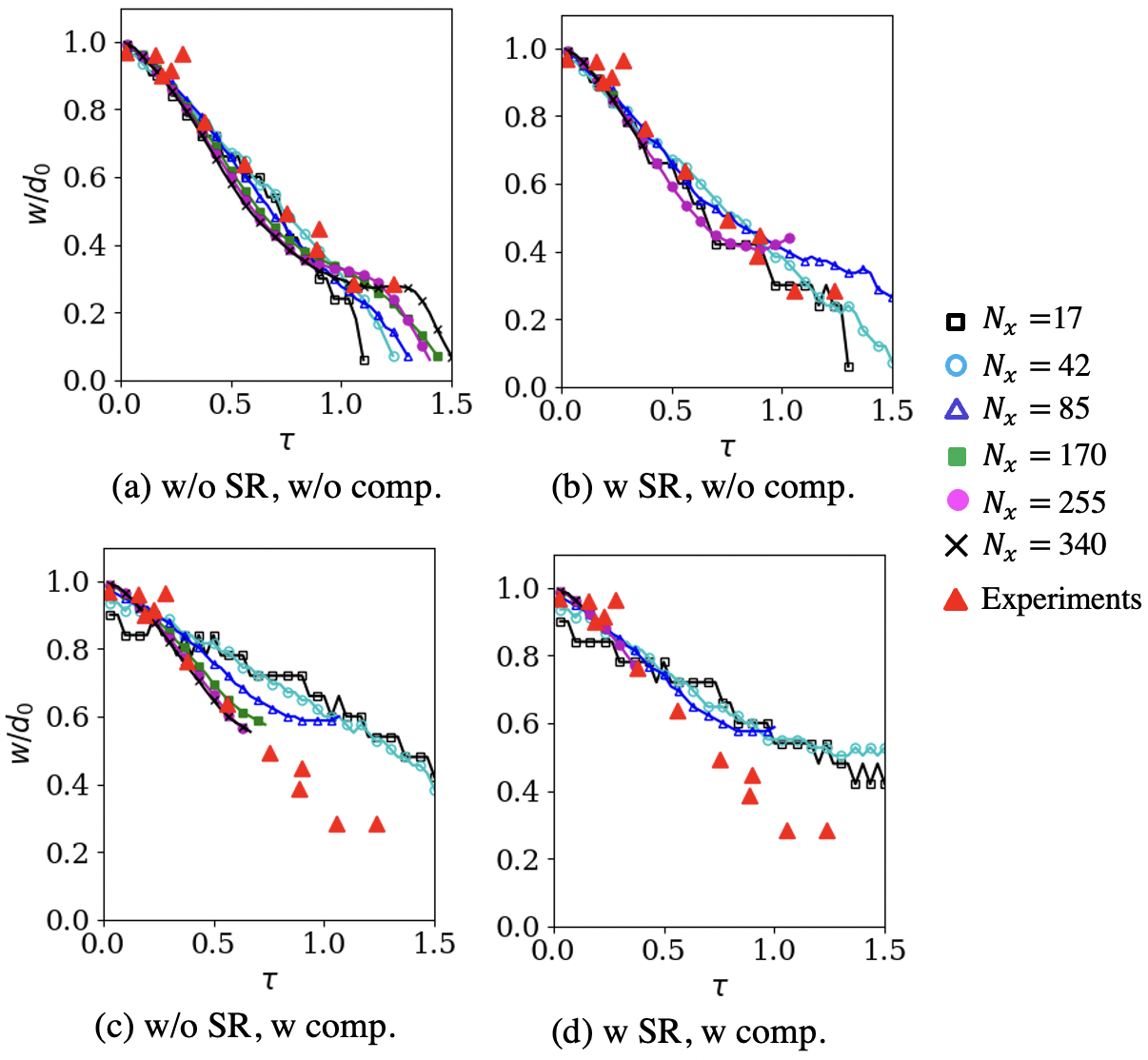}}
       \caption{Time-evolution of the droplet width ($d$) compared against the experimental measurements~\cite{igra_as_2002}. }
   \label{fig:2d-shockw}
  \end{center}
\end{figure}

\begin{figure}
   \begin{center}
     \subfigure{\includegraphics[width=4in]{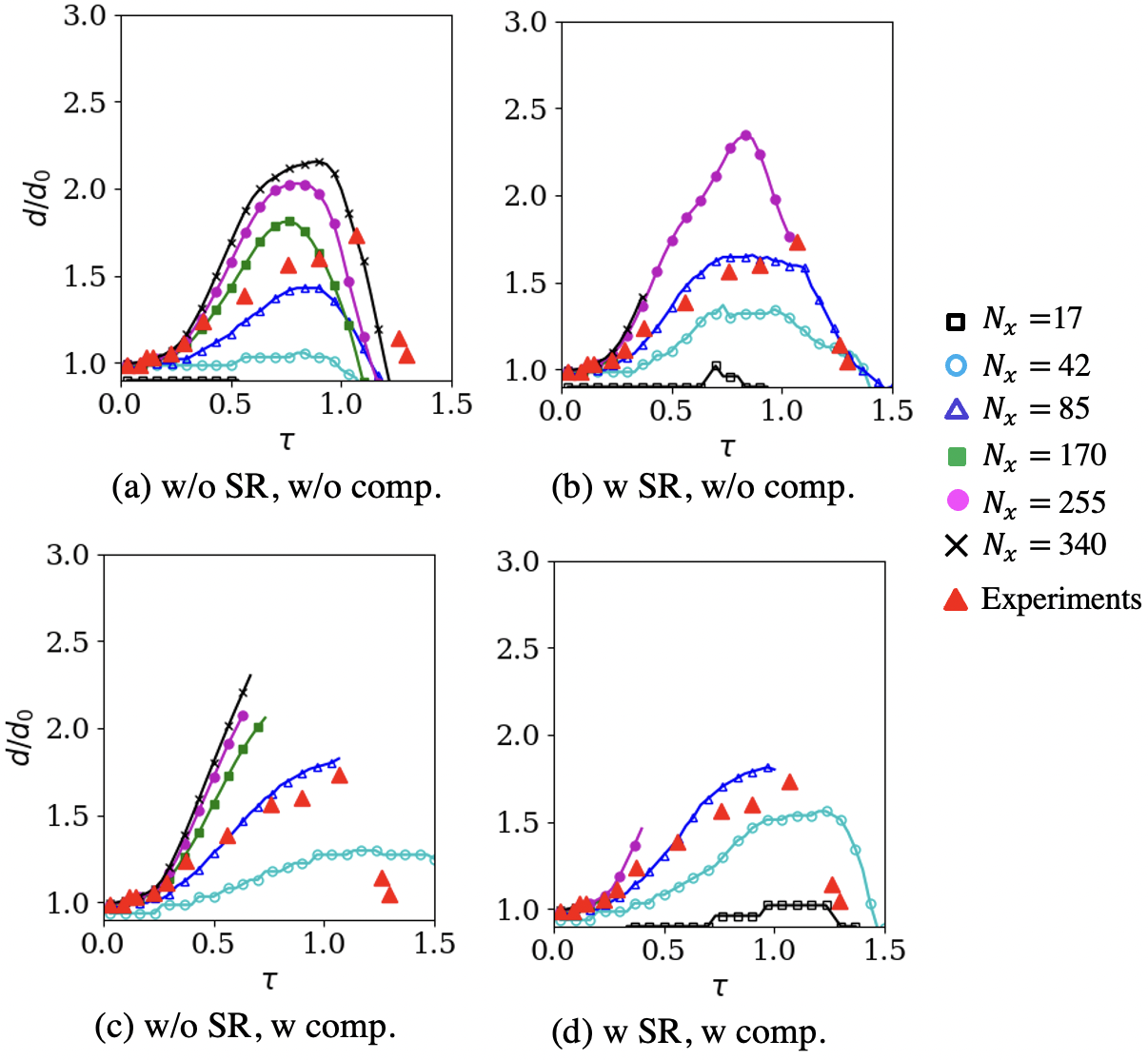}}
       \caption{Time-evolution of the droplet height ($d$) compared against the experimental measurements~\cite{igra_as_2002}. }
   \label{fig:2d-shockd}
  \end{center}
\end{figure}

\begin{figure}
   \begin{center}
     \subfigure{\includegraphics[width=4in]{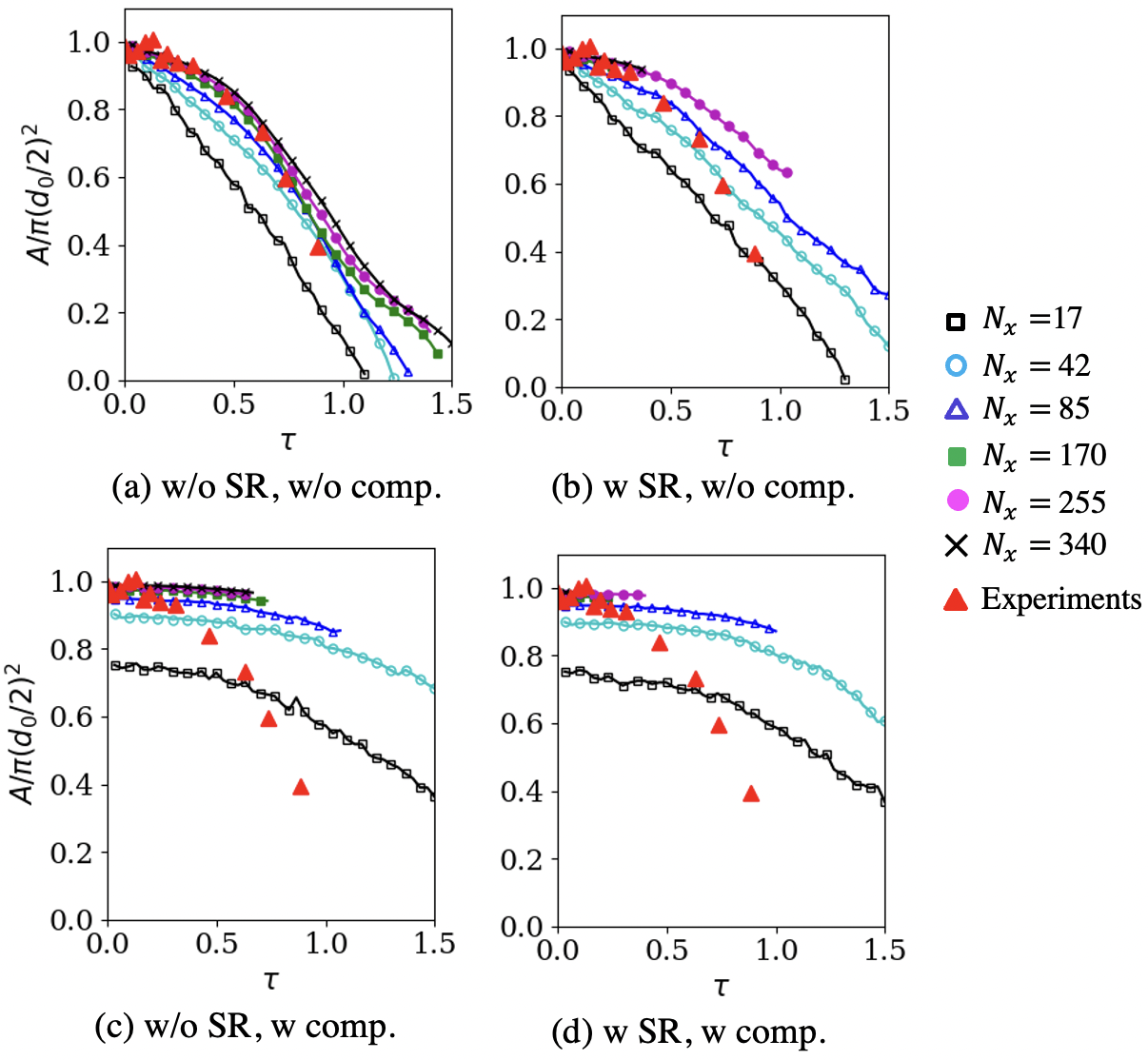}}
       \caption{Time-evolution of the droplet area ($A$) compared against the experimental measurements~\cite{igra_as_2002}. }
   \label{fig:2d-shocka}
  \end{center}
\end{figure}

The centerline droplet width ($w$) decreases with time due to the droplet compression, and the spanwise droplet diameter ($d$) increases as the droplet gets elongated.
These are plotted in~\fig{2d-shockw} and~\fig{2d-shockd}, respectively.
Without the SR and the interface compression, the predicted droplet width $w$ matches well against the data ($<$5\% error) for all grid resolutions.
On the other hand, the droplet height $d$ is grossly underpredicted at coarser resolutions, and a grid convergence is obtained only for $N_x \ge 255$.
The fine grid results capture the same trend as the experimental measurements~\cite{igra_as_2002}. However, they show about 30-40\% overprediction.
Similar errors have also been observed in earlier numerical studies using the five-equation DIM, but the reasons are unclear.
The droplet area is expected to reduce with time for the experiments due to the stripping of small-sized droplets.
The simulations capture this without the SR and the interface compression as shown in~\fig{2d-shocka}.
The coarsest grid shows up to 25\% underprediction. However, a grid convergence is obtained for $N_x\ge 170$, and the finer grids' results match the measurements with <5\% error.

Regarding the stiff relaxation solver usage, similar to the leading-edge droplet predictions,  the predictions for all  three geometrical features ($d$, $w$, $A$) are similar with/without the SR.
The most significant differences are observed later in $d$ and $A$ predictions, which are still $<$10-15\%.
This further confirms the efficacy of the DEM in alleviating the need to use a separate SR solver, which, as demonstrated for this case, could be as much as 5.6-times as  costly.

The effect of the interface compression on the prediction is widely apparent on $w$, $d$, and $A$.
In reality, the droplet area would reduce with time due to the stripping of small-sized droplets.
One could estimate the size of these stripped droplets as $\sim6\ \mu$m based on $We\sim1$.
It is not possible to resolve these small droplets with our computational grid, and in the DIM simulations considered here,  they would be represented as intermediate volume fractions ($\epsilon<\alpha^{(2)}<1-\epsilon$).
The interface compression approach that is used here can only work with the resolved fluids, and it wrongly interprets the regions with $\epsilon<\alpha^{(2)}<1-\epsilon$ as numerically diffused regions, applying the compression there.
As a result of this, the results with the interface compression are insensitive to the threshold volume fraction $\alpha^{(2)}_T$,  but they do not match the data.
Transition to a dispersed phase Eulerian--Eulerian and Eulerian--we can consider lagrangian modeling approaches such as~\cite{panchal_jcp_2021} to model these regions with $\epsilon<\alpha^{(2)}<1-\epsilon$, however, that is not the focus of this work, and it would be considered in the future.

\subsection{Detonation-droplet interaction}
\label{sec:2ddeton}

One could encounter a detonation-droplet interaction in practical applications such as a liquid-fueled or liquid-assisted rotating detonation engine (RDE)~\cite{salvadori_aiaa_2022, kindracki_2015} or an energetic material packed with metal particles~\cite{gogulya_2004}.
Some recent efforts have focused on studying this phenomenon experimentally~\cite{dyson_aiaa_2022}.
However, the simulations here serve as a qualitative demonstration without detailed measurements for quantitative validation.
 
A 2D computational domain of size $(1500,100)$ mm is used.
A liquid droplet ($\kappa=2$) of diameter $d_0$ mm is initialized at $(x,y)=(750, 50)$ mm.
A uniform grid with $N_x=100$ cells across the droplet diameter is used.
The left and the right boundaries are set as supersonic outflows to let the internal waves pass through without reflection.
The top and the bottom boundaries are set as slip walls.
The entire domain is at rest initially ($u^{(1)}=u^{(2)}=0$) and the pressure are set as $p^{(1)}=p^{(2)}=10^5$ Pa. The densities are $\rho^{(1)}=1.204$ kg/m$^3$, and $\rho^{(2)}=1000$ kg/m$^3$. The gas-phase ($\kappa=1$) contains gaseous $\rm{H_2}$ and air at stoichiometry to mimic conditions in a typical rotating detonation engine~\cite{salvadori_ijhe_2022}.  

In order to initiate a 2D detonation wave,  $x<10$ mm is initially filled with detonation products at $T^{(1)}=3500$ K and $p^{(1)}=5\times 10^6$ Pa. 
A circular disturbance of radius 1 mm is asymmetrically added in the domain to trigger the transverse detonation waves~\cite{kailasanath_1985}.The droplet volume fraction $\alpha^{(2)}$ is initialized using~\eq{circle_init} with $\delta=0.5$.  
The droplet diameter $d_0=40$ mm is relatively large compared to real applications. However, it is chosen here considering the cost and the grid requirements for resolving a droplet.

The liquid phase is modeled using an SG EOS with $\gamma^{(2)}=4.1$ and $p^{0}=6\times10^{8}$, and the gas phase is modeled using a TPG EOS that uses NASA polynomial curve-fits for the species.
The droplet would eventually vaporize because of the high temperature behind the detonation. However, for the current conditions, the ratio of the  droplet vaporization time ($t_{drop}$) to the detonation passage time ($t_{deton}$) is estimated to be substantial, $t_{drop}/t_{deton} \sim 10^{6} $, suggesting that one can simulate the early stages of this detonation-droplet interaction without using any heat/mass-transfer models.
Similar to the previous test in~\sect{2dshock}, the inertia primarily drives the droplet deformation at these high-speed conditions. Therefore, the viscous and the surface tension effects are neglected for this test.

\begin{figure}
   \begin{center}
   \subfigure{\includegraphics[width=6.5in]{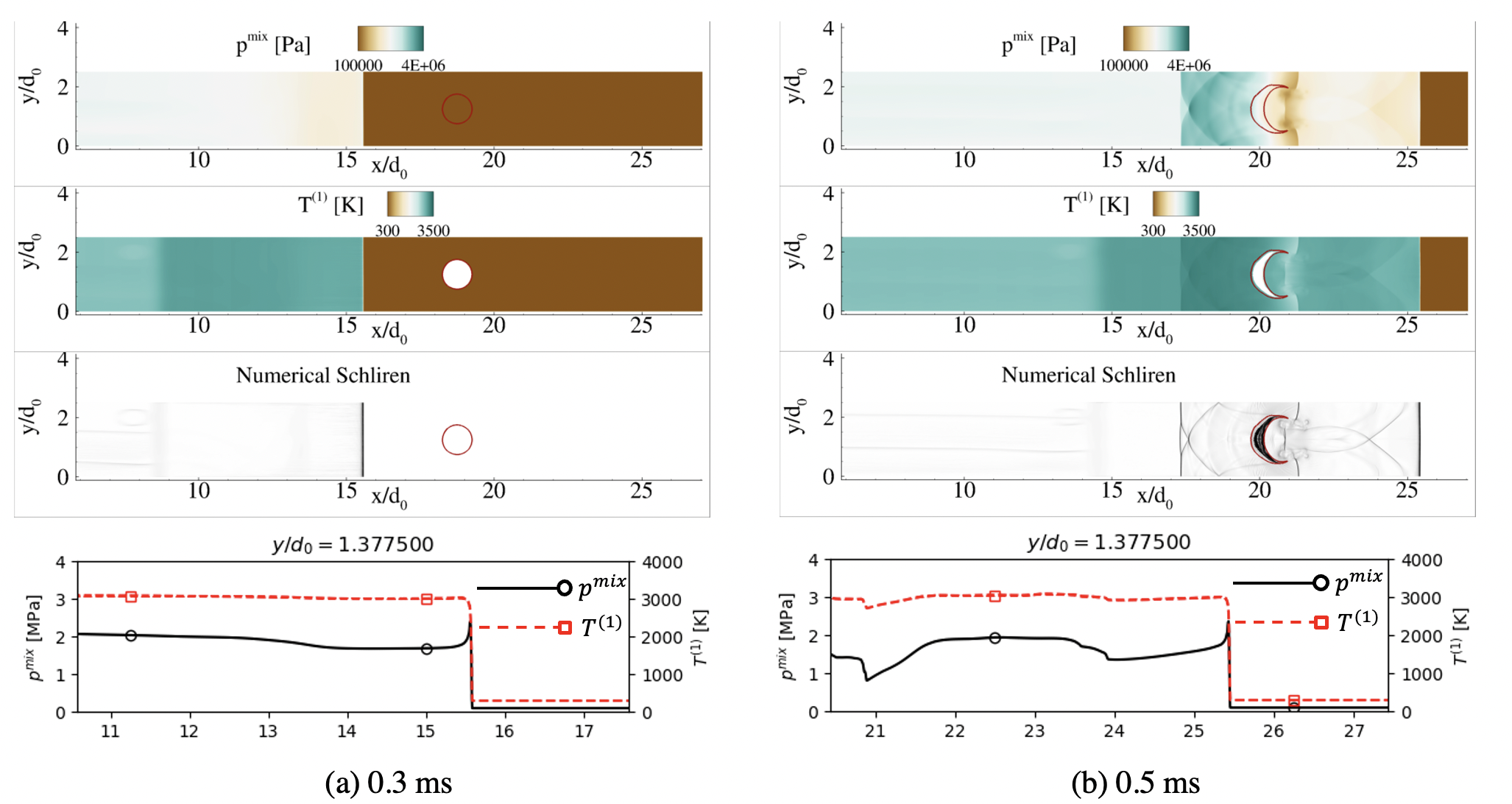}}
    \caption{Pressure contours (first row), temperature contours (second row),  numerical Schlieren (third row), and flow-field quantities along the centerline are shown for the 2D shock droplet interaction test. The droplet is identified using a $\alpha^{(2)}=0.5$ contour (thick red line).}
   \label{fig:2d-deton}
  \end{center}
\end{figure}

The simulation results at various time instances are shown in~\fig{2d-deton}. The results shown here are with the interface compression scheme and the SR. However, simulations without them have also been conducted.
As a result of the initiation, a 2D detonation wave propagates in the gas phase before reaching the droplet.
The structure of this detonation wave is confirmed by a pressure spike (of 2.55 MPa), an increase in the temperature (to 3078 K) due to coupled shock, and burning at the front.
The detonation speed before reaching the droplet is measured as 1968 m/s, which is extremely close to the Chapman–-Jouguet velocity of 1964 m/s computed using a Chemical Equilibrium Applications (CEA) code at the same conditions.

Similar to the shock-droplet interaction, as the detonation wave hits the droplet, it results in the transmission and reflection of non-planar waves.
A circular reflected shock is observed propagating back into the detonation products.
As the wave propagates after crossing the droplet, it eventually regains its reactive detonation structure.
The droplet remains circular at initial times. However, it deforms and elongates, as was also the case with the shock-droplet interaction.

\section{Conclusions}

A seven-equation diffused interface method (DIM) framework is developed in this work.
Unlike the commonly used reduced DIM models, the seven-equation model does not enforce a velocity/pressure equilibrium.
Alleviating a strict pressure relaxation allows this method for arbitrarily high pressure/density ratio and arbitrary EOS of the phases,  and alleviating the strict velocity equilibrium would enable it to be used in a dispersed phase regime  where the MPE are unresolved.
The seven-equation model has been used for simplistic 1D tests~\cite{abgrall_jcp_2003, saurel_jcp_1999}. However, it is extended to be used in 2D/3D with arbitrary EOS, surface tension, viscous effects, and multi-species reacting flows.
Various tests, i.e., multiphase shock tube, shock-material interaction, oscillating droplet, drag in a viscous medium, shock-droplet interaction, and detonation-droplet interaction, are conducted to validate different parts of the method.
Following are some novel contributions and conclusions of this work
\begin{itemize}
    \item A discrete equations method (DEM)~\cite{abgrall_jcp_2003} is used in this work to compute the non-conservative interphase exchange terms accurately.
    It is demonstrated via 1D and 2D tests that this relieves the need to use a stiff relaxation solver, which is otherwise required to maintain pressure--velocity equilibrium between the phases~\cite{saurel_jcp_1999} and could make the simulations as much as 5$\times$ slower.
    For the cases where the initial conditions are phase-specific, e.g., a shock-material interaction problem with the shock initialized in one of the phases,  the use of the pressure/velocities equilibrium at the start results in errors due to mismatching pre-/post-shock conditions between the phases.

    \item A stiff relaxation solver is still consistent with the previous works that used a seven-equation model, but certain modifications have been made.
    Adaptive-step gradient descent is used for pressure relaxation to make the scheme robust and useful for arbitrary EOS.
    The developed seven-equation framework is robust for arbitrarily  high density/pressure ratios ($\rho^{(2)}/\rho^{(1)}\sim1-10^{3}$,  $\rho^{(2)}/\rho^{(1)}\sim1-10^{5}$), and arbitrary EOS (SG, CPG, TPG, MG).

    \item To include surface tension, a modification is proposed in the inviscid fluxes of DEM and the corresponding HLLC Riemann solver. This proposed method can handle a surface-tension-related pressure jump across a material interface in an exact numerical manner.
    The pressure jump within a circular droplet is captured with $<2$\% error with numerically computed curvature.
    An oscillating droplet where the surface potential energy converts back and forth into the kinetic energy is simulated, and the oscillation period matches the exact value with a $<15$\% error.
    The subsequent peaks of kinetic energy reduce due to numerical diffusion. However, the results improve with grid refinement.

    \item Modifications for modeling 2D and 3D viscous effects within the DEM framework have been included.
    Droplet evolution in a viscous medium is simulated. The computed drag coefficient $C_D$ matches a steady-state drag correlation for a low Weber number case where the droplet does not deform.
    Vortex shedding is observed for a higher $Re$ case, but the droplet also deforms due to the higher $We$.

    \item An interface compression scheme initially used for the five-equation model is modified for the seven-equation model.
    The interface compression was necessary for the surface tension tests and the viscous drag computation. Otherwise, the droplet interface gets irregular, leading to difficulties in curvature computation and boundary and shear layer predictions surrounding the droplet surface.

    \item A shock-droplet interaction case is simulated at experimental conditions, and the computed droplet deformation parameters are shown to match against available data.
    This would create many small droplets due to the atomization process at such high $Re$ and $We$ in reality. However, since the grid does not resolve these, they manifest as regions of intermediate volume fractions ($\epsilon > \alpha^{(2)} > 1-\epsilon$) with DIM.
    As a result, the interface compression scheme was detrimental for this case since it would treat these regions as numerically diffused regions and apply compression.
    We can consider modeling these regions using a dispersed-phase approach in the future.

    \item Finally, the scheme has been demonstrated to work for multi-species, reacting flows via a detonation-droplet interaction problem.
    This test is a qualitative demonstration without any validation/verification data.
\end{itemize}

This work does not consider heat and mass transfer at the droplet interface, which is an obvious extension of this work.
Infinite~\cite{zein_jcp_2010, pelanti_ijmf_2019} or finite-rate~\cite{pelanti_2021} thermochemical relaxation solvers can be used.
A key benefit of the seven-equation model is its possible extension to dispersed-phase regions where an equilibrium between the velocities is not maintained.
It is shown in this work that with the use of DEM,  there is no need to enforce a stiff relaxation for resolved multiphase simulations, and the next step could be to use drag and heat and mass transfer laws for interphase exchange in the regions where there are unresolved droplets.
Such coupling of the seven-equation DIM with the dense-to-dilute dispersed model~\cite{panchal_jcp_2021} will be considered in the future.

\section*{Acknowledgment}

This work was supported by NASA Glenn Research Center [grant number NNX15AU91A]. 

\bibliographystyle{model1-num-names}
\bibliography{refs}

\begin{thebibliography}{87}
\expandafter\ifx\csname natexlab\endcsname\relax\def\natexlab#1{#1}\fi
\providecommand{\url}[1]{\texttt{#1}}
\providecommand{\href}[2]{#2}
\providecommand{\path}[1]{#1}
\providecommand{\DOIprefix}{doi:}
\providecommand{\ArXivprefix}{arXiv:}
\providecommand{\URLprefix}{URL: }
\providecommand{\Pubmedprefix}{pmid:}
\providecommand{\doi}[1]{\href{http://dx.doi.org/#1}{\path{#1}}}
\providecommand{\Pubmed}[1]{\href{pmid:#1}{\path{#1}}}
\providecommand{\bibinfo}[2]{#2}
\ifx\xfnm\relax \def\xfnm[#1]{\unskip,\space#1}\fi
\bibitem[{Lefebvre(1998)}]{lefebvre_1998}
\bibinfo{author}{A.~H. Lefebvre}, \bibinfo{title}{Gas turbine combustion},
  \bibinfo{publisher}{CRC press}, \bibinfo{year}{1998}.
\bibitem[{Sutton(2006)}]{sutton_2006}
\bibinfo{author}{G.~P. Sutton}, \bibinfo{title}{History of liquid propellant
  rocket engines}, \bibinfo{publisher}{AIAA}, \bibinfo{year}{2006}.
\bibitem[{Tarver et~al.(1996)Tarver, Chidester, and Nichols}]{tarver_1996}
\bibinfo{author}{C.~M. Tarver}, \bibinfo{author}{S.~K. Chidester},
  \bibinfo{author}{A.~L. Nichols},
\newblock \bibinfo{title}{Critical conditions for impact-and shock-induced hot
  spots in solid explosives},
\newblock \bibinfo{journal}{The Journal of Physical Chemistry}
  \bibinfo{volume}{100} (\bibinfo{year}{1996}) \bibinfo{pages}{5794--5799}.
\bibitem[{Vehring(2008)}]{vehring_2008}
\bibinfo{author}{R.~Vehring},
\newblock \bibinfo{title}{Pharmaceutical particle engineering via spray
  drying},
\newblock \bibinfo{journal}{Pharmaceutical research} \bibinfo{volume}{25}
  (\bibinfo{year}{2008}) \bibinfo{pages}{999--1022}.
\bibitem[{Gorokhovski and Herrmann(2008)}]{gorokhovski_annrev_2008}
\bibinfo{author}{M.~Gorokhovski}, \bibinfo{author}{M.~Herrmann},
\newblock \bibinfo{title}{Modeling primary atomization},
\newblock \bibinfo{journal}{Annu. Rev. Fluid Mech.} \bibinfo{volume}{40}
  (\bibinfo{year}{2008}) \bibinfo{pages}{343--366}.
\bibitem[{Saurel and Pantano(2018)}]{saurel_annrev_2018}
\bibinfo{author}{R.~Saurel}, \bibinfo{author}{C.~Pantano},
\newblock \bibinfo{title}{Diffuse-interface capturing methods for compressible
  two-phase flows},
\newblock \bibinfo{journal}{Annual Review of Fluid Mechanics}
  \bibinfo{volume}{50} (\bibinfo{year}{2018}) \bibinfo{pages}{105--130}.
\bibitem[{Balachandar and Eaton(2010)}]{balachandar_annrev_2010}
\bibinfo{author}{S.~Balachandar}, \bibinfo{author}{J.~K. Eaton},
\newblock \bibinfo{title}{Turbulent dispersed multiphase flow},
\newblock \bibinfo{journal}{Annual Review of Fluid Mechanics}
  \bibinfo{volume}{42} (\bibinfo{year}{2010}) \bibinfo{pages}{111--133}.
\bibitem[{Panchal and Menon(2021)}]{panchal_jcp_2021}
\bibinfo{author}{A.~Panchal}, \bibinfo{author}{S.~Menon},
\newblock \bibinfo{title}{A hybrid {Eulerian-Eulerian/Eulerian-Lagrangian}
  method for dense-to-dilute dispersed phase flows},
\newblock \bibinfo{journal}{Journal of Computational Physics}
  \bibinfo{volume}{439} (\bibinfo{year}{2021}) \bibinfo{pages}{110339}.
\bibitem[{Bryngelson et~al.(2019)Bryngelson, Schmidmayer, and
  Colonius}]{bryngelson19_IJMF}
\bibinfo{author}{S.~H. Bryngelson}, \bibinfo{author}{K.~Schmidmayer},
  \bibinfo{author}{T.~Colonius},
\newblock \bibinfo{title}{A quantitative comparison of phase-averaged models
  for bubbly, cavitating flows},
\newblock \bibinfo{journal}{International Journal of Multiphase Flow}
  \bibinfo{volume}{115} (\bibinfo{year}{2019}) \bibinfo{pages}{137--143}.
\bibitem[{Bryngelson et~al.(2021)Bryngelson, Fox, and Colonius}]{bryngelson21}
\bibinfo{author}{S.~H. Bryngelson}, \bibinfo{author}{R.~O. Fox},
  \bibinfo{author}{T.~Colonius},
\newblock \bibinfo{title}{Conditional moment methods for polydisperse
  cavitating flows},
\newblock \bibinfo{journal}{arXiv preprint arXiv: 2112.14172}
  (\bibinfo{year}{2021}).
\bibitem[{Sethian and Smereka(2003)}]{sethian_annrev_2003}
\bibinfo{author}{J.~A. Sethian}, \bibinfo{author}{P.~Smereka},
\newblock \bibinfo{title}{Level set methods for fluid interfaces},
\newblock \bibinfo{journal}{Annual Review of Fluid Mechanics}
  \bibinfo{volume}{35} (\bibinfo{year}{2003}) \bibinfo{pages}{341--372}.
\bibitem[{Sussman et~al.(2007)Sussman, Smith, Hussaini, Ohta, and
  Zhi-Wei}]{sussman_jcp_2007}
\bibinfo{author}{M.~Sussman}, \bibinfo{author}{K.~M. Smith},
  \bibinfo{author}{M.~Y. Hussaini}, \bibinfo{author}{M.~Ohta},
  \bibinfo{author}{R.~Zhi-Wei},
\newblock \bibinfo{title}{A sharp interface method for incompressible two-phase
  flows},
\newblock \bibinfo{journal}{Journal of computational physics}
  \bibinfo{volume}{221} (\bibinfo{year}{2007}) \bibinfo{pages}{469--505}.
\bibitem[{Hirt and Nichols(1981)}]{hirt_jcp_1981}
\bibinfo{author}{C.~W. Hirt}, \bibinfo{author}{B.~D. Nichols},
\newblock \bibinfo{title}{Volume of fluid {(VOF)} method for the dynamics of
  free boundaries},
\newblock \bibinfo{journal}{Journal of Computational Physics}
  \bibinfo{volume}{39} (\bibinfo{year}{1981}) \bibinfo{pages}{201--225}.
\bibitem[{Sussman and Puckett(2000)}]{sussman_jcp_2000}
\bibinfo{author}{M.~Sussman}, \bibinfo{author}{E.~G. Puckett},
\newblock \bibinfo{title}{A coupled level set and volume-of-fluid method for
  computing 3d and axisymmetric incompressible two-phase flows},
\newblock \bibinfo{journal}{Journal of Computational Physics}
  \bibinfo{volume}{162} (\bibinfo{year}{2000}) \bibinfo{pages}{301--337}.
\bibitem[{Brackbill et~al.(1992)Brackbill, Kothe, and
  Zemach}]{brackbill_jcp_1992}
\bibinfo{author}{J.~Brackbill}, \bibinfo{author}{D.~B. Kothe},
  \bibinfo{author}{C.~Zemach},
\newblock \bibinfo{title}{A continuum method for modeling surface tension},
\newblock \bibinfo{journal}{Journal of Computational Physics}
  \bibinfo{volume}{100} (\bibinfo{year}{1992}) \bibinfo{pages}{335--354}.
\bibitem[{Fedkiw et~al.(1999)Fedkiw, Aslam, Merriman, and
  Osher}]{fedkiw_jcp_1999}
\bibinfo{author}{R.~P. Fedkiw}, \bibinfo{author}{T.~Aslam},
  \bibinfo{author}{B.~Merriman}, \bibinfo{author}{S.~Osher},
\newblock \bibinfo{title}{A non-oscillatory eulerian approach to interfaces in
  multimaterial flows (the ghost fluid method)},
\newblock \bibinfo{journal}{Journal of Computational Physics}
  \bibinfo{volume}{152} (\bibinfo{year}{1999}) \bibinfo{pages}{457--492}.
\bibitem[{Elghobashi(2019)}]{elghobashi_annrev_2019}
\bibinfo{author}{S.~Elghobashi},
\newblock \bibinfo{title}{Direct numerical simulation of turbulent flows laden
  with droplets or bubbles},
\newblock \bibinfo{journal}{Annual Review of Fluid Mechanics}
  \bibinfo{volume}{51} (\bibinfo{year}{2019}) \bibinfo{pages}{217--244}.
\bibitem[{Das and Udaykumar(2020)}]{das_jcp_2020}
\bibinfo{author}{P.~Das}, \bibinfo{author}{H.~Udaykumar},
\newblock \bibinfo{title}{A sharp-interface method for the simulation of
  shock-induced vaporization of droplets},
\newblock \bibinfo{journal}{Journal of Computational Physics}
  \bibinfo{volume}{405} (\bibinfo{year}{2020}) \bibinfo{pages}{109005}.
\bibitem[{Terashima and Tryggvason(2009)}]{terashima_jcp_2009}
\bibinfo{author}{H.~Terashima}, \bibinfo{author}{G.~Tryggvason},
\newblock \bibinfo{title}{A front-tracking/ghost-fluid method for fluid
  interfaces in compressible flows},
\newblock \bibinfo{journal}{Journal of Computational Physics}
  \bibinfo{volume}{228} (\bibinfo{year}{2009}) \bibinfo{pages}{4012--4037}.
\bibitem[{Barton et~al.(2011)Barton, Obadia, and Drikakis}]{barton_jcp_2011}
\bibinfo{author}{P.~T. Barton}, \bibinfo{author}{B.~Obadia},
  \bibinfo{author}{D.~Drikakis},
\newblock \bibinfo{title}{A conservative level-set based method for
  compressible solid/fluid problems on fixed grids},
\newblock \bibinfo{journal}{Journal of Computational Physics}
  \bibinfo{volume}{230} (\bibinfo{year}{2011}) \bibinfo{pages}{7867--7890}.
\bibitem[{Baer and Nunziato(1986)}]{baer1986}
\bibinfo{author}{M.~Baer}, \bibinfo{author}{J.~Nunziato},
\newblock \bibinfo{title}{A two-phase mixture theory for the
  deflagration-to-detonation transition ({DDT}) in reactive granular
  materials},
\newblock \bibinfo{journal}{International journal of multiphase flow}
  \bibinfo{volume}{12} (\bibinfo{year}{1986}) \bibinfo{pages}{861--889}.
\bibitem[{Liu et~al.(1994)Liu, Osher, and Chan}]{liu_jcp_1994}
\bibinfo{author}{X.-D. Liu}, \bibinfo{author}{S.~Osher},
  \bibinfo{author}{T.~Chan},
\newblock \bibinfo{title}{Weighted essentially non-oscillatory schemes},
\newblock \bibinfo{journal}{Journal of computational physics}
  \bibinfo{volume}{115} (\bibinfo{year}{1994}) \bibinfo{pages}{200--212}.
\bibitem[{Van~Leer(1979)}]{van_jcp_1979}
\bibinfo{author}{B.~Van~Leer},
\newblock \bibinfo{title}{Towards the ultimate conservative difference scheme.
  {V. A} second-order sequel to {Godunov}'s method},
\newblock \bibinfo{journal}{Journal of Computational Physics}
  \bibinfo{volume}{32} (\bibinfo{year}{1979}) \bibinfo{pages}{101--136}.
\bibitem[{Petitpas et~al.(2009)Petitpas, Massoni, Saurel, Lapebie, and
  Munier}]{petitpas_ijmf_2009}
\bibinfo{author}{F.~Petitpas}, \bibinfo{author}{J.~Massoni},
  \bibinfo{author}{R.~Saurel}, \bibinfo{author}{E.~Lapebie},
  \bibinfo{author}{L.~Munier},
\newblock \bibinfo{title}{Diffuse interface model for high speed cavitating
  underwater systems},
\newblock \bibinfo{journal}{International Journal of Multiphase Flow}
  \bibinfo{volume}{35} (\bibinfo{year}{2009}) \bibinfo{pages}{747--759}.
\bibitem[{Schmidmayer et~al.(2020)Schmidmayer, Petitpas, Le~Martelot, and
  Daniel}]{schmidmayer_cpm_2020}
\bibinfo{author}{K.~Schmidmayer}, \bibinfo{author}{F.~Petitpas},
  \bibinfo{author}{S.~Le~Martelot}, \bibinfo{author}{{\'E}.~Daniel},
\newblock \bibinfo{title}{{ECOGEN}: An open-source tool for multiphase,
  compressible, multiphysics flows},
\newblock \bibinfo{journal}{Computer Physics Communications}
  \bibinfo{volume}{251} (\bibinfo{year}{2020}) \bibinfo{pages}{107093}.
\bibitem[{Saurel et~al.(2016)Saurel, Boivin, and
  Le~M{\'e}tayer}]{saurel_cf_2016}
\bibinfo{author}{R.~Saurel}, \bibinfo{author}{P.~Boivin},
  \bibinfo{author}{O.~Le~M{\'e}tayer},
\newblock \bibinfo{title}{A general formulation for cavitating, boiling and
  evaporating flows},
\newblock \bibinfo{journal}{Computers \& Fluids} \bibinfo{volume}{128}
  (\bibinfo{year}{2016}) \bibinfo{pages}{53--64}.
\bibitem[{Rodio and Abgrall(2015)}]{rodio_jcp_2015}
\bibinfo{author}{M.~G. Rodio}, \bibinfo{author}{R.~Abgrall},
\newblock \bibinfo{title}{An innovative phase transition modeling for
  reproducing cavitation through a five-equation model and theoretical
  generalization to six and seven-equation models},
\newblock \bibinfo{journal}{International Journal of Heat and Mass Transfer}
  \bibinfo{volume}{89} (\bibinfo{year}{2015}) \bibinfo{pages}{1386--1401}.
\bibitem[{Saurel and Abgrall(1999)}]{saurel_jcp_1999}
\bibinfo{author}{R.~Saurel}, \bibinfo{author}{R.~Abgrall},
\newblock \bibinfo{title}{A multiphase godunov method for compressible
  multifluid and multiphase flows},
\newblock \bibinfo{journal}{Journal of Computational Physics}
  \bibinfo{volume}{150} (\bibinfo{year}{1999}) \bibinfo{pages}{425--467}.
\bibitem[{Kapila et~al.(2001)Kapila, Menikoff, Bdzil, Son, and
  Stewart}]{kapila_pof_2001}
\bibinfo{author}{A.~Kapila}, \bibinfo{author}{R.~Menikoff},
  \bibinfo{author}{J.~Bdzil}, \bibinfo{author}{S.~Son}, \bibinfo{author}{D.~S.
  Stewart},
\newblock \bibinfo{title}{Two-phase modeling of deflagration-to-detonation
  transition in granular materials: Reduced equations},
\newblock \bibinfo{journal}{Physics of Fluids} \bibinfo{volume}{13}
  (\bibinfo{year}{2001}) \bibinfo{pages}{3002--3024}.
\bibitem[{Allaire et~al.(2002)Allaire, Clerc, and Kokh}]{allaire_jcp_2002}
\bibinfo{author}{G.~Allaire}, \bibinfo{author}{S.~Clerc},
  \bibinfo{author}{S.~Kokh},
\newblock \bibinfo{title}{A five-equation model for the simulation of
  interfaces between compressible fluids},
\newblock \bibinfo{journal}{Journal of Computational Physics}
  \bibinfo{volume}{181} (\bibinfo{year}{2002}) \bibinfo{pages}{577--616}.
\bibitem[{Meng and Colonius(2015)}]{meng_sw_2015}
\bibinfo{author}{J.~Meng}, \bibinfo{author}{T.~Colonius},
\newblock \bibinfo{title}{Numerical simulations of the early stages of
  high-speed droplet breakup},
\newblock \bibinfo{journal}{Shock Waves} \bibinfo{volume}{25}
  (\bibinfo{year}{2015}) \bibinfo{pages}{399--414}.
\bibitem[{Liu et~al.(2019)Liu, Wang, Sun, Deiterding, and Wang}]{liu_ast_2019}
\bibinfo{author}{N.~Liu}, \bibinfo{author}{Z.~Wang}, \bibinfo{author}{M.~Sun},
  \bibinfo{author}{R.~Deiterding}, \bibinfo{author}{H.~Wang},
\newblock \bibinfo{title}{Simulation of liquid jet primary breakup in a
  supersonic crossflow under adaptive mesh refinement framework},
\newblock \bibinfo{journal}{Aerospace Science and Technology}
  \bibinfo{volume}{91} (\bibinfo{year}{2019}) \bibinfo{pages}{456--473}.
\bibitem[{Murrone and Guillard(2005)}]{murrone_jcp_2005}
\bibinfo{author}{A.~Murrone}, \bibinfo{author}{H.~Guillard},
\newblock \bibinfo{title}{A five equation reduced model for compressible two
  phase flow problems},
\newblock \bibinfo{journal}{Journal of Computational Physics}
  \bibinfo{volume}{202} (\bibinfo{year}{2005}) \bibinfo{pages}{664--698}.
\bibitem[{Petitpas et~al.(2007)Petitpas, Franquet, Saurel, and
  Le~Metayer}]{petitpas_jcp_2007}
\bibinfo{author}{F.~Petitpas}, \bibinfo{author}{E.~Franquet},
  \bibinfo{author}{R.~Saurel}, \bibinfo{author}{O.~Le~Metayer},
\newblock \bibinfo{title}{A relaxation-projection method for compressible
  flows. {Part II}: Artificial heat exchanges for multiphase shocks},
\newblock \bibinfo{journal}{Journal of Computational Physics}
  \bibinfo{volume}{225} (\bibinfo{year}{2007}) \bibinfo{pages}{2214--2248}.
\bibitem[{Perigaud and Saurel(2005)}]{perigaud_jcp_2005}
\bibinfo{author}{G.~Perigaud}, \bibinfo{author}{R.~Saurel},
\newblock \bibinfo{title}{A compressible flow model with capillary effects},
\newblock \bibinfo{journal}{Journal of Computational Physics}
  \bibinfo{volume}{209} (\bibinfo{year}{2005}) \bibinfo{pages}{139--178}.
\bibitem[{Abgrall and Rodio(2014)}]{abgrall_caf_2014}
\bibinfo{author}{R.~Abgrall}, \bibinfo{author}{M.~G. Rodio},
\newblock \bibinfo{title}{Discrete equation method ({DEM}) for the simulation
  of viscous, compressible, two-phase flows},
\newblock \bibinfo{journal}{Computers \& Fluids} \bibinfo{volume}{91}
  (\bibinfo{year}{2014}) \bibinfo{pages}{164--181}.
\bibitem[{Saurel et~al.(2008)Saurel, Petipas, and Abgrall}]{saurel_jfm_2008}
\bibinfo{author}{R.~Saurel}, \bibinfo{author}{F.~Petipas},
  \bibinfo{author}{R.~Abgrall},
\newblock \bibinfo{title}{Modelling phase transition in metastable liquids:
  application to cavitating and flashing flows},
\newblock \bibinfo{journal}{Journal of Fluid Mechanics} \bibinfo{volume}{607}
  (\bibinfo{year}{2008}) \bibinfo{pages}{313–350}.
\bibitem[{Bryngelson et~al.(2021)Bryngelson, Schmidmayer, Coralic, Meng, Maeda,
  and Colonius}]{bryngelson_2021}
\bibinfo{author}{S.~H. Bryngelson}, \bibinfo{author}{K.~Schmidmayer},
  \bibinfo{author}{V.~Coralic}, \bibinfo{author}{J.~C. Meng},
  \bibinfo{author}{K.~Maeda}, \bibinfo{author}{T.~Colonius},
\newblock \bibinfo{title}{{MFC}: An open-source high-order multi-component,
  multi-phase, and multi-scale compressible flow solver},
\newblock \bibinfo{journal}{Computer Physics Communications}
  \bibinfo{volume}{266} (\bibinfo{year}{2021}) \bibinfo{pages}{107396}.
\bibitem[{Schmidmayer et~al.(2020)Schmidmayer, Bryngelson, and
  Colonius}]{schmidmayer_jcp_2020}
\bibinfo{author}{K.~Schmidmayer}, \bibinfo{author}{S.~H. Bryngelson},
  \bibinfo{author}{T.~Colonius},
\newblock \bibinfo{title}{An assessment of multicomponent flow models and
  interface capturing schemes for spherical bubble dynamics},
\newblock \bibinfo{journal}{Journal of Computational Physics}
  \bibinfo{volume}{402} (\bibinfo{year}{2020}) \bibinfo{pages}{109080}.
\bibitem[{Wood(1930)}]{wood_1930}
\bibinfo{author}{A.~B. Wood}, \bibinfo{title}{A Textbook of Sound},
  \bibinfo{publisher}{G. Bell and Sons Ltd.}, \bibinfo{year}{1930}.
\bibitem[{Saurel et~al.(2009)Saurel, Petitpas, and Berry}]{saurel_jcp_2009}
\bibinfo{author}{R.~Saurel}, \bibinfo{author}{F.~Petitpas},
  \bibinfo{author}{R.~A. Berry},
\newblock \bibinfo{title}{Simple and efficient relaxation methods for
  interfaces separating compressible fluids, cavitating flows and shocks in
  multiphase mixtures},
\newblock \bibinfo{journal}{Journal of Computational Physics}
  \bibinfo{volume}{228} (\bibinfo{year}{2009}) \bibinfo{pages}{1678--1712}.
\bibitem[{Zein et~al.(2010)Zein, Hantke, and Warnecke}]{zein_jcp_2010}
\bibinfo{author}{A.~Zein}, \bibinfo{author}{M.~Hantke},
  \bibinfo{author}{G.~Warnecke},
\newblock \bibinfo{title}{Modeling phase transition for compressible two-phase
  flows applied to metastable liquids},
\newblock \bibinfo{journal}{Journal of Computational Physics}
  \bibinfo{volume}{229} (\bibinfo{year}{2010}) \bibinfo{pages}{2964--2998}.
\bibitem[{Pelanti and Shyue(2019)}]{pelanti_ijmf_2019}
\bibinfo{author}{M.~Pelanti}, \bibinfo{author}{K.-M. Shyue},
\newblock \bibinfo{title}{A numerical model for multiphase liquid--vapor--gas
  flows with interfaces and cavitation},
\newblock \bibinfo{journal}{International journal of multiphase flow}
  \bibinfo{volume}{113} (\bibinfo{year}{2019}) \bibinfo{pages}{208--230}.
\bibitem[{Demou et~al.(2022)Demou, Scapin, Pelanti, and
  Brandt}]{demou_jcp_2022}
\bibinfo{author}{A.~D. Demou}, \bibinfo{author}{N.~Scapin},
  \bibinfo{author}{M.~Pelanti}, \bibinfo{author}{L.~Brandt},
\newblock \bibinfo{title}{A pressure-based diffuse interface method for
  low-mach multiphase flows with mass transfer},
\newblock \bibinfo{journal}{Journal of Computational Physics}
  \bibinfo{volume}{448} (\bibinfo{year}{2022}) \bibinfo{pages}{110730}.
\bibitem[{Rodriguez et~al.(2021)Rodriguez, Bryngelson, Cao, and
  Colonius}]{rodriguez21}
\bibinfo{author}{M.~Rodriguez}, \bibinfo{author}{S.~H. Bryngelson},
  \bibinfo{author}{S.~Cao}, \bibinfo{author}{T.~Colonius},
\newblock \bibinfo{title}{Acoustically-induced bubble growth and phase change
  dynamics near compliant surfaces},
\newblock in: \bibinfo{booktitle}{11th International Symposium on Cavitation},
  \bibinfo{year}{2021}.
\bibitem[{Dorschner et~al.(2020)Dorschner, Biasiori-Poulanges, Schmidmayer,
  El-Rabii, and Colonius}]{dorschner_jfm_2020}
\bibinfo{author}{B.~Dorschner}, \bibinfo{author}{L.~Biasiori-Poulanges},
  \bibinfo{author}{K.~Schmidmayer}, \bibinfo{author}{H.~El-Rabii},
  \bibinfo{author}{T.~Colonius},
\newblock \bibinfo{title}{On the formation and recurrent shedding of ligaments
  in droplet aerobreakup},
\newblock \bibinfo{journal}{Journal of Fluid Mechanics} \bibinfo{volume}{904}
  (\bibinfo{year}{2020}).
\bibitem[{Coralic and Colonius(2014)}]{coralic_jcp_2014}
\bibinfo{author}{V.~Coralic}, \bibinfo{author}{T.~Colonius},
\newblock \bibinfo{title}{Finite-volume weno scheme for viscous compressible
  multicomponent flows},
\newblock \bibinfo{journal}{Journal of Computational Physics}
  \bibinfo{volume}{274} (\bibinfo{year}{2014}) \bibinfo{pages}{95--121}.
\bibitem[{Abgrall and Saurel(2003)}]{abgrall_jcp_2003}
\bibinfo{author}{R.~Abgrall}, \bibinfo{author}{R.~Saurel},
\newblock \bibinfo{title}{Discrete equations for physical and numerical
  compressible multiphase mixtures},
\newblock \bibinfo{journal}{Journal of Computational Physics}
  \bibinfo{volume}{186} (\bibinfo{year}{2003}) \bibinfo{pages}{361--396}.
\bibitem[{Chang and Liou(2007)}]{chang_jcp_2007}
\bibinfo{author}{C.-H. Chang}, \bibinfo{author}{M.-S. Liou},
\newblock \bibinfo{title}{A robust and accurate approach to computing
  compressible multiphase flow: Stratified flow model and {AUSM+}-up scheme},
\newblock \bibinfo{journal}{Journal of Computational Physics}
  \bibinfo{volume}{225} (\bibinfo{year}{2007}) \bibinfo{pages}{840--873}.
\bibitem[{Tokareva and Toro(2010)}]{tokareva_jcp_2010}
\bibinfo{author}{S.~Tokareva}, \bibinfo{author}{E.~F. Toro},
\newblock \bibinfo{title}{{HLLC}-type riemann solver for the {Baer-Nunziato}
  equations of compressible two-phase flow},
\newblock \bibinfo{journal}{Journal of Computational Physics}
  \bibinfo{volume}{229} (\bibinfo{year}{2010}) \bibinfo{pages}{3573--3604}.
\bibitem[{Nguyen and Dumbser(2015)}]{nguyen_amc_2015}
\bibinfo{author}{N.~T. Nguyen}, \bibinfo{author}{M.~Dumbser},
\newblock \bibinfo{title}{A path-conservative finite volume scheme for
  compressible multi-phase flows with surface tension},
\newblock \bibinfo{journal}{Applied Mathematics and Computation}
  \bibinfo{volume}{271} (\bibinfo{year}{2015}) \bibinfo{pages}{959--978}.
\bibitem[{Panchal(2022)}]{panchal2022}
\bibinfo{author}{A.~Panchal}, \bibinfo{title}{Modeling moderately dense to
  dilute multiphase flows}, Ph.D. thesis, Georgia Institute of Technology,
  \bibinfo{year}{2022}.
\bibitem[{Dyson et~al.(2022)Dyson, Vasu, Arakelyan, Berube, Briggs, Ramirez,
  Ninnemann, Thurmond, Kim, Green et~al.}]{dyson_aiaa_2022}
\bibinfo{author}{D.~Dyson}, \bibinfo{author}{S.~Vasu},
  \bibinfo{author}{A.~Arakelyan}, \bibinfo{author}{N.~Berube},
  \bibinfo{author}{S.~Briggs}, \bibinfo{author}{J.~Ramirez},
  \bibinfo{author}{E.~M. Ninnemann}, \bibinfo{author}{K.~Thurmond},
  \bibinfo{author}{G.~Kim}, \bibinfo{author}{W.~H. Green}, et~al.,
\newblock \bibinfo{title}{Detonation wave-induced breakup and combustion of
  {RP-2} fuel droplets},
\newblock in: \bibinfo{booktitle}{AIAA SciTech 2022 Forum},
  \bibinfo{year}{2022}, pp. \bibinfo{pages}{AIAA--2022--1453}.
\bibitem[{Shukla et~al.(2010)Shukla, Pantano, and Freund}]{shukla_jcp_2010}
\bibinfo{author}{R.~K. Shukla}, \bibinfo{author}{C.~Pantano},
  \bibinfo{author}{J.~B. Freund},
\newblock \bibinfo{title}{An interface capturing method for the simulation of
  multi-phase compressible flows},
\newblock \bibinfo{journal}{Journal of Computational Physics}
  \bibinfo{volume}{229} (\bibinfo{year}{2010}) \bibinfo{pages}{7411--7439}.
\bibitem[{Dodd and Ferrante(2016)}]{dodd_jfm_2016}
\bibinfo{author}{M.~S. Dodd}, \bibinfo{author}{A.~Ferrante},
\newblock \bibinfo{title}{On the interaction of {Taylor} length scale size
  droplets and isotropic turbulence},
\newblock \bibinfo{journal}{Journal of Fluid Mechanics} \bibinfo{volume}{806}
  (\bibinfo{year}{2016}) \bibinfo{pages}{356--412}.
\bibitem[{Fechter et~al.(2017)Fechter, Munz, Rohde, and
  Zeiler}]{fechter_jcp_2017}
\bibinfo{author}{S.~Fechter}, \bibinfo{author}{C.-D. Munz},
  \bibinfo{author}{C.~Rohde}, \bibinfo{author}{C.~Zeiler},
\newblock \bibinfo{title}{A sharp interface method for compressible
  liquid--vapor flow with phase transition and surface tension},
\newblock \bibinfo{journal}{Journal of Computational Physics}
  \bibinfo{volume}{336} (\bibinfo{year}{2017}) \bibinfo{pages}{347--374}.
\bibitem[{Schmidmayer et~al.(2017)Schmidmayer, Petitpas, Daniel, Favrie, and
  Gavrilyuk}]{schmidmayer_jcp_2017}
\bibinfo{author}{K.~Schmidmayer}, \bibinfo{author}{F.~Petitpas},
  \bibinfo{author}{E.~Daniel}, \bibinfo{author}{N.~Favrie},
  \bibinfo{author}{S.~Gavrilyuk},
\newblock \bibinfo{title}{A model and numerical method for compressible flows
  with capillary effects},
\newblock \bibinfo{journal}{Journal of Computational Physics}
  \bibinfo{volume}{334} (\bibinfo{year}{2017}) \bibinfo{pages}{468--496}.
\bibitem[{Pelanti and Shyue(2014)}]{pelanti_jcp_2014}
\bibinfo{author}{M.~Pelanti}, \bibinfo{author}{K.-M. Shyue},
\newblock \bibinfo{title}{A mixture-energy-consistent six-equation two-phase
  numerical model for fluids with interfaces, cavitation and evaporation
  waves},
\newblock \bibinfo{journal}{Journal of Computational Physics}
  \bibinfo{volume}{259} (\bibinfo{year}{2014}) \bibinfo{pages}{331--357}.
\bibitem[{Tanguy et~al.(2007)Tanguy, M{\'e}nard, and
  Berlemont}]{tanguy_jcp_2007}
\bibinfo{author}{S.~Tanguy}, \bibinfo{author}{T.~M{\'e}nard},
  \bibinfo{author}{A.~Berlemont},
\newblock \bibinfo{title}{A level set method for vaporizing two-phase flows},
\newblock \bibinfo{journal}{Journal of Computational Physics}
  \bibinfo{volume}{221} (\bibinfo{year}{2007}) \bibinfo{pages}{837--853}.
\bibitem[{Jain et~al.(2020)Jain, Mani, and Moin}]{jain_jcp_2020}
\bibinfo{author}{S.~S. Jain}, \bibinfo{author}{A.~Mani},
  \bibinfo{author}{P.~Moin},
\newblock \bibinfo{title}{A conservative diffuse-interface method for
  compressible two-phase flows},
\newblock \bibinfo{journal}{Journal of Computational Physics}
  \bibinfo{volume}{418} (\bibinfo{year}{2020}) \bibinfo{pages}{109606}.
\bibitem[{Chiapolino et~al.(2017)Chiapolino, Saurel, and
  Nkonga}]{chiapolino_jcp_2017sharpening}
\bibinfo{author}{A.~Chiapolino}, \bibinfo{author}{R.~Saurel},
  \bibinfo{author}{B.~Nkonga},
\newblock \bibinfo{title}{Sharpening diffuse interfaces with compressible
  fluids on unstructured meshes},
\newblock \bibinfo{journal}{Journal of Computational Physics}
  \bibinfo{volume}{340} (\bibinfo{year}{2017}) \bibinfo{pages}{389--417}.
\bibitem[{Akiki et~al.(2017)Akiki, Gallagher, and Menon}]{akiki_aiaa_2017}
\bibinfo{author}{M.~Akiki}, \bibinfo{author}{T.~P. Gallagher},
  \bibinfo{author}{S.~Menon},
\newblock \bibinfo{title}{Mechanistic approach for simulating hot-spot
  formations and detonation in polymer-bonded explosives},
\newblock \bibinfo{journal}{AIAA Journal} \bibinfo{volume}{55}
  (\bibinfo{year}{2017}) \bibinfo{pages}{585--598}.
\bibitem[{Salvadori et~al.(2022)Salvadori, Tudisco, Ranjan, and
  Menon}]{salvadori_ijhe_2022}
\bibinfo{author}{M.~Salvadori}, \bibinfo{author}{P.~Tudisco},
  \bibinfo{author}{D.~Ranjan}, \bibinfo{author}{S.~Menon},
\newblock \bibinfo{title}{Numerical investigation of mass flow rate effects on
  multiplicity of detonation waves within a {H2/Air} rotating detonation
  combustor},
\newblock \bibinfo{journal}{International Journal of Hydrogen Energy}
  \bibinfo{volume}{47} (\bibinfo{year}{2022}) \bibinfo{pages}{4155--4170}.
\bibitem[{Baurle et~al.(1994)Baurle, Alexopoulos, and Hassan}]{baurle_jpp_1994}
\bibinfo{author}{R.~Baurle}, \bibinfo{author}{G.~Alexopoulos},
  \bibinfo{author}{H.~Hassan},
\newblock \bibinfo{title}{Assumed joint probability density function approach
  for supersonic turbulent combustion},
\newblock \bibinfo{journal}{Journal of Propulsion and Power}
  \bibinfo{volume}{10} (\bibinfo{year}{1994}) \bibinfo{pages}{473--484}.
\bibitem[{Patel and Menon(2008)}]{patel_2008}
\bibinfo{author}{N.~Patel}, \bibinfo{author}{S.~Menon},
\newblock \bibinfo{title}{Simulation of spray–turbulence–flame interactions
  in a lean direct injection combustor},
\newblock \bibinfo{journal}{Combustion and Flame} \bibinfo{volume}{153}
  (\bibinfo{year}{2008}) \bibinfo{pages}{228--257}.
\bibitem[{Balakrishnan et~al.(2010)Balakrishnan, Nance, and
  Menon}]{balakrishnan_2010}
\bibinfo{author}{K.~Balakrishnan}, \bibinfo{author}{D.~Nance},
  \bibinfo{author}{S.~Menon},
\newblock \bibinfo{title}{Simulation of impulse effects from explosive charges
  containing metal particles},
\newblock \bibinfo{journal}{Shock Waves} \bibinfo{volume}{20}
  (\bibinfo{year}{2010}) \bibinfo{pages}{217--239}.
\bibitem[{Toro(2009)}]{toro_2009}
\bibinfo{author}{E.~F. Toro},
\newblock \bibinfo{title}{The hll and hllc riemann solvers},
\newblock in: \bibinfo{booktitle}{Riemann solvers and numerical methods for
  fluid dynamics}, \bibinfo{publisher}{Springer}, \bibinfo{year}{2009}, pp.
  \bibinfo{pages}{315--344}.
\bibitem[{Schmidmayer et~al.(2021)Schmidmayer, Caz{\'e}, Petitpas, Daniel, and
  Favrie}]{schmidmayer_2021}
\bibinfo{author}{K.~Schmidmayer}, \bibinfo{author}{J.~Caz{\'e}},
  \bibinfo{author}{F.~Petitpas}, \bibinfo{author}{E.~Daniel},
  \bibinfo{author}{N.~Favrie}, \bibinfo{title}{{Modelling interactions between
  waves and diffused interfaces}}, \bibinfo{year}{2021}. \URLprefix
  \url{https://hal.archives-ouvertes.fr/hal-03387818}, \bibinfo{note}{working
  paper or preprint}.
\bibitem[{Shyue and Xiao(2014)}]{shyue_jcp_2014}
\bibinfo{author}{K.-M. Shyue}, \bibinfo{author}{F.~Xiao},
\newblock \bibinfo{title}{An eulerian interface sharpening algorithm for
  compressible two-phase flow: the algebraic thinc approach},
\newblock \bibinfo{journal}{Journal of Computational Physics}
  \bibinfo{volume}{268} (\bibinfo{year}{2014}) \bibinfo{pages}{326--354}.
\bibitem[{Gryngarten and Menon(2013)}]{gryngarten_2013}
\bibinfo{author}{L.~D. Gryngarten}, \bibinfo{author}{S.~Menon},
\newblock \bibinfo{title}{A generalized approach for sub-and super-critical
  flows using the local discontinuous galerkin method},
\newblock \bibinfo{journal}{Computer Methods in Applied Mechanics and
  Engineering} \bibinfo{volume}{253} (\bibinfo{year}{2013})
  \bibinfo{pages}{169--185}.
\bibitem[{Jain et~al.(2021)Jain, Adler, West, Mani, Moin, and
  Lele}]{jain_arxiv_2021}
\bibinfo{author}{S.~S. Jain}, \bibinfo{author}{M.~C. Adler},
  \bibinfo{author}{J.~R. West}, \bibinfo{author}{A.~Mani},
  \bibinfo{author}{P.~Moin}, \bibinfo{author}{S.~K. Lele},
\newblock \bibinfo{title}{Assessment of diffuse-interface methods for
  compressible multiphase fluid flows and elastic-plastic deformation in
  solids},
\newblock \bibinfo{journal}{arXiv preprint arXiv:2109.09729}
  (\bibinfo{year}{2021}).
\bibitem[{{Poinsot} and {Lele}(1992)}]{poinsotlele_1992}
\bibinfo{author}{T.~J. {Poinsot}}, \bibinfo{author}{S.~K. {Lele}},
\newblock \bibinfo{title}{{Boundary conditions for direct simulations of
  compressible viscous flows}},
\newblock \bibinfo{journal}{Journal of Computational Physics}
  \bibinfo{volume}{101} (\bibinfo{year}{1992}) \bibinfo{pages}{104--129}.
\bibitem[{Sridharan et~al.(2015)Sridharan, Jackson, Zhang, and
  Balachandar}]{sridharan_jap_2015}
\bibinfo{author}{P.~Sridharan}, \bibinfo{author}{T.~L. Jackson},
  \bibinfo{author}{J.~Zhang}, \bibinfo{author}{S.~Balachandar},
\newblock \bibinfo{title}{Shock interaction with one-dimensional array of
  particles in air},
\newblock \bibinfo{journal}{Journal of Applied Physics} \bibinfo{volume}{117}
  (\bibinfo{year}{2015}) \bibinfo{pages}{075902}.
\bibitem[{Ghia et~al.(1982)Ghia, Ghia, and Shin}]{ghia_jcp_1982}
\bibinfo{author}{U.~Ghia}, \bibinfo{author}{K.~Ghia},
  \bibinfo{author}{C.~Shin},
\newblock \bibinfo{title}{{High-Re} solutions for incompressible flow using the
  navier-stokes equations and a multigrid method},
\newblock \bibinfo{journal}{Journal of Computational Physics}
  \bibinfo{volume}{48} (\bibinfo{year}{1982}) \bibinfo{pages}{387--411}.
\bibitem[{White and Corfield(2006)}]{white_2006}
\bibinfo{author}{F.~M. White}, \bibinfo{author}{I.~Corfield},
  \bibinfo{title}{Viscous fluid flow}, volume~\bibinfo{volume}{3},
  \bibinfo{publisher}{McGraw-Hill New York}, \bibinfo{year}{2006}.
\bibitem[{Igra et~al.(2002)Igra, Ogawa, and Takayama}]{igra_as_2002}
\bibinfo{author}{D.~Igra}, \bibinfo{author}{T.~Ogawa},
  \bibinfo{author}{K.~Takayama},
\newblock \bibinfo{title}{A parametric study of water column deformation
  resulting from shock wave loading},
\newblock \bibinfo{journal}{Atomization and Sprays} \bibinfo{volume}{12}
  (\bibinfo{year}{2002}).
\bibitem[{Chen(2008)}]{chen_aiaa_2008}
\bibinfo{author}{H.~Chen},
\newblock \bibinfo{title}{Two-dimensional simulation of stripping breakup of a
  water droplet},
\newblock \bibinfo{journal}{AIAA Journal} \bibinfo{volume}{46}
  (\bibinfo{year}{2008}) \bibinfo{pages}{1135--1143}.
\bibitem[{Rayleigh et~al.(1879)}]{rayleigh_1879}
\bibinfo{author}{L.~Rayleigh}, et~al.,
\newblock \bibinfo{title}{On the capillary phenomena of jets},
\newblock \bibinfo{journal}{Proc. R. Soc. London} \bibinfo{volume}{29}
  (\bibinfo{year}{1879}) \bibinfo{pages}{71--97}.
\bibitem[{Gallagher et~al.(2017)Gallagher, Akiki, Menon, Sankaran, and
  Sankaran}]{gallagher_2017}
\bibinfo{author}{T.~Gallagher}, \bibinfo{author}{M.~Akiki},
  \bibinfo{author}{S.~Menon}, \bibinfo{author}{V.~Sankaran},
  \bibinfo{author}{V.~Sankaran},
\newblock \bibinfo{title}{Development of the generalized maccormack scheme and
  its extension to low mach number flows},
\newblock \bibinfo{journal}{International Journal Numerical Methods in Fluids}
  \bibinfo{volume}{85} (\bibinfo{year}{2017}) \bibinfo{pages}{165--188}.
\bibitem[{Patel(2007)}]{patel_thesis_2007}
\bibinfo{author}{N.~V. Patel}, \bibinfo{title}{Simulation of hydrodynamic
  fragmentation from a fundamental and an engineering perspective}, Ph.D.
  thesis, Georgia Institute of Technology, \bibinfo{year}{2007}.
\bibitem[{Temkin and Kim(1980)}]{temkin_jfm_1980}
\bibinfo{author}{S.~Temkin}, \bibinfo{author}{S.~S. Kim},
\newblock \bibinfo{title}{Droplet motion induced by weak shock waves},
\newblock \bibinfo{journal}{Journal of Fluid Mechanics} \bibinfo{volume}{96}
  (\bibinfo{year}{1980}) \bibinfo{pages}{133--157}.
\bibitem[{Pilch and Erdman(1987)}]{pilch_ijmf_1987}
\bibinfo{author}{M.~Pilch}, \bibinfo{author}{C.~Erdman},
\newblock \bibinfo{title}{Use of breakup time data and velocity history data to
  predict the maximum size of stable fragments for acceleration-induced breakup
  of a liquid drop},
\newblock \bibinfo{journal}{International Journal of Multiphase Flow}
  \bibinfo{volume}{13} (\bibinfo{year}{1987}) \bibinfo{pages}{741--757}.
\bibitem[{Salvadori et~al.(2022)Salvadori, Panchal, Ranjan, and
  Menon}]{salvadori_aiaa_2022}
\bibinfo{author}{M.~Salvadori}, \bibinfo{author}{A.~Panchal},
  \bibinfo{author}{D.~Ranjan}, \bibinfo{author}{S.~Menon},
\newblock \bibinfo{title}{Numerical study of detonation propagation in {H2-Air}
  with kerosene droplets},
\newblock in: \bibinfo{booktitle}{AIAA SciTech 2022 Forum},
  \bibinfo{year}{2022}, pp. \bibinfo{pages}{AIAA--2022--0394}.
\bibitem[{Kindracki(2015)}]{kindracki_2015}
\bibinfo{author}{J.~Kindracki},
\newblock \bibinfo{title}{Experimental research on rotating detonation in
  liquid fuel--gaseous air mixtures},
\newblock \bibinfo{journal}{Aerospace Science and Technology}
  \bibinfo{volume}{43} (\bibinfo{year}{2015}) \bibinfo{pages}{445--453}.
\bibitem[{Gogulya et~al.(2004)Gogulya, Makhov, Dolgoborodov, Brazhnikov,
  Arkhipov, and Shchetinin}]{gogulya_2004}
\bibinfo{author}{M.~Gogulya}, \bibinfo{author}{M.~Makhov},
  \bibinfo{author}{A.~Y. Dolgoborodov}, \bibinfo{author}{M.~Brazhnikov},
  \bibinfo{author}{V.~Arkhipov}, \bibinfo{author}{V.~Shchetinin},
\newblock \bibinfo{title}{Mechanical sensitivity and detonation parameters of
  aluminized explosives},
\newblock \bibinfo{journal}{Combustion, Explosion, and Shock Waves}
  \bibinfo{volume}{40} (\bibinfo{year}{2004}) \bibinfo{pages}{445--457}.
\bibitem[{Kailasanath et~al.(1985)Kailasanath, Oran, Boris, and
  Young}]{kailasanath_1985}
\bibinfo{author}{K.~Kailasanath}, \bibinfo{author}{E.~Oran},
  \bibinfo{author}{J.~Boris}, \bibinfo{author}{T.~Young},
\newblock \bibinfo{title}{Determination of detonation cell size and the role of
  transverse waves in two-dimensional detonations},
\newblock \bibinfo{journal}{Combustion and Flame} \bibinfo{volume}{61}
  (\bibinfo{year}{1985}) \bibinfo{pages}{199--209}.
\bibitem[{Pelanti(2021)}]{pelanti_2021}
\bibinfo{author}{M.~Pelanti},
\newblock \bibinfo{title}{Arbitrary-rate relaxation techniques for the
  numerical modeling of compressible two-phase flows with heat and mass
  transfer},
\newblock \bibinfo{journal}{arXiv preprint arXiv:2108.00556}
  (\bibinfo{year}{2021}).

\end{thebibliography}

\end{document}